\documentclass[reprint,aps,pre,twocolumn]{revtex4-2} 

\usepackage{graphicx}
\usepackage{amsmath,amssymb,bm}
\usepackage{siunitx}
\usepackage{xcolor}
\usepackage{booktabs,multirow}
\usepackage{hyperref}
\hypersetup{hidelinks}
\usepackage{float}
\usepackage{gensymb}
\usepackage{array}
\usepackage{tabularx}
\DeclareUnicodeCharacter{03C4}{\ensuremath{\tau}}
\usepackage{placeins}
\usepackage{MnSymbol}
\makeatletter
\newcommand\equalcontrib{\frontmatter@footnote{These authors contributed equally.}}
\makeatother
\usepackage[left]{lineno}
\newcommand{\mathsfi}[1]{\mathsf{#1}}
\renewcommand{\vec}[1]{\mathbf{#1}}
\newcommand{\grad}{\boldsymbol{\nabla}}

\renewcommand{\tensor}[1]{\boldsymbol{\mathsfi{#1}}}
\newcommand{\ucd}[1]{\overset{\kern0.25em\smalltriangledown}{#1}}
\newcommand{\gsd}[1]{\overset{\kern0.25em\smallsquare}{#1}}
\newcommand{\bdot}{\boldsymbol{\cdot}}

\begin{document}
\title{Learning constitutive models and rheology from partial flow measurements}

\author{Alp M. Sunol\equalcontrib}
\email{asunol@seas.harvard.edu}
\affiliation{John A. Paulson School of Engineering and Applied Sciences, Harvard University, Cambridge, MA 02138, USA}
\affiliation{School of Engineering, Brown University, Providence, RI 02912, USA}

\author{James V. Roggeveen\equalcontrib}
\email{roggeveen@seas.harvard.edu}
\affiliation{John A. Paulson School of Engineering and Applied Sciences, Harvard University, Cambridge, MA 02138, USA}

\author{Mohammed G. Alhashim}
\affiliation{Research and Development Center, Saudi Aramco, Dhahran 31311, Saudi Arabia}

\author{Henry S. Bae}
\affiliation{John A. Paulson School of Engineering and Applied Sciences, Harvard University, Cambridge, MA 02138, USA}
\affiliation{Department of Physics, Harvard University, Cambridge, MA 02138, USA}

\author{Michael P. Brenner}
\email{brenner@seas.harvard.edu} 
\affiliation{John A. Paulson School of Engineering and Applied Sciences, Harvard University, Cambridge, MA 02138, USA}
\affiliation{Department of Physics, Harvard University, Cambridge, MA 02138, USA}

\date{\today}

\begin{abstract}
Constitutive laws relate fluid stress to deformation and underpin predictions of non-Newtonian behavior in industrial and biological fluids. Standard characterization relies on measurements in idealized flows that often miss physics relevant to complex geometries. Existing data-driven methods overfit sparse data, lack geometry portability, or presuppose constitutive forms. To unify measurement and constitutive discovery, we developed an end-to-end framework that leverages automatic differentiation through a full physics simulation. By embedding a frame-invariant tensor basis neural network (TBNN) within a differentiable non-Newtonian solver, we learn  constitutive laws from any flow observable without presupposing a specific model, spanning generalized Newtonian, viscoelastic, and yield-stress behavior. Unlike coordinate-dependent methods, learning local material response enables accurate flow predictions in unseen geometries and conditions without retraining. We then distill  the TBNN closure into symbolic form via automated  model selection using the Bayesian Information Criterion, extracting interpretable physical parameters. This work establishes a foundation for comprehensive characterization of complex fluids directly within their operating environment (“digital rheometry”) with broad applicability to constitutive discovery across engineering and the physical sciences.
\end{abstract}

\maketitle


\onecolumngrid
\vspace{-0.75em}
\noindent\textbf{Keywords:} rheology; differentiable simulations; inverse problems; constitutive modeling; physics-informed machine learning; fluid mechanics

\vspace{0.35em}
\noindent\textbf{Author Contributions:} A.M.S., J.V.R., and M.P.B. designed research; A.M.S. developed the differentiable non-Newtonian fluid solver and performed the tensor basis neural network training and analysis; J.V.R. developed the differentiable model-fitting framework and performed the bulk rheometry analysis; M.G.A. and H.S.B. assisted with initial implementations and numerical validation; and A.M.S., J.V.R., and M.P.B. wrote the paper.
\vspace{0.75em}
\twocolumngrid

Predicting how a fluid flows through complex environments, from drug delivery in micro-vessels to industrial oil extraction, requires an accurate mathematical description of the material's internal response to deformation. Although the principles of mass and momentum conservation are universal, they are insufficient to predict flow without a constitutive law: a material-specific ``fingerprint'' that relates local deformation  history to internal stress \cite{larson1998structure,morrison2001understanding,verdier2009rheological, frigaard2017bingham,zhou2023mechanics} and supplies the missing stress closure. For simple fluids like water, this relationship is linear and instantaneous. However, most biological and industrial fluids are ``complex,'' exhibiting non-linear \cite{carreau1972rheological,yasuda1981shear,herschel1926konsistenzmessungen}, history-dependent \cite{bird1987dynamics1,Oldroyd1950Formulation, phan1977new,Giesekus1982JNNFM} behaviors that arise from their underlying and evolving microstructure \cite{marchal1987new,lunsmann1993finite,debae1994practical}.

The utility of a constitutive model is two-fold: it can provide insight into a material's internal physics, such as the transition to a ``jammed'' state \cite{liu1998jamming,ovarlez2010three}, or serve as a tool for engineering design and prediction. Regardless of the objective, describing a fluid in terms of any constitutive model requires a multi-step process: collecting experimental data, selecting a mathematical model form, and inferring the corresponding material parameters. Bayesian approaches, for example, have been developed to quantify the relative credibility of candidate linear viscoelastic models \cite{Ewoldt2015Bayesian}. Traditionally, this characterization relies on benchtop rheometers to measure a fluid's bulk (average) response in simplified, approximately one-dimensional flows. These idealized tests, conducted far from a fluid's operating environment, often provide the primary data used to calibrate the models required to predict behavior in more complex, real-world environments \cite{denn1990issues, garg2025emerging}.

In practice, however, this traditional workflow faces key challenges in identifying the appropriate constitutive law for a particular fluid. For example, the model typically used to characterize a viscoelastic fluid’s response to sliding or shear deformations fails to predict the fluid’s strong resistance to stretching in contractions \cite{rothstein1999extensional, james2009boger, boyko2021pressure, hinch2024fast}. Fundamentally, such discrepancies arise because multiple constitutive laws often yield nearly identical responses under standard test conditions, only to diverge dramatically when subjected to more complex kinematics \cite{song2020evaluating, davoodi2022similarities,john2024comparison}. For this reason, common rheological experiments, relying on averaged bulk measurements like standard frequency sweeps, frequently lack the information necessary to constrain parameters within the regimes relevant to mixed shear and extension, motivating alternative protocols and measurements \cite{dealy1995official,mckinley2002filament,ewoldt2008new, pipe2009microfluidic,haward2012optimized}. Thus, models that fit simple flow data perfectly often fail to capture the second-order effects essential for predicting real-world systems, and this ambiguity is compounded by selecting a model a priori and fitting its parameters to data, rather than utilizing data to select the optimal model.

These challenges have motivated a growing body of work that moves away from classical constitutive models toward data-driven approaches for representing complex fluid behavior. One class of these methods has focused on obtaining constitutive relationships directly from averaged rheometer data using machine learning. These approaches, including Rheological Universal Differential Equations (RUDEs) \cite{Lennon2023PNAS} and Rheology Informed Neural Networks (RhINNs) \cite{Saadat2022,Dabiri2023,Ahn2023,Mahmoudabadbozchelou2021,Mahmoudabadbozchelou2024}, replace components of classical constitutive laws with neural networks trained on experimental data, which are then used to predict stresses in unseen flow conditions. Parallel efforts aim to use sparse symbolic regression to discover simple, interpretable models directly from the data  \cite{shanbhag2024sparse,sato2025rheo}. While promising for interpolating within fitted regimes, these models may struggle to generalize when trained solely on low-dimensional, averaged rheometry data, which can lead to non-unique or physically ungrounded solutions. For neural approaches in particular, this challenge is compounded by the need to infer thousands of network weights from such limited scalar measurements.

Motivated by the information bottleneck of bulk measurements, recent approaches infer constitutive  behavior from spatially resolved observations in complex flows. For example, SIMPLE uses scattering along Lagrangian flow histories to learn reduced-order microstructural dynamics and an associated stress mapping \cite{young2023scattering}. To instead infer constitutive behavior directly from velocity fields, a   prominent strategy employs Physics-Informed Neural Networks (PINNs) \cite{raissi2019physics} to solve coupled flow and constitutive equations while assimilating experimental data \cite{tartakovsky2020physics,Reyes2021,Thakur2024ViscoelasticNet,simavilla2025hammering}. While PINNs are powerful tools for model discovery, their reliance on soft physics constraints and specific boundary conditions often renders them geometry-specific and difficult to transfer. Alternatively, adjoint-based inverse methods fit specific constitutive models, such as the Carreau model, to velocity data \cite{Kontogiannis2025JFMRapids}. While rigorous, these methods limit physical expressivity to the chosen model and require formulating and solving an adjoint problem for each constitutive model considered. Performing broad model selection across multiple candidates thus becomes computationally prohibitive in complex geometries.

\begin{figure*}[thb!] 
 \centering
\includegraphics[width=\textwidth]{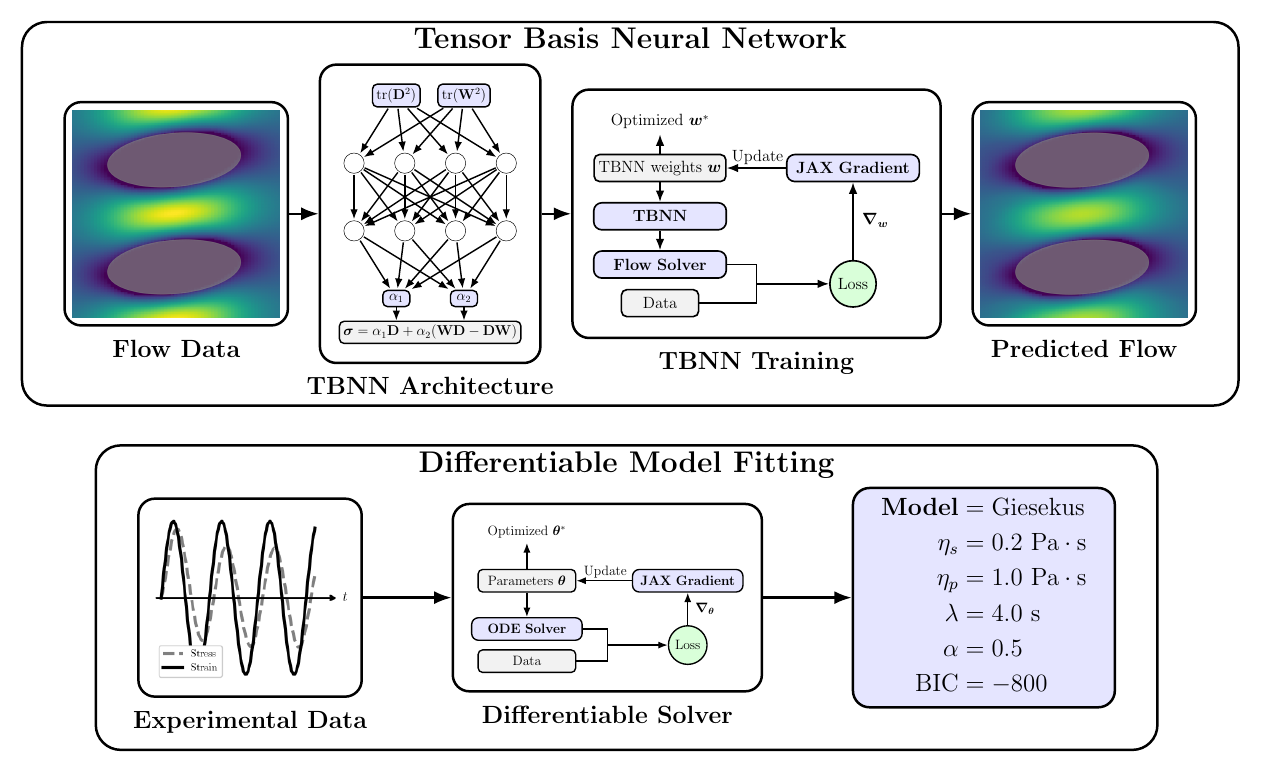}   \caption{\label{fig:system}\textbf{Learning rheological models from flow data.} \emph{Top:} A frame-invariant tensor basis neural network (TBNN) either maps scalar kinematic invariants directly to stress (instantaneous TBNN) or parameterizes the relaxation dynamics of an evolving conformation tensor (memory TBNN) within a differentiable flow solver implemented in JAX. Training on flow observations yields a TBNN closure that generalizes across conditions and geometries and enables flow prediction. \emph{Bottom:} For interpretability, we implement a differentiable ODE-based model-fitting framework that (i) fits classical constitutive laws (e.g.\ Giesekus) directly to rheometry time series and/or (ii) distills the learned TBNN response into an interpretable classical model; in both cases we select models and parameters using the Bayesian Information Criterion (BIC).}
\end{figure*}

To overcome these challenges, we present a novel approach that leverages automatic differentiation and differentiable simulations to learn fundamental constitutive relationships from local flow measurements (Fig.~\ref{fig:system}). Specifically, we develop a fully differentiable non-Newtonian fluid solver, building upon recent advances in differentiable computational fluid dynamics (CFD) \cite{kochkov2021machine, alhashim2025control}, by embedding   either an instantaneous or a memory tensor basis neural network (TBNN) \cite{ling2016reynolds} directly into the solver. Unlike black-box approaches, the TBNN maps scalar invariants in the flow directly to the stress or parameterizes the relaxation dynamics of an evolving conformation tensor, enforcing frame invariance and learning a geometry-agnostic material representation from the local kinematics that generalizes beyond the training geometry to predict flow in unseen environments. Further, this TBNN approach replaces the need to perform expensive inverse simulations on every possible constitutive model for a given fluid with a single training step.

Pairing this data-driven discovery with interpretability, we present a method that can efficiently fit families of physical constitutive laws. This second step borrows from RUDEs and RhINNs by fitting on a time-series of one-dimensional stresses like those in simplified rheometer flows to systematically identify the best model and its parameters. We first use this approach to interrogate the learned TBNN closures: by fitting classical models to  each closure's stress response, we identify the constitutive model and its physical parameters that best describe the fluid. We then further demonstrate this framework for direct interpretation of rheometer data, where we systematically identify the best model for a given set of data using the Bayesian Information Criterion (BIC), which captures the trade-off between model complexity and predictive power. To our knowledge, this is the first demonstration of automatic differentiation to fit and select among multiple classical constitutive families from rheometry time series using an information criterion.

Ultimately, our two-part framework turns complex flow measurements into  transferable constitutive closures and enables principled model selection, bridging the gap between data-driven flexibility and physically interpretable rheology for  characterization directly within a fluid's operating environment.

\section*{Learning instantaneous rheological models from velocimetry}

\begin{figure*}[!t]
  \centering
  \includegraphics[width=\textwidth]{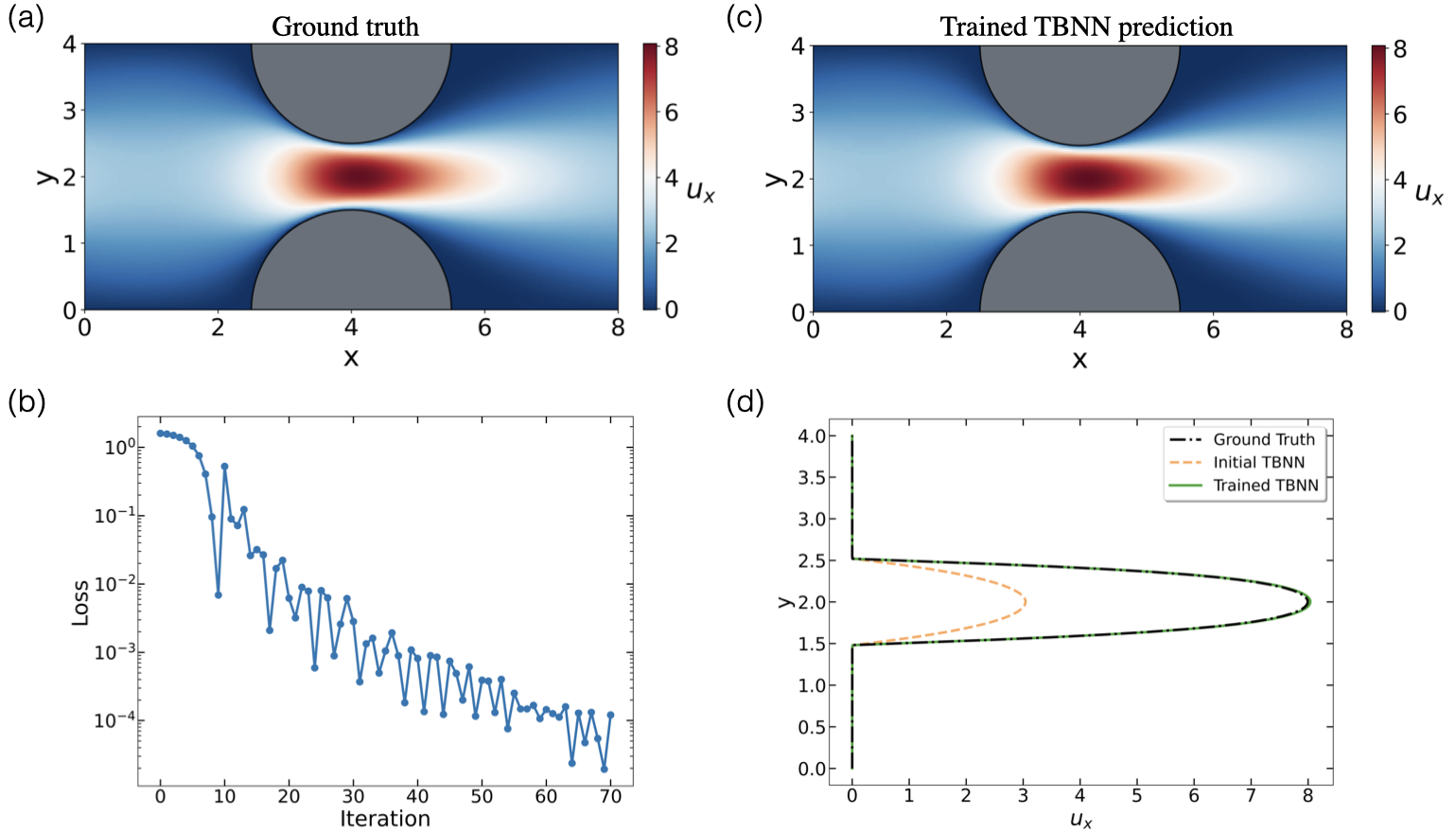}
    \caption{\textbf{Learning an instantaneous tensor basis neural network closure for stress.} (a) Ground-truth steady-state \(x\)-velocity for pressure-driven flow through a constriction (pressure gradient \(G=5\)). (b) Training loss versus iteration. (c) Steady-state \(x\)-velocity predicted by the simulation with the trained TBNN. (d) \(x\)-velocity at the constriction throat (\(x=4\)).}  \label{fig:tbnn_learning}
\end{figure*}

Learning constitutive behavior directly from complex flow fields requires a formulation that exposes how changes in material parameters propagate throughout the entire fluid domain. To this end, we developed a differentiable non-Newtonian flow solver capable of handling spatially varying viscosities (see Methods Sec.~\ref{sec:diff_non_newtonian_cfd}). Built within an automatic-differentiation framework, the solver provides exact gradients of any flow observable with respect to model parameters, enabling end-to-end optimization through the full simulation. Our immersed-boundary formulation \cite{alhashim2025control} allows us to simulate arbitrary geometries and boundary conditions, enabling training directly on complex or experimentally reconstructed flow domains. This differentiable formulation thus forms the foundation for data-driven inference of constitutive laws from flow data in complex and experimentally relevant geometries.

In principle, one could recover constitutive behavior by selecting a candidate model, embedding it in the differentiable flow solver, and optimizing its parameters to match the observed flow. However, testing each model separately would require repeated expensive gradient-based simulations and is therefore computationally inefficient. Instead, we employ a tensor basis neural network (TBNN) to learn the stress--strain-rate relationship directly from data. We first use an instantaneous TBNN, which, as detailed in Methods Sec.~\ref{sec:diff_non_newtonian_cfd}, encodes Galilean invariance by construction and, when supplied with the full invariant set and tensor bases, can represent any physically admissible stress  that is an algebraic function of the instantaneous kinematics (this restriction is lifted in the \hyperref[sec:memory]{memory-dependent section}). Thus, it  serves as a geometry-agnostic constitutive representation that can be embedded in any simulation to reproduce the measured flow behavior. Conceptually, this process is illustrated in the top panel of Fig.~\ref{fig:system}, where the TBNN learns from flow data to yield a transferable constitutive model.

To demonstrate this framework, we first consider a canonical shear-thinning Carreau--Yasuda (CY) model, a standard benchmark in non-Newtonian rheology, widely used to represent the smooth shear-thinning behavior characteristic of polymer solutions and biological fluids \cite{carreau1972rheological, yasuda1981shear, boyd2007analysis, lynch2022effects}. We use a fluid with parameters $\eta_0 = 1.0$, $\eta_\infty = 0.02$, $k = 5.0$, $n = 0.7$, and $a = 2.0$ as a baseline for validating the framework before relaxing these parameters and extending to other constitutive families. Rather than relying on simple shear, training is performed on an information-rich pressure-driven flow that spans orders of magnitude in local strain rate and thus a wide range of viscosities (Fig.~\ref{fig:sfig1}). Specifically, we simulate flow through a constriction--expansion channel containing a semi-circular obstacle under a nondimensional pressure gradient of $G = 5$ (see Methods Sec.~\ref{sec:diff_non_newtonian_cfd} for unit definitions). The resulting steady-state velocity field, shown in Fig.~\ref{fig:tbnn_learning}a, exhibits strong spatial variations in shear rate that provide an ideal dataset for learning a generalized constitutive relation.

We  train the TBNN on this dataset to learn the underlying stress law directly from the flow,
where  the neural network is formulated to predict a viscosity field as a function of local invariants (Methods Sec.~\ref{sec:tbnn_learning_eval}). The model is trained using the differentiable flow solver described in Methods Sec.~\ref{sec:diff_non_newtonian_cfd}, with gradients evaluated only after the flow reached steady state to minimize the loss between predicted and ground-truth velocity fields. During training, the loss decreases by more than four orders of magnitude (Fig.~\ref{fig:tbnn_learning}b), indicating convergence to a consistent constitutive representation. The velocity field reconstructed with the trained TBNN (Fig.~\ref{fig:tbnn_learning}c, \ref{fig:sfig2}) is visually indistinguishable from the ground truth, with the axial velocity profile at the constriction throat ($x = 4$) showing near-perfect agreement. These results demonstrate that a geometry-agnostic, invariant neural representation can recover the correct constitutive mapping purely from complex flow data.

\begin{figure}[t]
  \centering
  \includegraphics[width=\linewidth]{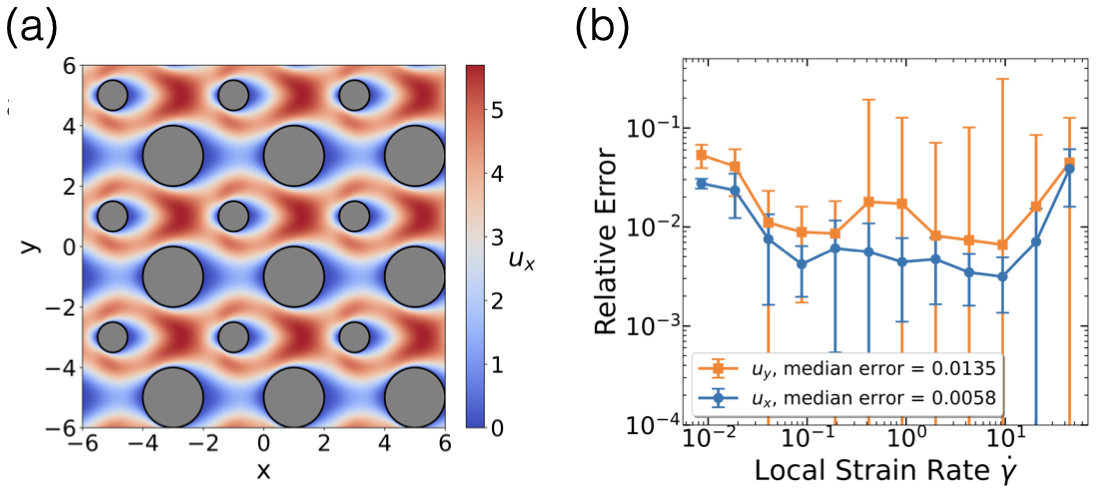}
    \caption{\textbf{Flow prediction in an unseen geometry.} (a) Steady-state \(x\)-velocity prediction for pressure-driven flow in a bidisperse porous medium with \(G = 7.5\). (b) Relative error compared to ground truth, binned as a function of local strain rate.}  \label{fig:diff_geometry}
\end{figure}

To evaluate whether the trained TBNN acts as a transferable constitutive model, we next test its ability to predict flows in conditions outside the training domain. We embed the trained model and the corresponding ground-truth CY fluid in a new geometry and increase the nondimensional pressure gradient to $G = 7.5$ (Methods Sec.~\ref{sec:diff_non_newtonian_cfd}). The TBNN-predicted velocity field (Fig.~\ref{fig:diff_geometry}a) closely matches the ground-truth simulation (Fig.~\ref{fig:sfig3}). Fig.~\ref{fig:diff_geometry}b shows the relative error (\ref{eq:strain_rate_relerr}) between the predicted and true velocity fields binned by local strain rate (Methods Sec.~\ref{sec:tbnn_learning_eval}), demonstrating  uniformly low error across all strain rates, with the smallest deviations in the range $10^{-1}<\dot{\gamma}<10^{1}$, where the training signal is strongest and most abundant.

We also verify that these results are robust to noise and degraded resolution (Methods Sec.~\ref{sec:tbnn_learning_eval}), demonstrating that accuracy is maintained even with an over tenfold decrease in resolution (Fig.~\ref{fig:sfig4}) and with correlated, spatially varying (heteroskedastic) velocity-field noise as high as 4\% of the 95th-percentile flow velocity (Figs.~\ref{fig:sfig5}--\ref{fig:sfig6}). Despite these degradations, the trained TBNN recovers the correct flow structure and constitutive mapping with low relative error, underscoring that a differentiable fluid solver provides strong physical regularization and enables learning even from coarse or noisy data (see \hyperref[sec:SI]{Supplementary Information} for detailed discussion).

\begin{figure}[t]
  \centering
  \includegraphics[width=\linewidth]{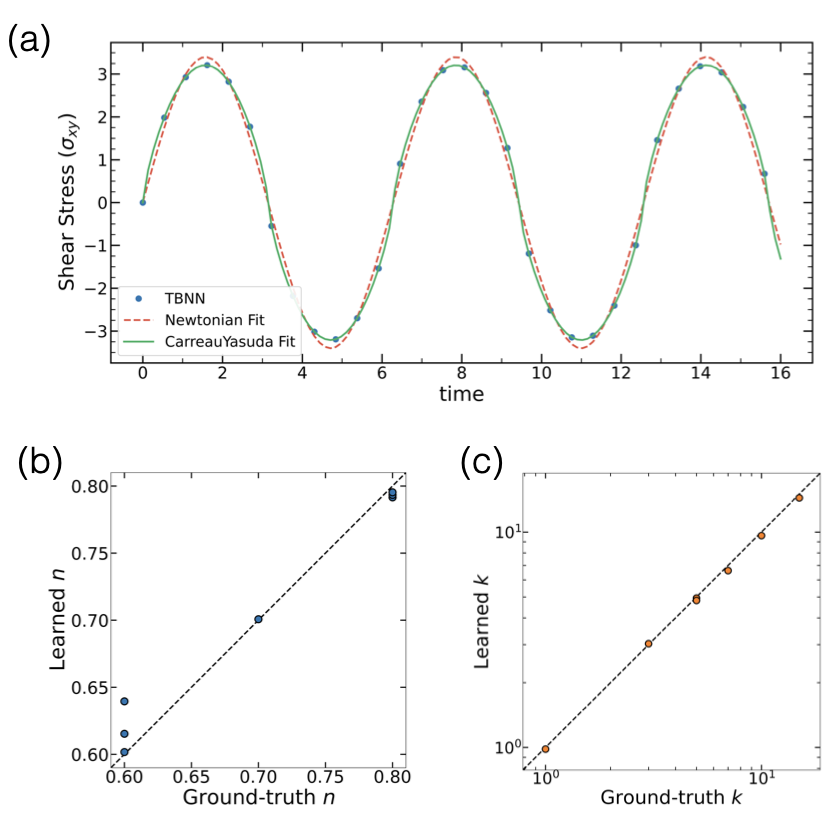}
    \caption{\textbf{Extracted Carreau--Yasuda parameters from an instantaneous TBNN.} (a) Representative oscillatory forcing: TBNN shear-stress output (points) with the best-fit Newtonian response (dashed) and Carreau--Yasuda (CY) response (solid) fit to the same trace; the CY model follows the waveform closely while the Newtonian fit misses the extrema. (b) Parity plot of the shear-thinning exponent \(n\) learned from the TBNN versus ground truth across seven runs; dashed line indicates \(y=x\). (c) Parity for the onset timescale \(k\) on log--log axes; points fall on the identity over more than an order of magnitude. In these runs \(\eta_0\) was fixed, and \(a\) and \(\eta_\infty\) are weakly constrained; numerical values for all parameters, the best-fit Newtonian viscosity, and model comparison statistics are given in Table~\ref{tab:tbnn-cy-extraction}. Across all runs the CY model is very strongly favored by BIC (\(\Delta\mathrm{BIC}=\mathrm{BIC}_{\mathrm{N}}-\mathrm{BIC}_{\mathrm{CY}}\gg 0\)).}
    \label{fig:cy_params_tbnn}
\end{figure}

\section*{Model interpretation and discovery from the TBNN}
Beyond predictive accuracy, a key advantage of our framework is interpretability: the trained TBNN can be interrogated to reveal the constitutive behavior it has learned and to identify classical models that best describe it. To accomplish this, we employ a differentiable model-fitting method (Fig.~\ref{fig:system}, bottom panel) that probes the trained TBNN under controlled deformation histories. In this setup, we apply prescribed kinematic forcings and compute the corresponding stress response predicted by the TBNN. Because this procedure evaluates the constitutive mapping directly without solving a full flow problem, it enables rapid exploration of parameter space and extrapolation to flow conditions beyond those used in training. We then fit a library of standard rheological models to the TBNN-generated stress data and select the one that provides the best statistical agreement using the Bayesian Information Criterion (BIC), which naturally penalizes model complexity: among models that fit the data comparably, BIC favors the most parsimonious. This provides a principled alternative to selecting a constitutive model a priori, instead letting the data determine both the model form and its parameters.

\begin{figure*}[!t]
    \centering
    \includegraphics[
        width=\textwidth,
        height=0.82\textheight,
        keepaspectratio
    ]{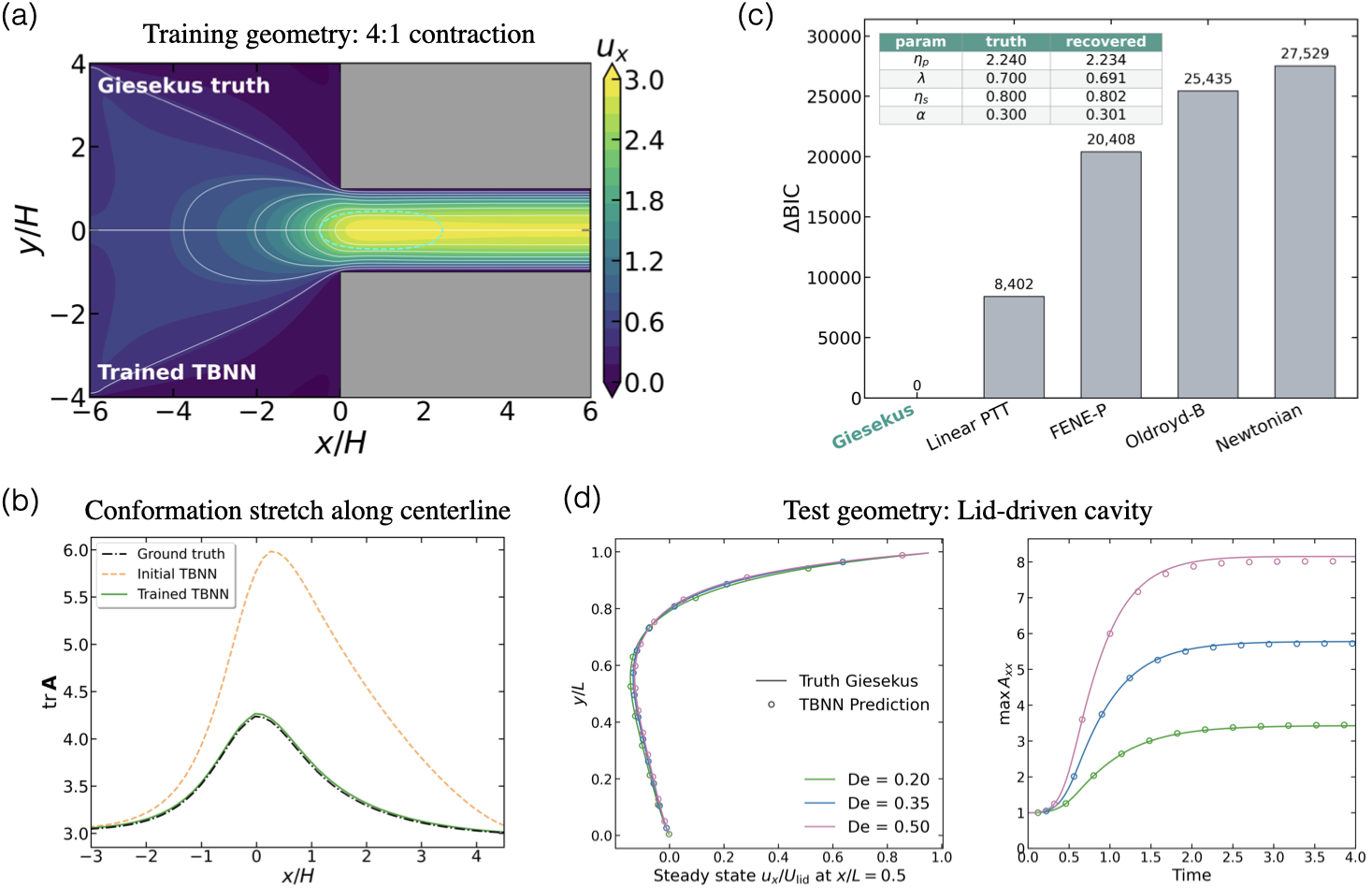}
    \caption{\textbf{Learning viscoelastic rheology with a memory TBNN.} (a) Steady streamwise velocity \(u_x\) in a \(4{:}1\) contraction at inlet speed \(U=0.5\) (\(\mathrm{De}=0.35\)) for the Giesekus ground truth (upper half) and the trained TBNN (lower half). (b) Centerline conformation stretch \(\mathrm{tr}\,\mathbf{A}\). Although conformation was not observed during training, the trained model reproduces the ground-truth profile. (c) \(\Delta\mathrm{BIC}\) relative to the best-fitting constitutive family obtained by probing the trained TBNN (Methods). Inset: ground-truth parameters and values recovered by jointly fitting the selected Giesekus model to the TBNN response. (d) Transfer of the contraction-trained closure to an unseen lid-driven cavity at \(\mathrm{De}=0.20\), \(0.35\), and \(0.50\): (left) steady streamwise-velocity profiles \(u_x/U_{\mathrm{lid}}\) along the vertical centerline \(x/L=0.5\), and (right) transient maximum conformation component \(\max A_{xx}\); shading marks the cosine ramp of the lid velocity. Agreement throughout the transient shows that the model has learned the fluid's underlying relaxation dynamics rather than a particular flow solution.}
    \label{fig:memory_tbnn}
\end{figure*}

We first apply this procedure to the TBNN trained on the constriction--expansion flow of Fig.~\ref{fig:tbnn_learning}. Using the digital rheometer (Methods Sec.~\ref{sec:digital_rheometer}), we impose sinusoidal shear-rate forcings with amplitude $f = 10$ and frequency $\omega = 1$ and record the corresponding stress response predicted by the TBNN. We then fit both Newtonian and Carreau--Yasuda (CY) models to the resulting stress curves (Fig.~\ref{fig:cy_params_tbnn}a). The Newtonian model fails to reproduce the nonlinear features of the response, whereas the CY model captures them accurately, yielding parameter values that closely match those used in the original simulation (first row of Table~\ref{tab:tbnn-cy-extraction}). Comparison of BIC values further confirms that the CY model provides a substantially better statistical description of the learned constitutive behavior than the Newtonian model, despite having more parameters.

We then repeat this analysis for six additional TBNNs trained on fluids with varying degrees of shear thinning, using power-law exponents $n = 0.6$ and $n = 0.8$ to represent stronger and weaker shear-thinning behavior, respectively. For each case, we also vary the transition parameter $k$, which determines the shear rate at which thinning begins. The CY parameters recovered from the digital rheometer closely match the corresponding ground-truth values across all cases (Table~\ref{tab:tbnn-cy-extraction}). To visualize the consistency of parameter recovery, Fig.~\ref{fig:cy_params_tbnn}b shows the similarity between the learned and true values of $n$ and $k$, demonstrating that the TBNN accurately reproduces both the magnitude and trend of the shear-thinning response.

\section*{Learning memory-dependent constitutive models}\label{sec:memory}

Although the instantaneous TBNN can represent any physically admissible stress that is algebraic in the instantaneous kinematics when supplied with the complete set of invariants and tensor bases, it cannot describe fading memory or other history dependence because it contains no evolving internal state. To extend our framework to such materials, we introduce the conformation tensor ($\mathbf{A}$) as an evolving internal state that provides a coarse-grained measure of microstructural stretch and orientation. We then construct the memory TBNN’s invariant features and tensor basis from ($\mathbf{A}$), allowing it to learn, in a frame-invariant form, how the material stores and relaxes stress (Methods Secs.~\ref{sec:visco_solver} and~\ref{sec:memory_methods}). The resulting closure is thermodynamically consistent by construction: the log-conformation formulation maintains a physically admissible conformation state, and non-negative dissipation follows from its mobility parameterization. We retain the polymer modulus \(G_p\), relaxation time \(\lambda\), and solvent viscosity \(\eta_s\) as explicit scalar parameters fitted jointly with the neural network, separating the characteristic stress and time scales from the learned functional form. Although no finite internal-state architecture can span every possible history-dependent behavior, this formulation encompasses a broad family of constitutive dynamics, including Oldroyd-B, Giesekus, FENE-P, and Saramito-type \cite{saramito2007new} yield-stress behavior, while allowing relaxation laws more general than any of these individual models.

We first test the memory TBNN on the canonical \(4{:}1\) contraction flow of a Giesekus fluid, a viscoelastic model with deformation-dependent polymer mobility that produces nonlinear stress relaxation and shear thinning. We train the model using planar velocity fields alone. The inlet speed is \(U=0.5\), giving a Deborah number \(\mathrm{De}=\lambda U/H=0.35\), which describes the ratio of the material relaxation time to the characteristic residence time \(H/U\). The trained memory TBNN reproduces the steady streamwise-velocity field throughout the contraction almost exactly (Fig.~\ref{fig:memory_tbnn}a), with the cross-stream velocity and corresponding error fields shown in Supplementary Fig.~\ref{fig:sfig_giesekus_fields}. Although the conformation tensor is not observed during training, the model also recovers the conformation stretch \(\mathrm{tr}\,\mathbf{A}\), with relative \(L_2\) errors of \(0.581\%\) over the full fluid domain and \(0.819\%\) within the weighted high-stretch region (Fig.~\ref{fig:memory_tbnn}b and Supplementary Fig.~\ref{fig:sfig_giesekus_fields}).

Using the same procedure as for the instantaneous TBNN, we probe the trained memory TBNN under oscillatory shear and fit classical constitutive models to its response using the rheometric ``readback'' described in Methods Sec.~\ref{sec:digital_rheometer}. BIC unambiguously selects the correct underlying Giesekus model, while the functionally similar Linear Phan--Thien--Tanner (PTT) model is the closest competitor but remains strongly disfavored (Fig.~\ref{fig:memory_tbnn}c).  The selected model also recovers the physical parameters nearly exactly (Fig.~\ref{fig:memory_tbnn}c, inset), including the mobility parameter \(\alpha=0.298\pm0.003\) across three optimization schedules applied to the same network initialization (Supplementary Figs.~\ref{fig:sfig_giesekus_training} and~\ref{fig:sfig_giesekus_rheometry}).

To test whether the learned constitutive dynamics transfer beyond the training flow, we embed the contraction-trained memory TBNN in an unseen lid-driven cavity without retraining. The model reproduces both the steady centerline velocity profiles and the transient evolution of \(\max A_{xx}\) at \(\mathrm{De}=0.20\), \(0.35\), and \(0.50\) (Fig.~\ref{fig:memory_tbnn}d). At \(\mathrm{De}=0.50\), the rms velocity error is \(2.90\times10^{-4}U_{\mathrm{lid}}\), and the relative \(L_2\) error in \(A_{xx}\) is \(0.0101\) (Supplementary Fig.~\ref{fig:sfig_cavity_fields}). This excellent agreement, maintained throughout the transient despite moving from an inlet-driven contraction to a closed, moving-wall-driven recirculating flow, demonstrates that the model has learned the fluid's underlying relaxation dynamics rather than a single flow solution.

We next also consider a finitely extensible nonlinear elastic–Peterlin (FENE-P) fluid in the same contraction geometry, testing whether the same TBNN can represent bounded polymer stretch, a mechanism qualitatively distinct from the nonlinear mobility of Giesekus (Supplementary Fig.~\ref{fig:sfig_fenep_recovery}). BIC correctly selects FENE-P over every alternative by at least \(11{,}580\), although the TBNN parameter estimates are less precise: for example, \(\lambda\) is approximately \(17\%\) below its ground-truth value (Supplementary Figs.~\ref{fig:sfig_fenep_recovery} and~\ref{fig:sfig_fenep_identifiability}). However, when the FENE-P form is specified, direct calibration to the same velocity data recovers all four parameters across four distinct initializations, showing that the lower readback accuracy reflects the difficulty of simultaneously learning the family and its parameters rather than insufficient data once the model form is known (Supplementary Fig.~\ref{fig:sfig_fenep_direct} and Supplementary Sec.~\ref{sec:si_fenep}). Thus, when more precise parameters are required, model discovery can be followed by direct calibration of the selected classical law against the original flow data.

\begin{figure}[t]
    \centering
    \includegraphics[width=\linewidth]{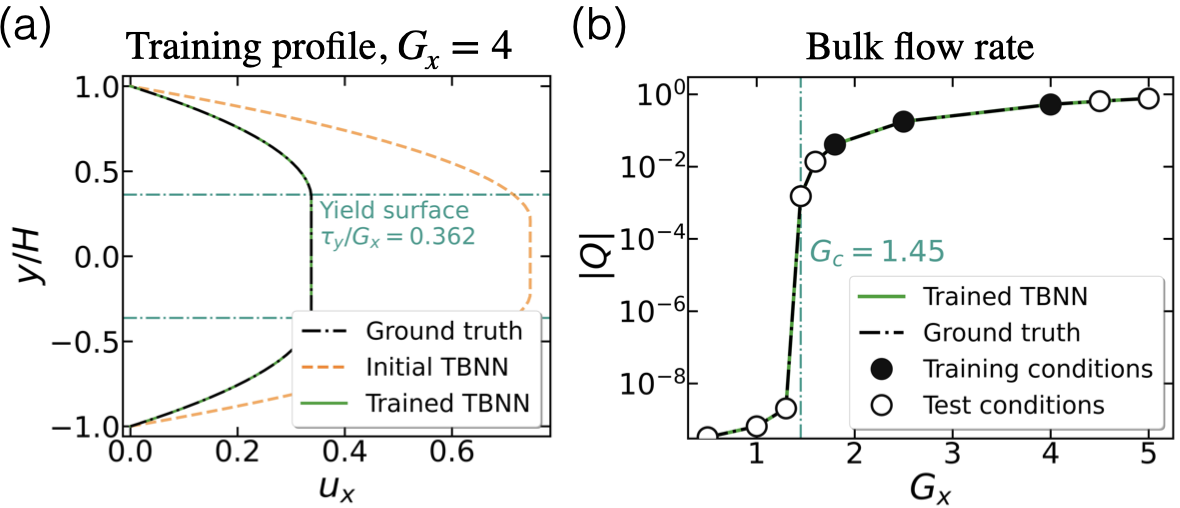}    \caption{\textbf{Learning elastoviscoplastic rheology.} (a) Streamwise-velocity profile \(u_x(y)\) in a plane channel driven at \(G_x=4\) for the Saramito ground truth and the initial/trained memory TBNN. Dash-dotted lines mark the yield surfaces, \(y/H=\pm\tau_y/G_x=\pm0.362\). (b) Bulk flow rate \(|Q|\) versus \(G_x\), with critical drive \(G_c=1.45\). The trained curve is the mean of five seeds (spread below plot resolution); training used \(G_x=1.8\), \(2.5\), and \(4\), all other points are unseen tests. The learned closure reproduces the flow curve across nine decades, including sub-yield arrest and the yielding transition.}
    \label{fig:evp_tbnn}
\end{figure}

Finally, recovering yield-stress behavior provides a qualitatively stricter test because the closure must capture both flowing states and a transition to true arrest, rather than only a smooth memory-dependent response. We test this capability in a plane channel driven by the nondimensional streamwise pressure gradient \(G_x\), using a Saramito \cite{saramito2007new} elastoviscoplastic fluid that responds as an elastic solid below its yield stress and flows as a viscoelastic liquid above it. Although yielding could in principle be represented directly through the learned mobility, for training efficiency and interpretability we augment the memory TBNN with the yield stress \(\tau_y\) as an explicit fourth material scalar (Methods Sec.~\ref{sec:memory_methods}). Training on transient velocity fields and the final bulk flow rate at \(G_x=1.8\), \(2.5\), and \(4\) recovers the velocity profiles and yield surfaces and accurately predicts the unseen \(G_x=5\) profile  (Fig.~\ref{fig:evp_tbnn}a and Supplementary Fig.~\ref{fig:sfig_evp_profiles_arrest}). Across five independent network initializations, the learned closure reproduces the flow curve over approximately nine decades, including the critical drive \(G_c= \tau_y/H =1.45\), the yielding transition, and sub-yield arrest (Fig.~\ref{fig:evp_tbnn}b and Supplementary Figs.~\ref{fig:sfig_evp_profiles_arrest} and~\ref{fig:sfig_evp_training}).

Together, the Giesekus, FENE-P, and Saramito results show that the same memory TBNN architecture and differentiable training framework can represent nonlinear relaxation, finite extensibility, and yield-stress behavior from velocity data. Furthermore, the learned dynamics remain interpretable, as we can identify a classical constitutive family and its physical parameters, extending the advantages of the instantaneous TBNN to memory-dependent closures.

\begin{figure*}
    \centering
    \includegraphics[width=0.8\linewidth]{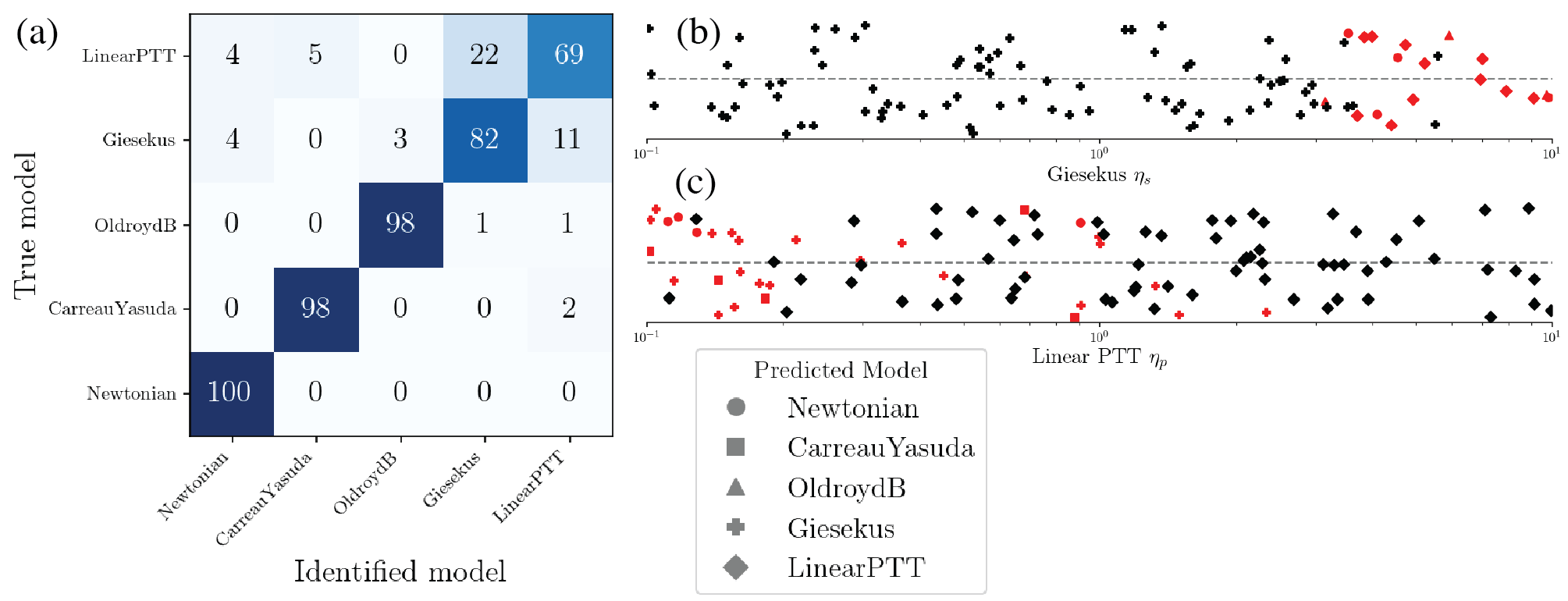}
    \caption{\textbf{Model selection and parameter identification.} (a) Confusion plot of our model-fitting approach when tested against 100 random instantiations of five common constitutive models, demonstrating that our approach achieves high identification accuracy across model families. (b) and (c) show that for models with less than $90\%$ accuracy, misidentification is strongly correlated with specific parameter regimes in which the complex model's response is statistically indistinguishable from a simpler law, as the applied forcing protocol does not adequately excite the distinguishing rheological behavior.}
    \label{fig:fittingresponse}
\end{figure*}

\section*{Model selection from bulk rheometry}

The differentiable model-fitting machinery used above to distill a TBNN into a classical constitutive law applies equally to any stress time-series, including direct experimental measurements. Standard rheometric protocols probe simple, low-dimensional flows whose information content is inherently limited. In this setting, parsimonious classical models, with a handful of physically meaningful parameters, are better suited than high-dimensional neural representations, provided one can systematically determine which model best describes the data.  Prior Bayesian inference has addressed this problem by quantifying how strongly rheological data support one candidate linear viscoelastic model over another while accounting for uncertainty in fitted parameters \cite{Ewoldt2015Bayesian}. Here, we pursue a complementary strategy for efficient selection across candidate constitutive families spanning nonlinear and time-dependent behavior, applying our differentiable fitting framework to noisy bulk rheometer data and using the Bayesian Information Criterion (BIC) to balance model complexity against predictive power.

To systematically evaluate this approach, we generated noisy synthetic rheometer datasets, each comprising 12 large-amplitude oscillatory shear (LAOS) traces, for 100 random instantiations across five common constitutive models (Newtonian, Carreau-Yasuda, Oldroyd-B, Giesekus, and Linear PTT). For every dataset, we fit each of the five candidates jointly to all 12 traces using a differentiable ODE solver via gradient descent (Methods Sec.~\ref{sec:digital_rheometer}). We then selected the best-fit model using BIC. A representative fit and held-out prediction are shown in Supplementary Fig.~\ref{fig:demonstration}.

The results of this test are summarized in Fig.~\ref{fig:fittingresponse}(a). Without optimizing the initial guesses or optimizer parameters for a given model, our approach is able to correctly identify the ground truth model in nearly all cases for the simpler constitutive models. For the most complex model, Linear PTT, identification succeeds 69\% of the time, and the failure cases themselves are informative. 

We can understand these cases in more detail by looking at where in the sampled parameter ranges our approach failed. In Figs.~\ref{fig:fittingresponse}(b) and (c) we plot a key parameter from each of the two models with less than 90\% identification accuracy. We see that in all cases, consistent failure, as denoted by the red symbols, is associated with strong clustering at particular values of the parameters. In these parameter regimes, the complex model's response under the applied forcing is statistically indistinguishable from a simpler constitutive law. This is not a failure of the fitting method but a statement about  what the applied forcing and available data can support: when BIC selects a simpler model, it correctly identifies that the additional parameters of the more complex law are not justified by the available data. A model that cannot be distinguished from a simpler alternative under realistic experimental conditions is, for practical purposes, that simpler model. 

This result also highlights the critical role of experimental design: the identifiability of a constitutive model is conditioned on the forcing protocol. In this case, we generally followed the protocol for fitting rheometer-like data laid out in other data-driven approaches \cite{Lennon2023PNAS}, but even these extended forcings did not fully resolve the most complex models.   Conversely, restricting the protocol to a single forcing condition substantially reduces identification accuracy for all models except the Newtonian baseline (Supplementary Sec.~\ref{sec:si_minimal_forcing} and Fig.~\ref{fig:sfig7}). Thus, more generally, model selection and experimental design are fundamentally coupled, and our differentiable framework is uniquely positioned to exploit this connection through gradient-based optimization of forcing protocols.

Beyond model selection, the fits themselves yield physical parameter estimates. Table~\ref{tab:median-factors} summarizes the geometric median factor (estimate/true) for each parameter across correctly identified models. For Newtonian, Oldroyd-B, and Giesekus fluids, the median factor is within $1\%$ of the true value across all parameters (with occasional outliers when true parameters are extremely small). Carreau–Yasuda models show larger deviations, particularly for $\eta_\infty$, which is governed by high-shear behavior that is not well sampled by sinusoidal forcings, consistent with the identifiability limitations discussed above. Rerunning with a smaller learning rate and more epochs reduces errors to below $0.3\%$ for all CY parameters except $\eta_\infty$, supporting the conclusion that the remaining deviations reflect forcing design rather than a fundamental limitation of the method.

In addition to synthetic benchmarks, we applied the same model-selection machinery directly to experimental measurements. Fitting the classical model library to large-amplitude oscillatory shear data from a metal-crosslinked hydrogel~\cite{Lennon2023PNAS}, BIC selects a White--Metzner model, in which viscosity and relaxation time vary with shear rate. This model matches a trained RUDE closure in-sample and extrapolates more reliably to stress amplitudes beyond the training data (Supplementary Sec.~\ref{sec:gel_si}), showing that, on information-limited bulk data, a few physical parameters can outperform a heavily parameterized neural closure.

Together, these results demonstrate that differentiable model fitting can robustly identify constitutive laws and parameters when the data are informative. Just as importantly, they reveal that model identifiability depends critically on whether the measurement protocol probes the relevant rheological regimes.

\section*{Discussion and Conclusion}\label{Discussion}

This work establishes a framework for learning constitutive behavior directly from flow measurements through differentiable simulations. By embedding  either an instantaneous stress closure or memory-dependent constitutive dynamics within a fluid solver, we demonstrated that constitutive laws can be inferred from sparse, noisy, or indirect observations -- turning rheological inference into a gradient-based optimization problem. Together, the instantaneous and memory TBNNs extend this framework from generalized Newtonian fluids to viscoelastic and elastoviscoplastic materials, including nonlinear relaxation, finite extensibility, and true yield-stress behavior. The ability of the tensor basis neural network (TBNN) to recover the correct stress response from limited data underscores that the governing equations themselves provide a powerful inductive bias: physical structure reduces the need for large datasets and increases robustness to noise, enabling the learned constitutive model to be directly interrogated and compared against classical laws. 

A central implication is that rheological characterization need not be confined to idealized benchtop geometries. Because the TBNN learns a local, frame-invariant  closure, the resulting constitutive representation is portable across conditions and geometries and can be embedded in new simulations to predict flow in unseen environments. While we trained here primarily on velocity fields  (augmented in selected tests by pressure and bulk flow-rate observations), the framework generalizes naturally to any measurable flow quantity that a forward model can predict, such as tracer trajectories,  time-resolved flow-rate  signals, or combinations  of spatial and integral observations. This flexibility provides a route to turning process measurements into a rheometer, enabling in-line or on-chip characterization, or digital rheometry, in settings where sampling or laboratory rheometry is impractical or would alter the material, such as emulsions, polymer melts, or multiphase suspensions.

At the same time, our results clarify a practical limitation of conventional bulk model fitting that is often under-emphasized: identifiability is conditioned on the forcing protocol and the rheological regimes it excites. In our bulk-rheometry tests, misidentification clustered in parameter regimes where the response of a more complex model is statistically indistinguishable from a simpler law under the applied sinusoidal forcings, and certain parameters were weakly constrained when the protocol did not adequately sample the relevant regimes (e.g., parameters governing high-shear behavior in Carreau--Yasuda fits). Accordingly, when an information criterion favors a simpler model, it can correctly indicate that additional parameters are not justified by the available data -- a conclusion that is especially relevant for highly parameterized neural closures, where thousands of parameters are often inferred from the same limited scalar measurements. The FENE-P contraction-flow results show that this identifiability constraint extends to flow-field inference: model-agnostic inference identifies the correct constitutive family but yields less precise parameter estimates, whereas direct calibration recovers all four parameters from the same velocity data.

Because our solver is fully differentiable, it offers a path to overcome these limits through optimized experimental design. The same gradients used to fit model parameters can be used to quantify how informative a forcing protocol or geometry is and to design inputs that maximally discriminate between competing constitutive hypotheses or minimize parameter uncertainty. We did not perform such protocol optimization here; rather, our results motivate it as a natural next step that would close the loop between measurement and theory and automate what has historically been an intuitive, labor-intensive aspect of rheological characterization.

Finally,  ongoing and future work will extend the framework to multi-mode relaxation spectra, additional internal variables for thixotropy and aging,  interfacial multiphase flows, fully three-dimensional flows (with encouraging closure-evaluation overhead in the present benchmarks, see Methods Sec.~\ref{sec:computational_cost}), and applications using experimental flow data, broadening the range of materials and operating environments accessible to this approach. Looking further ahead, we envision differentiable fluid simulations combined with invariant, data-driven closures as a route to model discovery when no classical constitutive law suffices: learned responses can be distilled into compact analytical forms via model selection, symbolic regression, or mechanistic constraints, while the same differentiable infrastructure can support inverse problems in the forward direction by optimizing material parameters or operating conditions for targeted system behavior. Together, these results mark a step toward automated, in-operando rheological characterization: differentiable simulations unify measurement, modeling, and optimization in a single framework that learns, tests, and designs constitutive descriptions.

\section*{Materials and Methods}
\subsection{Governing equations}
We model incompressible, time-dependent flow with the Cauchy momentum balance and a general deviatoric stress:
\begin{align}
\nabla\cdot\mathbf{u} &= 0, \label{eq:continuity} \\
\rho\big(\partial_t\mathbf{u}+(\mathbf{u}\cdot\nabla)\mathbf{u}\big)
&=-\nabla p+\nabla\cdot\tensor{\sigma}+\mathbf{f}, \label{eq:momentum}
\end{align}
where \(\mathbf{u}\) is velocity, \(p\) pressure, \(\tensor{\sigma}\) the deviatoric stress, and \(\mathbf{f}\) an imposed body force (e.g., a uniform pressure gradient or immersed boundary force). 

Constitutive laws and rheological models relate a fluid's stress tensor $\tensor{\sigma}$ to the local rate-of-strain, $\tensor{D} = 1/2 (\grad \vec{u}+(\grad \vec{u})^\mathrm{T})$. The simplest such model is a Newtonian fluid, where the viscosity $\eta$ is a simple scalar factor relating the two,
\begin{equation}
    \tensor{\sigma} = 2\eta \tensor{D}.
\end{equation}
Increasing in complexity is a class of constitutive laws known as Generalized Newtonian Fluids (GNFs). Such fluids are characterized by a viscosity $\eta$ that becomes a function of the strain rate $\dot{\gamma} = ||\tensor{D}|| \equiv \sqrt{2\tensor{D}:\tensor{D}}$. One such model is a power-law fluid, with a viscosity $\eta(\dot{\gamma})$ given by
\begin{equation}
\eta(\dot{\gamma}) = K\dot{\gamma}^{n-1}, \label{eq:powerlaw_stress} 
\end{equation}
where \(K\) and \(n\) are the consistency and shear-thinning index, respectively. 
Another such model is the Carreau--Yasuda model \cite{carreau1972rheological, yasuda1981shear}, given by
\begin{equation}
\eta(\dot{\gamma})
=\eta_\infty+(\eta_0-\eta_\infty)\Big[1+(k\,\dot{\gamma})^{a}\Big]^{\frac{n-1}{a}}, \label{eq:carreau_yasuda_eta}
\end{equation}
where $k$ is the characteristic timescale, $a$ is the transition sharpness, \(n\) is the shear-thinning index, and $\eta_0$ and $\eta_\infty$ are the zero- and infinite-shear viscosity respectively.

When studying polymeric fluids, consisting of a long polymer dissolved in a solvent, constitutive models are written in terms of an extra stress $\tensor{\tau}$, such that the total fluid stress $\tensor{\sigma}$ is given by
\begin{equation}
    \tensor{\sigma} = 2\eta_s \tensor{D} + \tensor{\tau}.
\end{equation}
Here, $\eta_s$ is called the solvent viscosity and describes the Newtonian contribution to the stress. The simplest viscoelastic model that is often used in describing fluids is the Oldroyd-B model \cite{Oldroyd1950Formulation}, whose extra stress obeys an evolution equation
\begin{equation}
    \lambda\ucd{\tensor{\tau}}+\tensor{\tau}=2\eta_p\tensor{D},
\end{equation}
where
\begin{equation}
    \ucd{\tensor{\tau}} = \frac{\partial \tensor{\tau}}{\partial t} + \vec{u}\bdot\grad\tensor{\tau} - \tensor{\tau}\bdot\grad\vec{u}-\grad\vec{u}^\mathrm{T}\bdot\tensor{\tau},
\end{equation}
is the upper-convected derivative, $\lambda$ is the relaxation time and $\eta_p$ is the polymer viscosity.

We consider several constitutive extensions of Oldroyd-B. These include the
Giesekus model \cite{Giesekus1982JNNFM} whose extra stress evolves like
\begin{equation}
    \lambda\ucd{\tensor{\tau}}+\tensor{\tau}+\frac{\alpha\lambda}{\eta_p}\boldsymbol{\tau}\bdot\boldsymbol{\tau}=2\eta_p\tensor{D},
\end{equation}
and the Linear Phan--Thien--Tanner model \cite{phan1977new}, given by
\begin{equation}
    \lambda\gsd{\tensor{\tau}}+\left(1+\frac{\varepsilon\lambda}{\eta_p}\mathrm{tr}(\tensor{\tau})\right)\tensor{\tau}=2\eta_p\tensor{D}.
\end{equation}
Here,
\begin{equation}
    \gsd{\tensor{\tau}} = \ucd{\tensor{\tau}} + \zeta(\tensor{\tau}\bdot\tensor{D}+\tensor{D}\bdot\tensor{\tau})
\end{equation}
is known as the Gordon-Schowalter derivative.

The White--Metzner model \cite{white1963development} retains the Oldroyd-B stress evolution but makes the polymer viscosity and relaxation time functions of strain rate,
\begin{align}
    \lambda(\dot{\gamma})\ucd{\tensor{\tau}}
    +\tensor{\tau}
    &=2\eta_p(\dot{\gamma})\tensor{D},
    \nonumber\\
    \eta_p(\dot{\gamma})
    &=\eta_{p,0}
    \left[1+(K\dot{\gamma})^a\right]^{(n-1)/a},
    \nonumber\\
    \lambda(\dot{\gamma})
    &=\lambda_0
    \left[1+(L\dot{\gamma})^b\right]^{(m-1)/b}.
    \label{eq:white_metzner}
\end{align}
Here, \(\eta_{p,0}\) and \(\lambda_0\) are the zero-rate polymer viscosity and relaxation time, \(K\) and \(L\) are their respective timescales, \(n\) and \(m\) are rate-dependence exponents, and \(a\) and \(b\) control the transition sharpness.

The Saramito elastoviscoplastic model \cite{saramito2007new} introduces a yield stress \(\tau_y\) by gating stress relaxation through the yield factor \(\kappa_y\),
\begin{align}
    \lambda\ucd{\tensor{\tau}}
    +\kappa_y\tensor{\tau}
    &=2\eta_p\tensor{D},
    \nonumber\\
    \kappa_y
    &=\left\langle
    1-\frac{\tau_y}{\lvert\tensor{\tau}_d\rvert}
    \right\rangle_+,
    \nonumber\\
    \tensor{\tau}_d
    &=\tensor{\tau}
    -\frac{\mathrm{tr}(\tensor{\tau})}{3}\tensor{I},
    \qquad
    \lvert\tensor{\tau}_d\rvert
    =\sqrt{\frac{1}{2}
    \tensor{\tau}_d:\tensor{\tau}_d}.
    \label{eq:saramito_stress}
\end{align}
For differentiability, the positive-part operator is evaluated as \(\delta_y\operatorname{softplus}(x/\delta_y)\), with \(\delta_y=10^{-3}\), and \(\lvert\tensor{\tau}_d\rvert\) is floored at \(10^{-12}\) before division. Below yield, \(\kappa_y\) rapidly approaches zero, suppressing relaxation and permitting arrest; above yield, viscoelastic flow resumes. Setting \(\tau_y=0\) recovers Oldroyd-B.

\subsubsection*{Conformation-tensor forms}

The symmetric positive-definite conformation tensor \(\tensor{A}\) measures polymer stretch and orientation, with \(\tensor{A}=\tensor{I}\) at equilibrium. Defining \(G_p=\eta_p/\lambda\), Oldroyd-B is written
\begin{align}
    \ucd{\tensor{A}}
    &=-\frac{1}{\lambda}
    \left(\tensor{A}-\tensor{I}\right),
    \nonumber\\
    \tensor{\tau}
    &=G_p\left(\tensor{A}-\tensor{I}\right).
    \label{eq:oldroyd_b_conformation}
\end{align}

The equivalent Giesekus form is
\begin{align}
    \ucd{\tensor{A}}
    &=-\frac{1}{\lambda}\left[
    \tensor{A}-\tensor{I}
    +\alpha\left(\tensor{A}-\tensor{I}\right)^2\right],
    \nonumber\\
    \tensor{\tau}
    &=G_p\left(\tensor{A}-\tensor{I}\right).
    \label{eq:giesekus_conformation}
\end{align}

FENE-P limits polymer extension through the Peterlin factor \(f\) \cite{bird1987dynamics1},
\begin{align}
    \ucd{\tensor{A}}
    &=-\frac{1}{\lambda}
    \left(f\tensor{A}-a\tensor{I}\right),
    \nonumber\\
    \tensor{\tau}
    &=G_p\left(f\tensor{A}-a\tensor{I}\right),
    \nonumber\\
    f
    &=\frac{L^2}{L^2-\mathrm{tr}(\tensor{A})},
    \qquad a=\frac{L^2}{L^2-3}.
    \label{eq:fenep_conformation}
\end{align}

Saramito retains the Oldroyd-B stress readout, \(\tensor{\tau}=G_p(\tensor{A}-\tensor{I})\), but gates conformation relaxation according to
\begin{equation}
    \ucd{\tensor{A}}
    =-\frac{\kappa_y}{\lambda}
    \left(\tensor{A}-\tensor{I}\right).
    \label{eq:saramito_conformation}
\end{equation}

Numerically, we evolve the log-conformation \(\tensor{\Psi}=\log\tensor{A}\), as described in Sec.~\ref{sec:visco_solver}.

\subsection{Differentiable  generalized Newtonian fluid solver}\label{sec:diff_non_newtonian_cfd}

\subsubsection*{Differentiable solver and numerics}\label{subsec:diff_cfd_solver}
We solve the incompressible equations in a differentiable JAX-based framework that follows established components from JAX-CFD and immersed-boundary (IB) methods \cite{kochkov2021machine, alhashim2025control}. 
At a high level, we use a staggered Cartesian grid, second-order central differences for diffusive terms, a conservative upwind discretization for advection, and a projection step that corrects a provisional velocity via a pressure Poisson solve to enforce \(\nabla\cdot\mathbf{u}=0\). 

Spatially varying viscosity introduces significant additional stiffness that standard explicit schemes cannot handle.  To address this, we extended the solver to support fully implicit (backward Euler) integration for wall-bounded flows and semi-implicit (IMEX) schemes for unbounded or periodic domains.  The resulting linear subproblems are solved iteratively (BiCGSTAB) to tight tolerances on divergence and kinetic-energy drift, and boundary conditions are imposed either directly (no-slip walls, pressure inlets/outlets) or via IB forcing for complex geometries. We iterate residuals until \(\|\nabla\cdot\mathbf{u}\|_2 < 10^{-8}\) and relative changes in kinetic energy fall below \(10^{-10}\).

All solver operations — state updates, pressure projection, and non-Newtonian stress evaluation — are expressed as pure JAX transformations, enabling exact reverse-mode differentiation through the full computation.  Sensitivities are propagated through each iterative update without an explicit adjoint PDE derivation, yielding gradients of any scalar objective (e.g., velocity-field losses) with respect to constitutive parameters or neural network weights.

This end-to-end differentiable pipeline provides the map from constitutive parameters to flow observables and their gradients, and it is the backbone used to train the tensor basis neural network (TBNN) closure.

\subsubsection*{Solver validation}
To validate the solver, we implemented classical generalized-Newtonian models, where the local viscosity is governed by the local strain rate, including the power-law and Carreau--Yasuda (CY) fluids. 
Both models were benchmarked against OpenFOAM simulations of steady Poiseuille flow, showing quantitative agreement in velocity and pressure profiles.  All training data used in this work were thus generated by forward simulations from our differentiable solver.

\subsubsection*{Instantaneous constitutive closure via a tensor basis neural network}\label{subsec:tbnn_description}
While analytical constitutive models such as the power-law or Carreau--Yasuda form can be fit individually, doing so for every flow type or geometry rapidly becomes cumbersome and inflexible. 
Instead, we adopt a general tensorial representation of the stress based on a tensor basis neural network (TBNN), which expresses the stress as a sum over invariant tensorial bases weighted by scalar functions of the flow invariants \cite{ling2016reynolds, Lennon2023PNAS}:
\begin{align}
\tensor{\sigma}(\nabla\mathbf{u})
&= \sum_{i=1}^{N} \alpha_i(\mathcal{I})\,\tensor{B}_i(\nabla\mathbf{u}),
\label{eq:tbnn_general}
\end{align}
where \(\alpha_i(\mathcal{I})\) are scalar coefficient functions of an invariant set \(\mathcal{I}=\{I_1,I_2,\dots,I_K\}\), and \(\tensor{B}_i\) are tensor bases formed from \(\nabla\mathbf{u}\) and its symmetric and antisymmetric parts. 
This construction guarantees Galilean invariance and provides a systematic, data-driven extension of classical constitutive laws.

For two-dimensional incompressible flow, the complete basis set reduces to
\begin{align}
\tensor{D} &= \tfrac{1}{2}\big(\nabla\mathbf{u}+\nabla\mathbf{u}^{\top}\big), \label{eq:D_def}\\
\tensor{W} &= \tfrac{1}{2}\big(\nabla\mathbf{u}-\nabla\mathbf{u}^{\top}\big), \label{eq:W_def}\\
I_1 &= \mathrm{tr}(\tensor{D}^2), \quad I_2 = \mathrm{tr}(\tensor{W}^2), \label{eq:invariants}\\
\tensor{B}_1 &= \tensor{D}, \quad \tensor{B}_2 = \tensor{W}\tensor{D}-\tensor{D}\tensor{W}. \label{eq:bases}
\end{align}

Because the training data are generated from the Carreau–Yasuda model, which produces purely extensional stresses without rotation-induced components, we omit the second invariant and the antisymmetric basis. We verified this approximation by including the full set of bases and invariants in a separate training, which learned nearly zero dependence on the second invariant and antisymmetric term, confirming that their contribution is negligible. Accordingly, the simplified closure used for the final results still adheres to the TBNN framework while reducing to a generalized-Newtonian form with a data-driven viscosity:
\begin{align}
\tensor{\sigma} 
&= 2\,\eta(I_1)\,\tensor{D}, 
\label{eq:gn_tbnn}
\end{align}

We parameterize \(\eta\) with a monotone, bounded head whose parameters are generated by a neural network over invariants. 
\begin{align}
\eta(I_1) 
&= \eta_\infty\Big[1 + r\,\big(1 - F(I_1)\big)\Big], 
\qquad r \equiv \frac{\eta_0}{\eta_\infty}-1, 
\label{eq:eta_logadd_compact}\\
z(I_1) 
&\equiv \log\left(\frac{\sqrt{2\,I_1}}{\dot{\gamma}_{\mathrm{ref}}}\right), 
\label{eq:z_from_I1}\\
F(I_1) 
&= \sum_{m=1}^{M} \alpha_m(I_1)\,
\mathrm{sigm}\!\left(\frac{z(I_1)-\mu_m(I_1)}{s_m(I_1)}\right),
\qquad M = 12,
\label{eq:F_mixture}
\end{align}
where \(F\) is a mixture of \(M{=}12\) logistic modes, with \(\mathrm{sigm}(x)=1/(1+e^{-x})\). 
The mode parameters \(\{\alpha_m(I_1),\mu_m(I_1),s_m(I_1)\}\) are produced by a 16-unit Multi-Layer Perceptron (MLP), with \(\alpha_m\) normalized and \(s_m>0\), yielding a total of 646 trainable parameters. 
In these expressions, \(\eta_0\) and \(\eta_\infty\) denote the zero- and infinite-shear viscosities, respectively, and \(\dot{\gamma}_{\mathrm{ref}}\) is a fixed reference shear-rate scale used for nondimensionalization, which we set as \(\dot{\gamma}_{\mathrm{ref}} = 1.0\). 

This form ensures positivity of \(\eta\) while allowing sufficient flexibility to capture nonlinear shear-thinning and thickening responses, providing a smooth, differentiable constitutive closure compatible with the TBNN framework.

\subsubsection*{Nondimensionalization and units}\label{subsec:units}
All quantities are reported in pressure-driven viscous units. 
Lengths are scaled by a characteristic length scale \(H_{\mathrm{ref}}\), velocities by \(U_{\mathrm{ref}} = G_{\mathrm{ref}}H_{\mathrm{ref}}^2/\eta_0\), times by \(T_{\mathrm{ref}} = \eta_0/(G_{\mathrm{ref}}H_{\mathrm{ref}})\), stresses and pressures by \(\tau_{\mathrm{ref}} = G_{\mathrm{ref}}H_{\mathrm{ref}}\), and shear rates by \(\dot{\gamma}_{\mathrm{ref}} = G_{\mathrm{ref}}H_{\mathrm{ref}}/\eta_0\).
Unless otherwise stated, the zero-shear viscosity is fixed at \(\eta_0=1\) and the reference pressure gradient at \(G_{\mathrm{ref}}=1\).

All training of the instantaneous TBNN is performed in the constriction geometry, where the reference length corresponds to the gap width \(H_{\mathrm{gap}}=1\), and the imposed nondimensional pressure gradient is \(G^*=G/G_{\mathrm{ref}}=5\).  As a generalization test, we additionally evaluate the trained model in a bidisperse porous-medium geometry, where \(H_{\mathrm{ref}}\) is taken as the radius of the larger circular obstacle and the imposed gradient is \(G^*=7.5\).

\subsection{Learning and evaluation protocol for the instantaneous TBNN}\label{sec:tbnn_learning_eval}

\subsubsection*{Reference data and PIV emulation}\label{subsec:adding_piv_res_noise}
Ground-truth fields are generated by forward simulations of the Carreau--Yasuda model. 
For Figs.~\ref{fig:tbnn_learning} and \ref{fig:cy_params_tbnn}, we train directly on the native simulation grid (no observation operator).

To mimic experimental measurements, in Figs.~\ref{fig:sfig4}--\ref{fig:sfig6}, we apply a PIV observation operator that first performs Hann-windowed, separable smoothing over interrogation windows and then samples at vector-center locations:
\begin{align}
\mathbf{y}
&= \mathcal{S}\,\mathcal{H}\,\mathbf{u}_{\mathrm{true}} \;+\; \boldsymbol{\varepsilon},
\label{eq:piv_obs_hann}
\end{align}
where \(\mathcal{H}\) denotes Hann-windowed averaging and \(\mathcal{S}\) denotes sampling on the coarser PIV grid.

The measurement noise \(\boldsymbol{\varepsilon}\) is \emph{correlated and heteroskedastic}. 
Correlation is introduced by a separable Gaussian smoothing on the vector grid with widths tied to the interrogation-window geometry:
\begin{align}
\boldsymbol{\varepsilon}
&= \Sigma(\mathbf{x}) \,\odot\, (\mathcal{G} * \boldsymbol{\xi})\;+\; 
\varsigma_{\mathrm{bias}}\,(\mathcal{G} * \mathbf{b}),
\qquad \boldsymbol{\xi}\sim\mathcal{N}(\mathbf{0},\,\mathbf{I}),
\label{eq:piv_noise_corr}\\
\mathcal{G}&=\mathcal{G}(\varsigma_x,\varsigma_y),\qquad
\varsigma_x=\tfrac{1}{2}\,\chi_{\mathrm{corr}}\,\frac{W_x}{s_x},\quad
\varsigma_y=\tfrac{1}{2}\,\chi_{\mathrm{corr}}\,\frac{W_y}{s_y},
\label{eq:gauss_widths}
\end{align}
with \(W_x, W_y\) the interrogation-window sizes and \(s_x, s_y\) the strides (vector spacings). 
The nondimensional factor \(\chi_{\mathrm{corr}}\) sets the correlation width of the synthetic noise relative to the vector spacing, 
and \(\varsigma_x,\varsigma_y\) are the Gaussian smoothing widths applied along each axis of the PIV vector grid.

Heteroskedasticity is modeled by a spatially varying scale field
\begin{align}
\Sigma(\mathbf{x})
&= \varsigma_{\mathrm{base}}\Big[1+\beta_{\nabla}\,\widehat{G}(\mathbf{x})\Big], 
\qquad 
\varsigma_{\mathrm{base}}=\alpha\,U_{95},
\label{eq:varsigma_field}
\end{align}
where \(U_{95}\) is the \(95^{\mathrm{th}}\)-percentile speed on the \emph{downsampled} field, 
\(\widehat{G}\) is a normalized speed-gradient magnitude on the vector grid, and 
\(\varsigma_{\mathrm{bias}}\) adds a low-frequency background offset via the same kernel \(\mathcal{G}\).

For runs where we explicitly decreased resolution and/or added noise, we used a Hann kernel for \(\mathcal{H}\), fixed \(\chi_{\mathrm{corr}}=0.35\), \(\beta_{\nabla}=0.5\), and \(\varsigma_{\mathrm{bias}}=0.1\,\varsigma_{\mathrm{base}}\), while varying the interrogation-window size (and thus the effective vector spacing) and the noise amplitude \(\alpha\).

\subsubsection*{Training objective and schedule}\label{subsec:tbnn_training}

We train the TBNN closure by matching the \emph{observed} velocity field---either the full simulation grid or its PIV-processed counterpart---while applying mild regularization. 
Let \(\mathbf{u}_{\mathrm{pred}}\) be the solver output and \(\mathcal{O}\in\{\mathrm{Id},\,\mathcal{S}\mathcal{H}\}\) the observation operator (identity for full-grid training; Hann-windowed averaging and sampling for PIV). 
The data-fidelity term is
\begin{align}
\mathcal{L}_{\mathrm{data}}
&= \frac{1}{N_{\mathrm{obs}}}\sum_{j=1}^{N_{\mathrm{obs}}}
\big\|\big(\mathcal{O}\,\mathbf{u}_{\mathrm{pred}}\big)(\mathbf{x}_j) 
- \mathbf{y}(\mathbf{x}_j)\big\|_2^2 ,
\label{eq:loss_data}
\end{align}
where, for training directly on the simulation grid, \(\mathbf{y}=\mathbf{u}_{\mathrm{true}}\). For the native-grid trainings (Figs.~\ref{fig:tbnn_learning} and~\ref{fig:cy_params_tbnn}; Table~\ref{tab:tbnn-cy-extraction}), the sum in Eq.~\eqref{eq:loss_data} runs over a fixed subset of grid cells that omits the slowly varying channel core upstream and downstream of the constriction, concentrating the objective on the kinematically informative region (cells inside the solid obstacle carry no signal, as both velocity fields vanish there); the PIV-processed trainings use all observed vectors on the PIV grid.

A scale-invariant shape term emphasizes matching flow patterns independently of magnitude. We first compute the optimal scalar alignment
\begin{align}
a
&=
\frac{\displaystyle
\sum_{j=1}^{N_{\mathrm{obs}}}
\big(\mathcal{O}\mathbf{u}_{\mathrm{pred}}\big)(\mathbf{x}_j)
\cdot\mathbf{y}(\mathbf{x}_j)}
{\displaystyle
\sum_{j=1}^{N_{\mathrm{obs}}}
\big\|
\big(\mathcal{O}\mathbf{u}_{\mathrm{pred}}\big)(\mathbf{x}_j)
\big\|_2^2},
\notag\\
\mathcal{L}_{\mathrm{shape}}
&=
\frac{1}{N_{\mathrm{obs}}}
\sum_{j=1}^{N_{\mathrm{obs}}}
\big\|
a\big(\mathcal{O}\mathbf{u}_{\mathrm{pred}}\big)(\mathbf{x}_j)
-\mathbf{y}(\mathbf{x}_j)
\big\|_2^2 .
\label{eq:loss_shape}
\end{align}
The alignment \(a\) is treated as constant during differentiation.

The viscosity-head prior discourages (i) excessive shear thinning and (ii) overly bumpy profiles in log-viscosity space in order to maintain stability of the forward simulation:
\begin{align}
\mathcal{L}_{\mathrm{slope}}
&= \sum_{q=1}^{N_z}
  \Big(\max\{0,\ |\partial_z \log\eta(z_q)|
  - s_{\mathrm{th}}\}\Big)^{2},
\qquad s_{\mathrm{th}} = 0.60,
\label{eq:loss_slope}
\\[6pt]
\mathcal{L}_{\mathrm{curv}}
&= \sum_{q=1}^{N_z}
  \Big(\max\{0,\ |\partial_{zz}\log\eta(z_q)|
  - c_{\mathrm{th}}\}\Big)^{2},
\label{eq:loss_curv}
\end{align}
with \(z=\log(\sqrt{2I_1}/\dot{\gamma}_{\mathrm{ref}})\) and \(c_{\mathrm{th}}=0.5\).
Both penalties remained zero throughout the representative native-grid training.

The total objective is
\begin{align}
\mathcal{L}
&= \lambda_{\mathrm{data}}\mathcal{L}_{\mathrm{data}}
+ \lambda_{\mathrm{shape}}\mathcal{L}_{\mathrm{shape}}
\\
&\quad+ \lambda_{\mathrm{slope}}\mathcal{L}_{\mathrm{slope}}
+ \lambda_{\mathrm{curv}}\mathcal{L}_{\mathrm{curv}},
\label{eq:loss_total}
\end{align}
where 
\(\lambda_{\mathrm{data}}=1\), 
\(\lambda_{\mathrm{shape}}=0.5\),
\(\lambda_{\mathrm{slope}}=10^{-3}\), and
\(\lambda_{\mathrm{curv}}=3\times10^{-4}\).

We employ a two-stage training schedule. In the first stage, the network weights are held fixed while \(\eta_\infty\) is adjusted until the sign of its gradient flips (typically within \(8\)–\(12\) iterations). In the second stage, \(\eta_\infty\) is frozen and the curvature and shape parameters of the viscosity head are trained. To avoid rapid convergence to a trivial Newtonian local minimum that lowers the loss, we fix \(\eta_0=1\) throughout training. In practice, \(\eta_0\) could  be treated as a hyperparameter or unfrozen after convergence of the second stage. Adam learning rates were \(5\times10^{-1}\) for \(\eta_\infty\) in the first stage and \(5\times10^{-2}\) for the viscosity-head parameters in the second stage. For runs 3 and 4 in Table~\ref{tab:tbnn-cy-extraction}, training was more stable when the mixture centers in \(F(I_1)\) were also held fixed. Additional training choices and practical considerations are discussed in Supplementary Sec.~\ref{sec:si_lessons}.

\subsubsection*{Evaluation metrics}\label{subsec:eval_metrics}
Except for the training loss, all evaluation metrics are computed against the \emph{ground-truth simulation fields}, not the window-averaged or noisy observations.

In Fig.~\ref{fig:diff_geometry}, we report a strain-rate–binned relative error, which quantifies the pointwise deviation normalized by the local ground-truth velocity magnitude and averaged within bins of strain rate:
\begin{align}
\mathrm{RelErr}(\dot{\gamma})
&= \Bigg\langle
\frac{\big\|\mathbf{u}_{\mathrm{pred}}(\mathbf{x}) - \mathbf{u}_{\mathrm{true}}(\mathbf{x})\big\|_2}
{\|\mathbf{u}_{\mathrm{true}}(\mathbf{x})\|_2}
\Bigg\rangle_{\dot{\gamma}\,\text{bin}}.
\label{eq:strain_rate_relerr}
\end{align}

In Figs.~\ref{fig:sfig4}a and~\ref{fig:sfig5}a, we report the domain-level relative root-mean-squared error (RRMSE), which measures the global difference between predicted and true velocity fields across all grid points:
\begin{align}
\mathrm{RMSE}_{\mathbf{u}}
&=\left(\frac{1}{N_{\mathrm{grid}}}\sum_{j=1}^{N_{\mathrm{grid}}}
\big\|\mathbf{u}_{\mathrm{pred}}(\mathbf{x}_j)
- \mathbf{u}_{\mathrm{true}}(\mathbf{x}_j)\big\|_2^2\right)^{1/2},
\label{eq:rmse}\\[4pt]
\mathrm{RRMSE}_{\mathbf{u}}
&=\frac{\mathrm{RMSE}_{\mathbf{u}}}{
\left(\frac{1}{N_{\mathrm{grid}}}\sum_{j=1}^{N_{\mathrm{grid}}}
\|\mathbf{u}_{\mathrm{true}}(\mathbf{x}_j)\|_2^2\right)^{1/2}}.
\label{eq:rrmse}
\end{align}

The strain-rate–binned relative error (Fig.~\ref{fig:diff_geometry}b) highlights local performance across different flow regimes, whereas the RRMSE (Figs.~\ref{fig:sfig4}a and~\ref{fig:sfig5}a) provides a single aggregate measure of overall predictive accuracy.

\subsection{Differentiable viscoelastic fluid solver}\label{sec:visco_solver}

\subsubsection*{Differentiable solver and numerics}\label{subsec:visco_numerics}

The viscoelastic solver uses the same staggered Cartesian grid, second-order central discretization of solvent diffusion, boundary treatment, and pressure-projection framework described for the generalized-Newtonian solver in Sec.~\ref{sec:diff_non_newtonian_cfd}. The reported viscoelastic simulations used conservative first-order upwind momentum advection and a backward-Euler IMEX update: the Newtonian solvent contribution was treated implicitly, whereas momentum advection and the divergence of the updated polymer stress were treated explicitly. The implicit velocity systems were solved using BiCGSTAB. As above, the complete update was implemented using pure JAX operations to permit reverse-mode differentiation through the transient simulation.

We represent material memory by the symmetric positive-definite conformation tensor \(\mathbf{A}\) and perform its transport using the log-conformation variable \(\boldsymbol{\Psi}=\log\mathbf{A}\), with \(\mathbf{A}=\exp\boldsymbol{\Psi}\). All viscoelastic models reported here used the Becker--Knechtges eigenvalue-free kernel, which evaluates these transformations through two-dimensional Cayley--Hamilton expressions without an eigendecomposition \cite{becker2023eigenvalue}. The conformation update was operator split into a Fattal--Kupferman closed-form upper-convected stretch, van Leer TVD advection of \(\boldsymbol{\Psi}\) with SSP-RK2 time integration, and a frozen-coefficient affine relaxation step integrated by a matrix exponential \cite{fattal2004constitutive}.

For the memory TBNN's cavity-transfer calculations, steadiness was evaluated over the final quarter of each trajectory by requiring the relative standard deviation of the kinetic energy to be below \(5\times10^{-3}\) and the normalized temporal trends in both \(\max A_{xx}\) and the minimum streamfunction \(\psi_{\min}\) to be below \(10^{-2}\). All reported cavity-transfer calculations satisfied this stress-inclusive criterion.

\subsubsection*{Solver validation}\label{subsec:visco_validation}

We validated the viscoelastic implementation against reduced-flow solutions that separately exercise transient relaxation, upper-convected transport, and polymer-stress evaluation. For Oldroyd-B fluids, startup Couette flow was compared with the Maxwell shear-stress response, the first normal-stress difference was tested over a range of Weissenberg numbers, and plane Poiseuille flow was compared with the corresponding analytical velocity, shear-stress, and normal-stress profiles. The Giesekus implementation was compared with independent steady-simple-shear reference solutions for shear thinning and normal stresses and was verified to recover Oldroyd-B behavior as \(\alpha\rightarrow0\). FENE-P and linear PTT were likewise checked against steady-shear reference solutions and their Oldroyd-B limits, while the Saramito implementation was subjected to yielded-branch, zero-yield-stress, and cessation regression tests.

The lid-driven-cavity velocity--pressure solver was independently benchmarked using Newtonian centerline-velocity profiles at \(\mathrm{Re}=100\), \(400\), and \(1000\) from Ghia et al.\ and the \(\mathrm{Re}=1000\) streamfunction and vortex diagnostics of Botella and Peyret \cite{ghia1982high,botella1998benchmark}.

\subsubsection*{Constitutive closure via a memory tensor basis neural network}\label{subsec:memory_tbnn_description}

We parameterize the constitutive response through a dimensionless elastic free-energy potential \(\phi(\mathbf{A})\) and a positive-semidefinite isotropic mobility \(\boldsymbol{\mathcal{M}}(\mathbf{A})\). The potential determines the polymer stress and the thermodynamic force driving relaxation, while the mobility determines how rapidly that force is dissipated:
\begin{align}
\mathbf{K}(\mathbf{A})
&= 2\mathbf{A}
\frac{\partial\phi}{\partial\mathbf{A}},
\label{eq:memory_tbnn_K}\\
\boldsymbol{\tau}_p
&= G_p\,\mathbf{K}(\mathbf{A}),
\label{eq:memory_tbnn_stress}\\
\ucd{\mathbf{A}}
&= -\frac{1}{\lambda}\,
\boldsymbol{\mathcal{M}}(\mathbf{A})
\mathbf{K}(\mathbf{A}).
\label{eq:memory_tbnn_relax}
\end{align}
Here, \(G_p\) sets the polymer-stress scale, \(\lambda\) the relaxation timescale, and \(\eta_s\) the Newtonian solvent contribution; these physical scalars are fitted jointly with the neural closure. The potential--mobility architecture makes nonnegative dissipation a structural property rather than a learned constraint. Because \(\boldsymbol{\mathcal{M}}\) is positive semidefinite and coaxial with the positive-definite conformation tensor, the continuous constitutive law satisfies
\begin{equation}
\begin{aligned}
\mathcal{D}_{\mathrm{rel}}
&\equiv
-\left.
\frac{\mathrm{d}(G_p\phi)}{\mathrm{d}t}
\right|_{\mathrm{relax}},\\
\mathcal{D}_{\mathrm{rel}}
&=
\frac{G_p}{2\lambda}\,
\mathbf{K}:
\left(\mathbf{A}^{-1}
\boldsymbol{\mathcal{M}}\mathbf{K}\right),\\
\mathcal{D}_{\mathrm{rel}}
&\geq 0.
\end{aligned}
\label{eq:memory_tbnn_dissipation}
\end{equation}
Likewise, constructing the scalar functions from invariants of \(\mathbf{A}\) and the tensorial outputs from isotropic functions of \(\mathbf{A}\) ensures material frame indifference by construction.

The scalar invariant inputs are centered so that the undeformed state \(\mathbf{A}=\mathbf{I}\) corresponds to the origin:
\begin{equation}
\begin{aligned}
x_1
&= \mathrm{tr}(\mathbf{A})-3,\\
x_2
&= \mathrm{tr}(\mathbf{A}^2)-3,\\
x_3
&= \ln\det\mathbf{A}.
\end{aligned}
\label{eq:memory_tbnn_invariants}
\end{equation}
This affine centering does not change the functions that the network can represent, but makes the equilibrium anchoring explicit and improves numerical conditioning near the undeformed state.

We represent the elastic free energy by a general invariant neural potential. To enforce a stress-free undeformed state while retaining nonlinear freedom away from equilibrium, we use the anchored parameterization
\begin{equation}
\begin{aligned}
\phi(\mathbf{x})
&= N_{\phi}(\mathbf{x})
-N_{\phi}(\mathbf{0})\\
&\quad-\mathbf{x}\cdot
\left[\nabla N_{\phi}(\mathbf{0})
-\mathbf{q}\right],\\
\mathbf{q}
&= \left(
\tfrac{1}{2},0,-\tfrac{1}{2}
\right).
\end{aligned}
\label{eq:memory_tbnn_phi}
\end{equation}
The anchoring fixes \(\phi(\mathbf{I})=0\) and \(\mathbf{K}(\mathbf{I})=\mathbf{0}\), while \(N_{\phi}\) learns the nonlinear dependence of stored elastic energy on the conformation invariants. Differentiation of the potential gives
\begin{equation}
\begin{aligned}
\mathbf{K}
&= 2\phi_1\mathbf{A}
+4\phi_2\mathbf{A}^2
+2\phi_3\mathbf{I},\\
\phi_i
&\equiv
\frac{\partial\phi}{\partial x_i}.
\end{aligned}
\label{eq:memory_tbnn_K_basis}
\end{equation}

The mobility is learned independently using the first two isotropic tensor powers:
\begin{equation}
\begin{aligned}
\widetilde{\boldsymbol{\mathcal{M}}}
(\mathbf{A})
&= m_0(\mathbf{x})\mathbf{I}
+m_1(\mathbf{x})\mathbf{A},\\
m_0(\mathbf{x})
&= \operatorname{softplus}
\!\left[N_0(\mathbf{x})\right],\\
m_1(\mathbf{x})
&= N_1(\mathbf{x}).
\end{aligned}
\label{eq:memory_tbnn_mobility}
\end{equation}
A smooth spectral constraint produces the positive-semidefinite mobility \(\boldsymbol{\mathcal{M}}\) used in Eq.~\eqref{eq:memory_tbnn_relax}; smooth determinant and coercivity floors likewise preserve numerical admissibility during optimization. The three scalar functions \(N_{\phi}\), \(N_0\), and \(N_1\) are independent multilayer perceptrons with two hidden layers of 32 \(\tanh\) units and a scalar linear output, giving 3,651 trainable network parameters in total, excluding the fitted physical scalars.

The learned potential and mobility together define the constitutive law, with classical conformation-tensor models recovered as particular choices of this pair. Oldroyd-B is recovered exactly for \(N_{\phi}=0\), \(m_0=1\), and \(m_1=0\). Giesekus behavior retains the Hookean potential but introduces a tensor-dependent mobility; upper-convected linear PTT uses a state-dependent scalar mobility; and FENE-P places the finite-extensibility nonlinearity in the learned potential. Equations~\eqref{eq:memory_tbnn_K_basis} and~\eqref{eq:memory_tbnn_mobility} constitute a memory TBNN: invariant scalar functions weight the stress basis \(\{\mathbf{I},\mathbf{A},\mathbf{A}^2\}\) and mobility basis \(\{\mathbf{I},\mathbf{A}\}\). 

We note, however, that because the explicit scalars are optimized together with the neural potential and mobility, constitutive scales can be distributed between the scalars and network weights; their trained values are therefore internal parameters of the TBNN representation and should not be identified directly with the correspondingly named parameters of the classical models described above.

This minimal basis was chosen to separate nonlinear stored elasticity from nonlinear relaxation and reduce ambiguity when identifying FENE-P-like behavior from flow observations. The construction remains systematically extensible through additional invariant inputs, higher-order isotropic mobility operators, or additional internal states while retaining the potential--mobility principle and its thermodynamic guarantee. In particular, additional internal variables and coupled positive-semidefinite mobilities could represent Rolie--Poly-like entanglement dynamics~\cite{likhtman2003simple} in thermodynamically consistent form~\cite{dolata2023thermodynamically,drucker2024definite}. Fully unconstrained source heads would broaden the hypothesis class further to that used by RUDEs~\cite{Lennon2023PNAS}, but thermodynamic consistency is not guaranteed by that architecture.

For the yield-stress fluid experiments, we augment the potential--mobility closure with an explicit fitted yield stress \(\tau_y\). Although yielding could in principle be encoded through a learned mobility that vanishes below yield, introducing \(\tau_y\) directly improves identifiability and training efficiency. A smooth Saramito factor gates the complete constrained mobility:
\begin{equation}
\begin{aligned}
\boldsymbol{\mathcal{M}}_y
&= \kappa_y\boldsymbol{\mathcal{M}},\\
\kappa_y
&= \delta_y
\operatorname{softplus}
\!\left(\frac{g_y}{\delta_y}\right),\\
g_y
&= 1-\frac{\tau_y}
{\lVert\boldsymbol{\tau}_d^{\,H}\rVert}.
\end{aligned}
\label{eq:memory_tbnn_yield}
\end{equation}
The yield criterion is evaluated using the fixed Hookean stress ruler
\begin{equation}
\begin{aligned}
\boldsymbol{\tau}_d^{\,H}
&= \operatorname{dev}
\!\left[
G_p(\mathbf{A}-\mathbf{I})
\right],\\
\lVert\boldsymbol{\tau}_d^{\,H}\rVert
&= \left[
\tfrac{1}{2}
\boldsymbol{\tau}_d^{\,H}:
\boldsymbol{\tau}_d^{\,H}
\right]^{1/2}.
\end{aligned}
\label{eq:memory_tbnn_yield_norm}
\end{equation}
Here, \(\delta_y=10^{-3}\) is the smoothing width introduced above for the Saramito truth model. Using the fixed Hookean stress prevents compensating changes in the learned stress map, \(G_p\), and \(\tau_y\) from shifting the yield surface and obscuring the distinction between yielded and arrested states. Below \(\tau_y\), the yield factor drives the complete relaxation mobility toward zero, producing sub-yield arrest while preserving reversible elastic deformation. Setting \(\tau_y=0\) recovers the non-yielding closure. The extension was disabled for the Giesekus and FENE-P trainings and enabled for the Saramito trainings.

\subsection{Learning and evaluation protocols for the memory TBNN}\label{sec:memory_methods}

\subsubsection*{Reference data}\label{subsec:memory_reference}

Ground-truth trajectories were generated using the conformation-tensor models and differentiable viscoelastic solver described above. Giesekus and FENE-P training used a \(4{:}1\) planar contraction extending \(6H\) upstream and \(12H\) downstream of the contraction plane, with upstream and downstream channel heights of \(8H\) and \(2H\), respectively. The inlet speed was \(U=0.5\), and both fluids used \(G_p=3.2\), \(\lambda=0.7\), and \(\eta_s=0.8\). The Giesekus truth model used \(\alpha=0.3\), whereas FENE-P used \(L^2=12\). Saramito training used startup plane-channel flows with the same \(G_p\), \(\lambda\), and \(\eta_s\), together with \(\tau_y=1.45\), at pressure-gradient magnitudes \(G_x=1.8\), \(2.5\), and \(4\). All fits used the native simulated trajectories without an additional observation operator or synthetic degradation. Both planar velocity components were observed throughout each transient; the Saramito fits additionally used the final bulk flow rate. Conformation, stress, constitutive-family labels, and yield surfaces were not supplied during training.

\subsubsection*{Contraction-flow training}\label{subsec:memory_contraction_training}

The contraction flows were initialized from rest, with the inlet velocity increased using the cosine ramp implemented by the forward solver,
\begin{equation}
\begin{aligned}
s(t)
&= \operatorname{clip}
\!\left(\frac{t}{t_r},0,1\right),\\
r(t)
&= \frac{1-\cos[\pi s(t)]}{2},\\
U_{\mathrm{in}}(t)
&= U\,r(t).
\end{aligned}
\label{eq:memory_inlet_ramp}
\end{equation}
The ramp time \(t_r\) is therefore the time required to reach the prescribed inlet speed. We used \(t_r=0.7\) for Giesekus and \(t_r=1.0\) for FENE-P, and integrated both trajectories to \(T=2\).

To emphasize the mixed shear and extension surrounding the contraction without using the ground-truth conformation, we applied the static geometric weight
\begin{equation}
\begin{aligned}
w(x,y)
&= \frac{w_{\mathrm{raw}}(x,y)}
{\left\langle w_{\mathrm{raw}}\right\rangle_{\Omega}},\\
w_{\mathrm{raw}}(x,y)
&= 1+\kappa\,
S\!\left(\frac{x-x_{\mathrm{on}}}{\ell}\right)
S\!\left(\frac{x_{\mathrm{off}}-x}{\ell}\right)\\
&\quad\times
\exp\!\left[
-\frac{(y-y_c)^2}{2\sigma_y^2}
\right],\\
x_{\mathrm{on}}
&= x_c-a,\\
x_{\mathrm{off}}
&= x_c+c\,\overline{U}\lambda_{\mathrm{ref}},
\qquad
\overline{U}=4U,\\
S(z)
&= \frac{1}{1+\exp(-z)}.
\end{aligned}
\label{eq:memory_roi_weight}
\end{equation}
Here, \(\langle\cdot\rangle_{\Omega}\) denotes the spatial mean. We used \(x_c=y_c=0\), \(a=0.5\), \(c=1.75\), \(\ell=0.4\), \(\sigma_y=0.4\), \(\kappa=4\), and the fixed reference value \(\lambda_{\mathrm{ref}}=0.7\), giving \(x_{\mathrm{on}}=-0.5\) and \(x_{\mathrm{off}}=2.45\). The reference relaxation time sets only the downstream extent of the observation region; the weight is independent of the truth conformation and polymer stress.

For each contraction experiment, the network was trained against both velocity components at all saved times using
\begin{equation}
\begin{aligned}
\mathcal{L}_{\mathrm{ctr}}
&= \sum_{n=1}^{N_t}
\sum_{j=1}^{N_x}
w(\mathbf{x}_j)
\left\|
\mathbf{u}_{\theta}(\mathbf{x}_j,t_n)
-\mathbf{u}^{*}(\mathbf{x}_j,t_n)
\right\|_2^2.
\end{aligned}
\label{eq:memory_contraction_loss}
\end{equation}
The headline Giesekus and FENE-P trainings used 24 alternating blocks. In each block, \(G_p\), \(\lambda\), and \(\eta_s\) were first optimized with bounded L-BFGS-B for at most 20 iterations while the network was fixed, followed by 25 Adam updates of the network while the physical scalars were fixed, for a total of 600 network updates. The Adam learning rate was \(10^{-4}\) for Giesekus and \(5\times10^{-4}\) for FENE-P. Two alternative partitions of the same 600 Giesekus network updates are reported in Supplementary Sec.~\ref{sec:si_giesekus}, with additional training choices discussed in Supplementary Sec.~\ref{sec:si_lessons}.

The physical scalars were initialized to unity, and the neural potential and mobility were initialized at their exact Oldroyd-B representation. Thus, the TBNN  was initially provided with no nonlinear mobility or finite-extensibility response; these distinct nonlinear mechanisms had to be inferred from the velocity trajectories. The reported FENE-P statistics used five independent hidden-layer initializations, whereas the Giesekus schedule comparison used a common initialization to isolate the effect of the optimization schedule.

\subsubsection*{Contraction-flow evaluation and transfer}\label{subsec:memory_contraction_evaluation}

Velocity errors were evaluated against the undegraded ground-truth fields using the RMSE and RRMSE definitions in Eqs.~\eqref{eq:rmse} and~\eqref{eq:rrmse}. Conformation was used only after training and was compared through componentwise and centerline profiles together with relative \(L^2\) errors. Agreement in these unobserved fields therefore tests whether velocity matching recovered the underlying material deformation rather than only the observed flow field. The trained closures were additionally interrogated using the rheometric readback and model-selection procedure described in the following section.

As a geometric-transfer test, the Giesekus closure obtained with the 24-block schedule was frozen and applied without retraining to lid-driven cavities at
\begin{equation}
\mathrm{De}
=\frac{\lambda U_{\mathrm{lid}}}{L}
\in\{0.20,0.35,0.50\}.
\label{eq:memory_cavity_de}
\end{equation}
The lid speed used the cosine ramp in Eq.~\eqref{eq:memory_inlet_ramp} with \(t_r=0.7\), and each trajectory was integrated to \(T=4\). We compared the steady normalized centerline velocity \(u_x/U_{\mathrm{lid}}\) and the transient maximum polymer stretch \(\max A_{xx}\). The former tests transfer of the predicted flow field, whereas the latter directly probes the learned material memory throughout startup. Complete cavity configurations and auxiliary field errors are given in Supplementary Sec.~\ref{sec:si_cavity}.

As a separate FENE-P identifiability control, the known-form FENE-P model was calibrated directly against the same \(U=0.5\) contraction velocities from four distinct parameter initializations. This control isolates parameter estimation with the constitutive family fixed from simultaneous learning of the family and its parameters; its protocol is given in Supplementary Sec.~\ref{sec:si_fenep}.

\subsubsection*{Yield-stress channel training and evaluation}\label{subsec:memory_evp_training_evaluation}

The Saramito closure was trained simultaneously against startup channel flows at \(G_x=1.8\), \(2.5\), and \(4\), each integrated to \(T=2.1=3\lambda\). Because the velocity magnitude varies strongly with the applied pressure gradient, the transient velocity contributions were balanced using
\begin{equation}
\begin{aligned}
W_i
&= \sum_{n,j}
\left\|
\mathbf{u}^{*}_{i}(\mathbf{x}_j,t_n)
\right\|_2^2,\\
\beta_i
&= \frac{W_{\max}}{W_i},
\qquad
W_{\max}=\max_i W_i.
\end{aligned}
\label{eq:memory_evp_weights}
\end{equation}
For the reported \(3\lambda\) training set, the resulting weights were \(\beta_{1.8}=5.895\), \(\beta_{2.5}=2.996\), and \(\beta_{4}=1\).

The full objective combined the weighted transient velocity mismatch with the final bulk flow rate,
\begin{equation}
\begin{aligned}
\mathcal{L}_{\mathrm{EVP}}
&= \sum_i\beta_i
\mathcal{L}_{\mathrm{vel},i}
+\lambda_Q\sum_i
\left[
\frac{Q_{\theta,i}(T)-Q_i^{*}(T)}
{Q_{s,i}}
\right]^2,\\
\mathcal{L}_{\mathrm{vel},i}
&= \sum_{n,j}
\left\|
\mathbf{u}_{\theta,i}(\mathbf{x}_j,t_n)
-\mathbf{u}^{*}_{i}(\mathbf{x}_j,t_n)
\right\|_2^2,\\
Q_i(t)
&= \int
\left\langle u_{x,i}(y,t)\right\rangle_x
\,\mathrm{d}y,\\
Q_{s,i}
&= \max\!\left[
\max_t\lvert Q_i^{*}(t)\rvert,
10^{-9}
\right].
\end{aligned}
\label{eq:memory_evp_loss}
\end{equation}
The coefficient \(\lambda_Q\) was fixed from the initial model by balancing the summed velocity and flow-rate contributions; for the reported \(3\lambda\) training set, \(\lambda_Q=221.67\). The bulk observable helps distinguish true arrest from a small but persistent residual flow near the yield threshold.

The potential and mobility networks were again initialized at their Oldroyd-B representation, while \(G_p\), \(\lambda\), \(\eta_s\), and \(\tau_y\) were initialized to unity. The network was first held fixed while the four physical scalars were optimized by bounded L-BFGS-B for at most 60 iterations. Training then proceeded through four cycles of 60 Adam network updates at a learning rate of \(5\times10^{-4}\), followed by an L-BFGS-B scalar re-optimization of at most 30 iterations. This initialization supplies the existence of a Saramito yield threshold through the explicit scalar \(\tau_y\), while leaving its magnitude and the remaining potential--mobility response to be inferred from the flow observations.

The five independently initialized trained closures were subsequently evaluated over a fixed drive ladder spanning \(0.5\leq G_x\leq5\), with only \(G_x=1.8\), \(2.5\), and \(4\) used during training. Yielded and near-yield cases were integrated for \(15\lambda\), while the unseen sub-yield cases \(G_x=0.5\), \(1\), and \(1.3\) were continued to \(30\lambda\) to distinguish damped elastic ring-down and arrest from persistent creep. Evaluation included velocity profiles, bulk flow rates, yield surfaces, and unobserved conformation fields. The complete drive ladder and robustness tests across alternative training drives and horizons are reported in Supplementary Sec.~\ref{sec:si_evp}.

\subsection{Differentiable model fitting}\label{sec:digital_rheometer}
\subsubsection*{Model simplification}
For inelastic constitutive laws such as those learned by the instantaneous TBNN, the stress is an algebraic function of the instantaneous kinematics, so probing the trained model under prescribed deformation histories requires no temporal integration. The memory TBNN and Oldroyd-B type models, however, are all partial differential equations that depend on space and time. When fitting to one-dimensional shear rheometer data, we can simplify these models into ODEs following the example of \cite{Lennon2023PNAS} by assuming that since the flow should be purely azimuthal $\vec{u}=z\vec{e}_\theta$ while varying only in the vertical direction, the stress $\tensor{\sigma}$ should also have no variations in the azimuthal direction. This means that the terms $\vec{u}\bdot\grad\tensor{\tau} =\vec{u}\bdot\grad\tensor{D} = \vec{0}$ and $\tensor{D}$ becomes independent of any spatial coordinates, leaving the extra-stress equation as an ODE only in terms of time. This assumption inherently assumes that the Weissenberg number is small enough that out-of-plane instabilities do not develop, which would break the assumption of spatial homogeneity. 

\subsubsection*{Rheometric fitting}\label{digital_rheometer_methods}
Under the quasi-two-dimensional flow assumption of a shear rheometer, the gradient of the flow $\grad \vec{u} = \dot{\gamma}(t)\vec{e}_{12}$ only has one non-zero component in the shear direction characterized by a potentially time-dependent shear rate $\dot{\gamma}$, which is spatially homogeneous across the entire rheometer. In a shear rheometer, this shear rate $\dot{\gamma}$ may be prescribed and the resulting shear stress $\sigma_{12}$ measured by averaging the torsional resistance encountered by the top plate. Alternatively, the shear stress may be prescribed and the strain measured directly. We will use the former case, though the approach would work equally well in the latter case. While the shear strain is assumed to be unidirectional, the stress $\tensor{\sigma}$ has six independent components which can be non-zero due to nonlinear coupling between the different stress terms in viscoelastic constitutive laws.

Under our previous assumptions, for a particular constitutive model we simulate the stress response to a prescribed strain rate $\dot{\gamma}$ by integrating the (time-only) constitutive equations forward in time to obtain all six stress components for a given set of material parameters. Most experimental data takes the form of a measured shear stress, $\hat{\sigma}_{12}(t_n)$, while the other five components of the ground truth are unknown. Here the $t_n$ are discrete measurement times. The standard loss function for fitting this data is the mean squared error,
\begin{equation}
    \mathcal{L} = \frac{1}{N}\sum_{n=1}^N \left(\hat{\sigma}_{12}(t_n) - \sigma_{12}(t_n; \theta)\right)^2,
    \label{eq:loss_fn}
\end{equation}
where $\sigma_{12}$ is our predicted shear stress and $\theta$ the vector of material parameters that defines the model. Thus, fitting a given model to the data takes the form of a nonlinear minimization problem over $\theta$, where the minimization must be performed through the solution to the governing equations which determine $\sigma_{12}$ for a given $\dot{\gamma}$.

We perform the integration using a differentiable solver Diffrax \cite{kidger2021on}, which allows us to automatically take gradients of the loss function, Eq. (\ref{eq:loss_fn}), with respect to the parameters $\theta$. We can use gradient-descent algorithms to efficiently find a set of parameters $\theta$ that best describe the data by minimizing the loss.

Because classical constitutive models have only a handful of parameters, we can efficiently fit an ensemble of candidate models to the same dataset, in contrast to neural-network-based approaches that rely on thousands of non-physical parameters. Assuming that the errors in the fit are independent and normally distributed, the log-likelihood of the particular estimate is given by
$$\hat{\mathcal{L}}_{\text{log-like}} = -\frac{N}{2} \left( \ln(2\pi\mathcal{L}) + 1 \right),$$
where $\mathcal{L}$ is the mean squared error defined in Eq. (\ref{eq:loss_fn}). The Bayesian information criterion for a particular fit to a model is then
\begin{equation}
    \text{BIC} = k \ln(N) - 2 \hat{\mathcal{L}}_{\text{log-like}}.
\end{equation}
Here, $k$ represents the number of parameters in the particular model, which is given simply as the length of the vector $\theta$. Under a Bayesian framework, the model which best describes the system \textit{given the observed data} has the smallest BIC, a measure which naturally penalizes more complicated models with higher degrees of freedom. This comparison thus ranks the specified candidate models under the assumed error model, but unlike a full Bayesian treatment \cite{Ewoldt2015Bayesian}, it does not provide posterior model probabilities or parameter credible intervals.

\subsubsection*{Model fitting protocol}
We apply this approach by generating 100 random models for each of Newtonian, Carreau-Yasuda, Oldroyd-B, Giesekus, and Linear PTT constitutive laws by randomly sampling material parameters for each model. The exact ranges of material parameters we used for this sample are given in Table~\ref{tab:parameter-ranges}. We then generate a set of ground truth data for each model by simulating synthetic shear stress data under the forcing function
\begin{equation}
    \dot{\gamma} = f\sin(\omega t)\label{eq:forcing}
\end{equation}
for every combination of $f\in[0.01,0.1,1.0,10.0]$ and $\omega\in[0.33,1.0,2.0]$. We add Gaussian noise with an amplitude of 0.03 onto the ground truth data to simulate experimental uncertainty. We fit this ground truth data back onto each of the five models, with initial guesses of one for each material parameter. We used an ADAM optimizer to perform the nonlinear optimization with a learning rate of $0.1$ over 1000 epochs. An example of this fitting process is shown in Supplementary Fig.~\ref{fig:demonstration} for a set of Giesekus ground truth data. 

We reran the Carreau-Yasuda parameter identification tests on a new sample of 24 models with a learning rate of 0.001 and 50,000 epochs, which led to much better convergence onto the ground-truth parameters.

\subsection{Computational cost and scaling}\label{sec:computational_cost}

Computational benchmarks are summarized in Supplementary Sec.~\ref{sec:si_computational_benchmarks} and Supplementary Table~\ref{tab:si_computational_cost}. The principal result is that closure complexity did not translate proportionally into solver cost: despite increasing the differentiated dimension from \(4\)--\(5\) analytic parameters to \(646\)--\(3{,}651\) network weights, forward-objective and value-and-gradient evaluations of the TBNNs required only \(1.22\)--\(1.60\) times those of the matched analytic models on a single NVIDIA A100-SXM4-80GB. For scale, the longest training supporting a main-text figure was the representative Giesekus run, which required \(34.40\) h. The principal TBNN tradeoff was adjoint memory, which increased by approximately \(2.7\)-fold for the memory closures and \(5.57\)-fold for the instantaneous closure, whose rematerialized blocks contained more physical timesteps. At comparable resolution per direction, extending an \(N_x\times N_y\) planar grid to \(N_x\times N_y\times N_z\) increases the leading cell-local work and state storage by approximately a factor of \(N_z\) and introduces additional velocity and conformation components. However, spatial distribution across accelerators can mitigate the cell-count growth, while more frequent rematerialization -- recomputing intermediate states rather than retaining them throughout the reverse pass -- trades peak adjoint memory for a controllable amount of forward computation. Moreover, although the general instantaneous TBNN requires an expanded three-dimensional tensor basis, the memory TBNN already uses three-dimensional conformation invariants and isotropic tensor bases and therefore requires neither a larger parameterization nor a reformulation merely to operate in three dimensions.

\section*{Data availability}

The data supporting the findings of this study, including the data underlying all display items, are available on Zenodo at \url{https://doi.org/10.5281/zenodo.22184699} and via the code repository listed below. The previously published hydrogel rheometry data analyzed in Supplementary Sec.~\ref{sec:gel_si} are available in ref.~\cite{Lennon2023PNAS}.

\section*{Code availability}

The differentiable non-Newtonian flow solver, TBNN training and analysis code, and the differentiable model-fitting framework are openly available at \url{https://github.com/asunol/jax_rheology}.

\section*{Acknowledgments}
We thank Kaylie Hausknecht, Erin Crawley, Randy Ewoldt, Gareth McKinley, Dave Weitz and Kaushik 
Bhattacharya for important discussions.
This work was supported by the Office of Naval Research (ONR N00014-23-1-2654) and the NSF AI Institute of Dynamic Systems (2112085).

\bibliographystyle{apsrev4-2}
\bibliography{arXiv_template}

\clearpage
\onecolumngrid
\appendix

\setcounter{section}{0}
\setcounter{subsection}{0}
\setcounter{figure}{0}
\setcounter{table}{0}
\setcounter{equation}{0}

\renewcommand{\thesection}{S\arabic{section}}
\renewcommand{\thesubsection}{S\arabic{subsection}}
\renewcommand{\thefigure}{S\arabic{figure}}
\renewcommand{\thetable}{S\arabic{table}}
\renewcommand{\theequation}{S\arabic{equation}}

\makeatletter
\renewcommand{\p@subsection}{}
\renewcommand{\theHsubsection}{S\arabic{subsection}}
\renewcommand{\theHfigure}{S\arabic{figure}}
\renewcommand{\theHtable}{S\arabic{table}}
\renewcommand{\theHequation}{S\arabic{equation}}
\makeatother

\begin{center}
{\large\bfseries Supplementary Information for:}\\[6pt]
{\Large\bfseries Learning constitutive models and rheology from partial flow measurements}\\[10pt]
Alp M. Sunol,$^{1,2,*}$ James V. Roggeveen,$^{1,*}$ Mohammed G. Alhashim,$^{3}$ Henry S. Bae,$^{1,4}$ and Michael P. Brenner$^{1,4,\dagger}$\\[6pt]
{\small
$^{1}$John A. Paulson School of Engineering and Applied Sciences, Harvard University, Cambridge, MA 02138, USA\\
$^{2}$School of Engineering, Brown University, Providence, RI 02912, USA\\
$^{3}$Research and Development Center, Saudi Aramco, Dhahran 31311, Saudi Arabia\\
$^{4}$Department of Physics, Harvard University, Cambridge, MA 02138, USA\\[4pt]
$^{*}$These authors contributed equally. \quad 
}
\end{center}
\vspace{6pt}

\phantomsection\label{sec:SI}

\subsection{Effect of training data resolution and noise}

Experimental flow measurements are typically lower in resolution and contain structured uncertainty. To assess how such factors influence learning, we synthetically degrade the training data to mimic micro-PIV conditions (Methods Sec.~\ref{sec:tbnn_learning_eval}). Specifically, the velocity field is down-sampled by Hann-windowed averaging over square interrogation windows of width $W_\mathrm{win}$, and measurement noise $\boldsymbol{\varepsilon}$ is added according to the correlated, heteroskedastic model described in the Methods. The noise amplitude scales with the 95$^{\mathrm{th}}$-percentile velocity $U_{95}$, and higher-shear regions receive proportionally larger perturbations, reproducing the anisotropic error structure of experimental velocimetry measurements. 

\begin{figure}[t!]
\centering
\includegraphics[width=\linewidth]{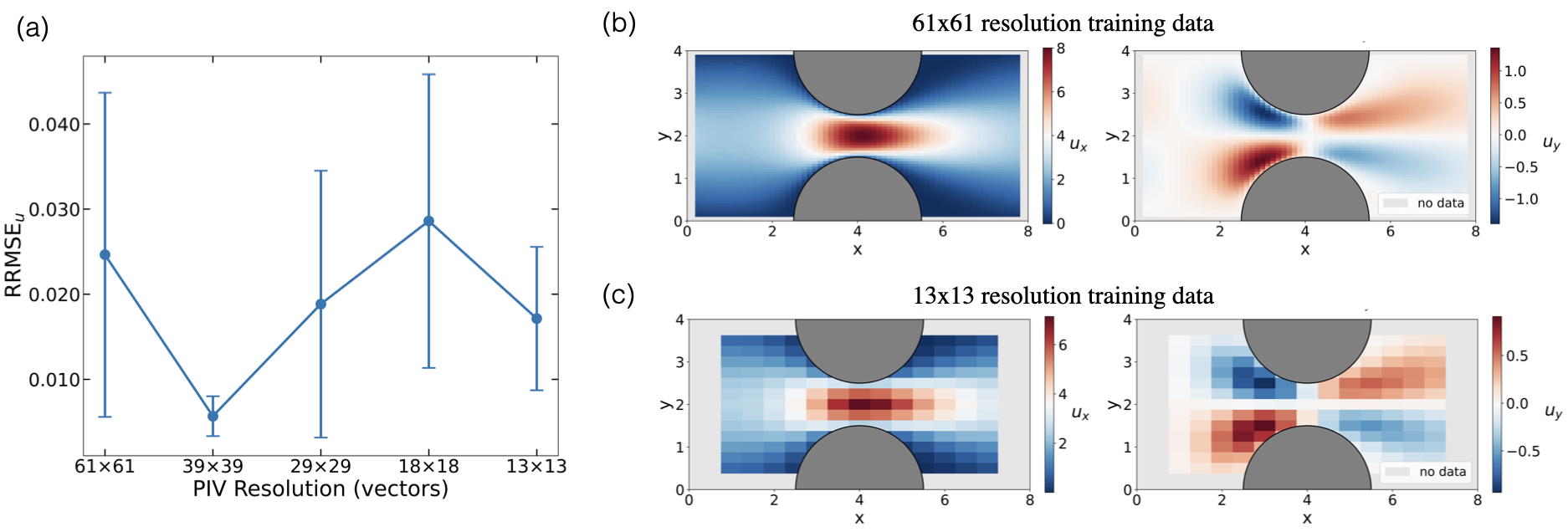}
\caption{\textbf{Robustness to coarse spatial resolution.}
(a) Relative root-mean-squared error (RRMSE$_u$) of predicted velocity fields as a function of synthetic PIV resolution, with error bars showing variation across three separate runs.
(b,c) Example downsampled velocity inputs used in training from the finest (61$\times$61) and coarsest (13$\times$13) resolutions.
Despite heavy downsampling, the trained model captures the dominant flow structures and reproduces velocity magnitudes with very low global error.}
\label{fig:sfig4}
\end{figure}

We first examine the effect of spatial resolution by varying $W_\mathrm{win}$ while keeping the noise parameters fixed. As shown in Fig.~\ref{fig:sfig4}a, the trained TBNN maintains its predictive accuracy, quantified by the relative root-mean-squared error (RRMSE$_u$), even when the resolution is reduced to $13\times13$ interrogation windows. This robustness demonstrates that spatially coarse flow measurements remain highly informative, containing sufficient kinematic diversity to recover the underlying constitutive behavior far beyond what is accessible from traditional bulk rheometry. Example  PIV-reconstructed velocity fields are shown in Fig.~\ref{fig:sfig4}b,c.

\begin{figure}[t!]
\centering
\includegraphics[width=\linewidth]{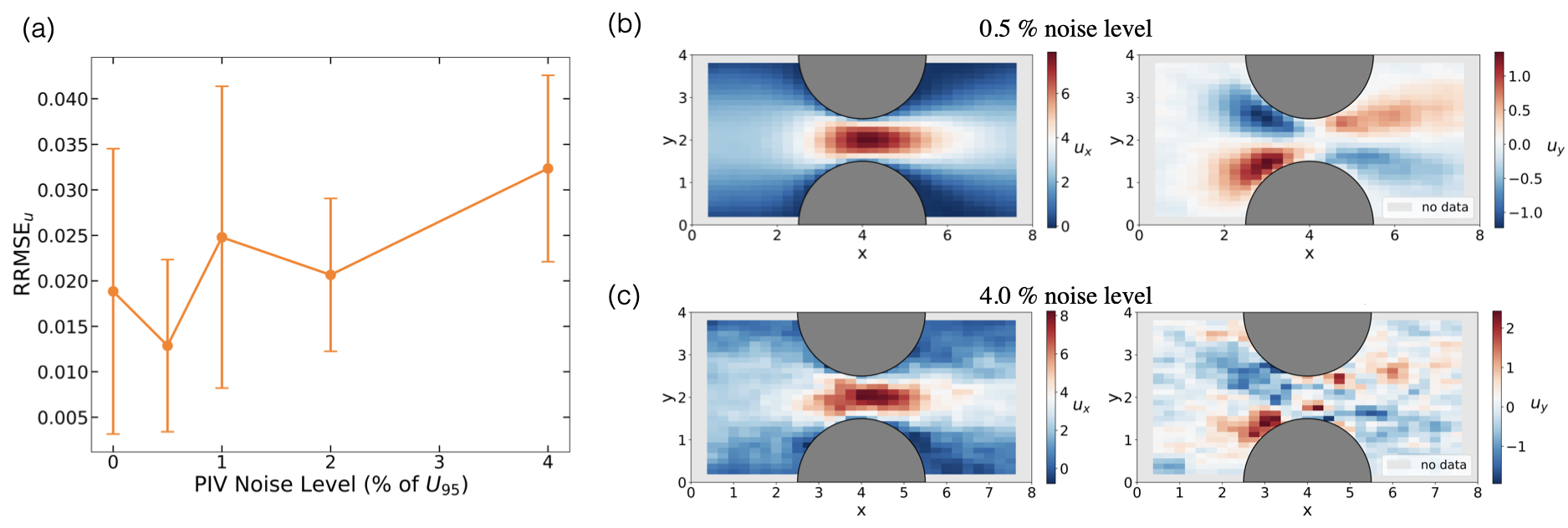}
\caption{\textbf{Robustness to measurement noise.}
(a) RRMSE$_u$ of TBNN-predicted velocity fields versus synthetic PIV noise level, with error bars showing variation across five separate runs except the no noise case where we performed three repetitions.
(b,c) Representative PIV velocity reconstructions used as training inputs for 0.5\% and 4.0\% noise amplitudes relative to \(U_{95}\).
Predictions remain stable and physically consistent even at high noise levels, confirming that the differentiable solver acts as a strong physical regularizer.}
\label{fig:sfig5}
\end{figure}

\begin{figure}[t!]
\centering
\includegraphics[width=\linewidth]{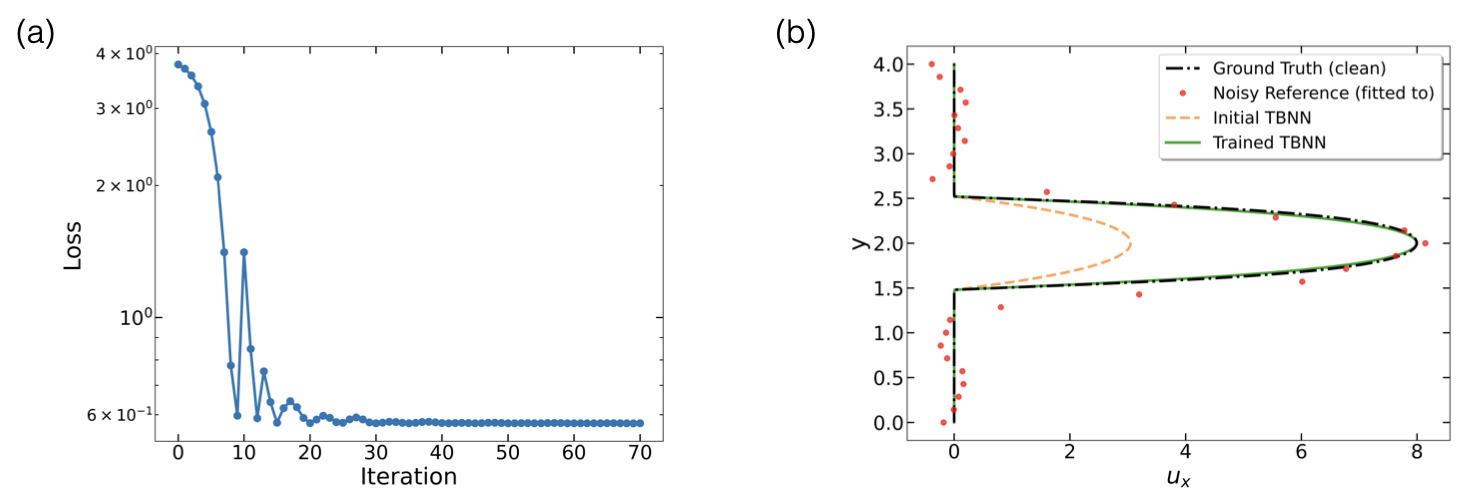}
\caption{\textbf{Training convergence and velocity-profile recovery for noisy data.}
(a) Evolution of the loss function during TBNN training for 4.0\% noise amplitude relative to $U_{95}$ shows a smaller decrease than the noise-free example, reflecting the irreducible mismatch introduced by measurement noise.
(b) Axial velocity profile at the constriction throat comparing the ground-truth field, noisy reference data, and TBNN predictions before and after training.
The trained model matches the ground-truth profile with minimal residual error despite noise.
Note that values within the solid obstacle (where the velocity should be zero) are excluded from training but shown here for visual continuity.}
\label{fig:sfig6}
\end{figure}

We next examine the effect of measurement noise using the same correlated, heteroskedastic model. Here the noise level on the horizontal axis of Fig.~\ref{fig:sfig5}a is defined as the amplitude of the base scale $\sigma_{\mathrm{base}}=\alpha\,U_{95}$, expressed as a percentage of the 95$^{\mathrm{th}}$-percentile velocity magnitude $U_{95}$ (computed on the $29\times 29$ downsampled vector grid, where the noise is added). As the noise amplitude increases, the RRMSE between the TBNN prediction and the ground-truth field rises slightly but remains low, even for perturbations as large as 4\%\, $U_{95}$. Although the total loss cannot decrease as much in these cases, reflecting the irreducible mismatch introduced by measurement noise (Fig.~\ref{fig:sfig6}a), the trained model still captures the underlying flow structure with high fidelity (Fig.~\ref{fig:sfig6}b). This robustness underscores a key advantage of our differentiable formulation: by enforcing the governing equations as hard constraints, the model resists overfitting noisy data and instead converges to the physically consistent constitutive relation, even when the available information is degraded. 

\subsection{Giesekus memory TBNN training and constitutive readback}\label{sec:si_giesekus}

We examine the training and interpretation of the memory TBNN for the Giesekus contraction experiment in greater detail. The ground-truth fluid had \(G_p=3.2\), \(\lambda=0.7\), \(\eta_s=0.8\), and \(\alpha=0.3\), corresponding to \(\eta_p=G_p\lambda=2.24\). Training used the \(4{:}1\) planar contraction at \(U=0.5\) (\(\mathrm{De}=0.35\)) on a \(128\times256\) computational grid. The timestep was \(10^{-4}\); 50 inner steps were taken between each of 400 saved states, giving a final time \(T=2\). The inlet velocity was increased from rest over \(t_r=0.7\) using the cosine ramp in Eq.~\eqref{eq:memory_inlet_ramp}. Both planar velocity components at every saved time entered the geometrically weighted loss in Eq.~\eqref{eq:memory_contraction_loss}; conformation, polymer stress, and pressure were not observed.

We compared three partitions of the alternating optimization schedule while holding the network initialization and total number of Adam updates fixed. Training~1 used six blocks of at most 40 L-BFGS-B scalar iterations followed by 100 Adam network updates; Training~2 used twelve blocks of 25 scalar iterations and 50 network updates; and Training~3 used twenty-four blocks of 20 scalar iterations and 25 network updates. Thus, every training contained 600 Adam updates and differed only in how frequently the physical scalars were re-optimized. The network learning rate was \(10^{-4}\), with a 20-update warmup, cosine decay over the complete training trajectory, and gradient clipping at unit global norm. The Adam state was retained across blocks. During the scalar steps, \(G_p\), \(\lambda\), and \(\eta_s\) were optimized with L-BFGS-B, with values constrained to \([0.02,20]\) and using relative function and gradient tolerances of \(10^{-9}\) and \(10^{-10}\), respectively. All three trainings began from \(G_p=\lambda=\eta_s=1\) and the same exact Oldroyd-B neural network initialization.

\begin{figure}[t!]
    \centering
    \includegraphics[width=\linewidth]{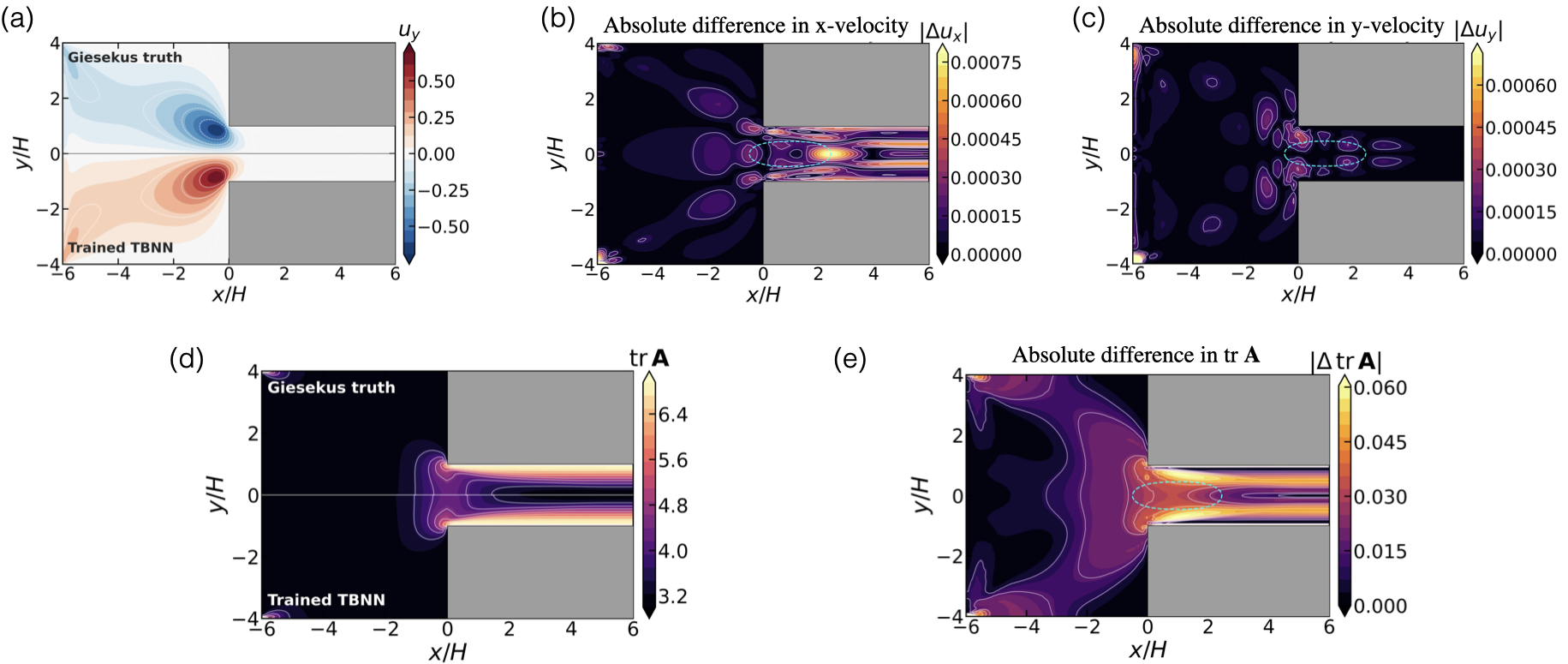}
    \caption{\textbf{Giesekus contraction fields and reconstruction errors.} Results are shown for the \(4{:}1\) contraction at \(U=0.5\) (\(\mathrm{De}=0.35\)). (a) Cross-stream velocity \(u_y\) for the Giesekus ground truth (upper half) and trained memory TBNN (lower half). (b,c) Absolute differences between the ground-truth and TBNN predictions for \(u_x\) and \(u_y\), respectively. (d) Conformation stretch \(\mathrm{tr}\,\mathbf{A}\) for the ground truth (upper half) and trained TBNN (lower half). (e) Absolute difference in \(\mathrm{tr}\,\mathbf{A}\). The cyan dashed outline in (b), (c), and (e) marks the high-stretch region emphasized by the training loss, spanning \(x/H=-0.50\) to \(2.45\). Although conformation was not observed during training, its relative \(L_2\) error is \(0.581\%\) over the full fluid domain and \(0.819\%\) within the weighted region.}
    \label{fig:sfig_giesekus_fields}
\end{figure}

\begin{figure}[t!]
    \centering
    \includegraphics[width=\linewidth]{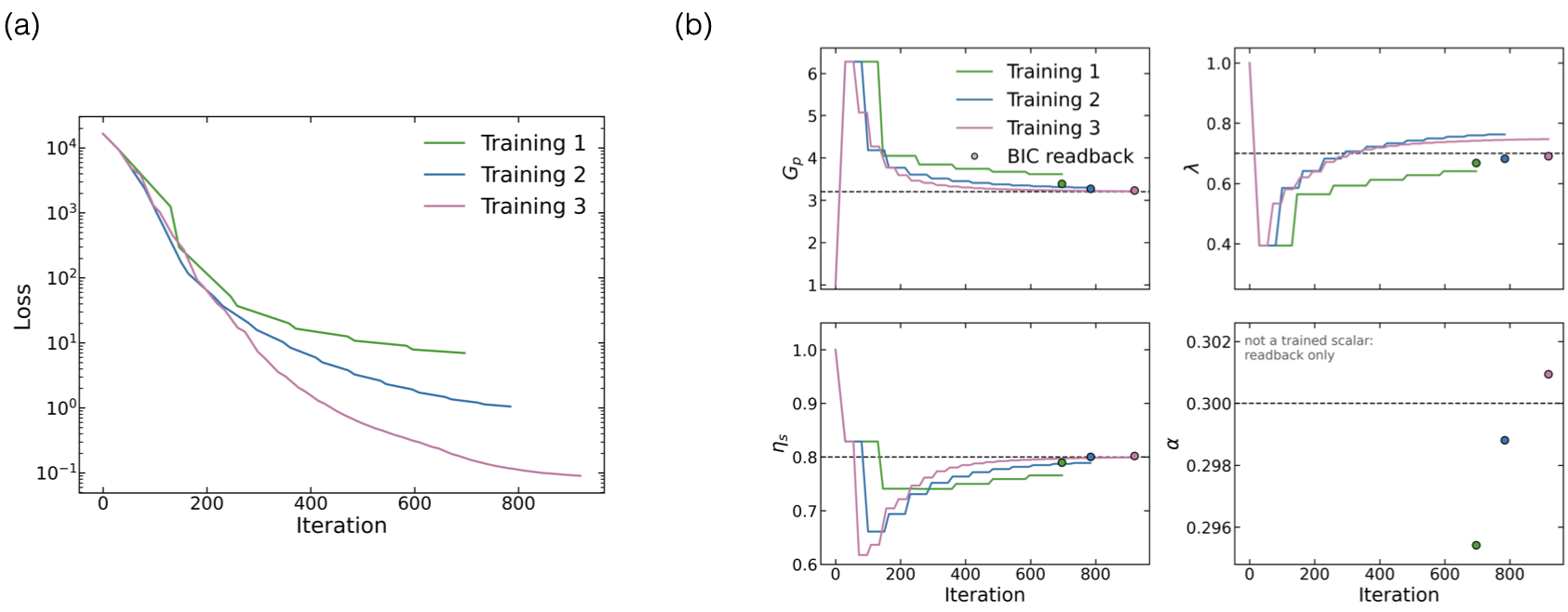}
    \caption{\textbf{Training and rheometric interrogation of the Giesekus memory TBNN.} (a) Training loss for three optimization schedules applied to the same network initialization. The schedules alternate optimization of the material scalars and network weights; curves show the accepted state at the end of each optimization block rather than rejected trial evaluations. (b) Solid lines are corresponding trajectories of the scalars \(G_p\), \(\lambda\), and \(\eta_s\) in the TBNN architecture. Dashed lines denote the ground-truth values \(G_p=3.2\), \(\lambda=0.7\), and \(\eta_s=0.8\). Circles mark values obtained by refitting the selected Giesekus model to each trained TBNN response; for the explicitly trained scalars, offsets from the solid-line endpoints reflect their coupling to the network weights in setting the learned constitutive scales. The Giesekus mobility parameter \(\alpha\) is represented implicitly by the network and is therefore not trained as a scalar; its recovered values are readbacks from the rheometric fits, with the ground truth \(\alpha=0.3\) shown by the dashed line.}
    \label{fig:sfig_giesekus_training}
\end{figure}

More frequent re-optimization of the physical scales improved convergence of the coupled inverse problem. The final velocity losses were \(6.99\), \(1.05\), and \(9.02\times10^{-2}\) for Trainings~1, 2, and 3, respectively (Supplementary Fig.~\ref{fig:sfig_giesekus_training}a). We therefore used Training~3 for the field comparisons in Fig.~\ref{fig:memory_tbnn} and Supplementary Fig.~\ref{fig:sfig_giesekus_fields}, and subsequently froze this closure for the cavity-transfer tests. Because the three trajectories share one network initialization, their spread measures sensitivity to the scalar--network alternation schedule rather than variability across random network seeds.

At the final contraction state, the trained closure reproduces both the observed velocity and the unobserved material deformation. Supplementary Fig.~\ref{fig:sfig_giesekus_fields} shows the cross-stream velocity, absolute errors in both velocity components, the conformation stretch \(\operatorname{tr}\mathbf{A}\), and its absolute error. Conformation recovery over a set \(\mathcal{J}\) of grid cells was quantified by
\begin{equation}
\begin{aligned}
\varepsilon(\mathcal{J})
&=
\frac{
\left[
\sum_{j\in\mathcal{J}}
\left(
\operatorname{tr}\mathbf{A}^{\mathrm{TBNN}}_j
-
\operatorname{tr}\mathbf{A}^{\star}_j
\right)^2
\right]^{1/2}
}{
\left[
\sum_{j\in\mathcal{J}}
\left(
\operatorname{tr}\mathbf{A}^{\star}_j
\right)^2
\right]^{1/2}
}.
\end{aligned}
\label{eq:si_giesekus_relative_l2}
\end{equation}
Here, \(j\) indexes grid cells with centers \(\mathbf{x}_j\), and \(\mathcal{J}_{\mathrm{fluid}}\) contains all cells in the fluid domain. The cyan contour bounds the binary ROI band
\(\mathcal{J}_{\mathrm{ROI}}
=
\{j\in\mathcal{J}_{\mathrm{fluid}}:
\chi(\mathbf{x}_j)\geq1/2\}\),
where
\(\chi=(w_{\mathrm{raw}}-1)/\kappa\)
is the ROI activation defined by Eq.~\eqref{eq:memory_roi_weight}. This band was used only to report a localized error; training used the continuous weight \(w(x,y)\), not a hard spatial crop. Although \(\mathbf{A}\) was absent from the loss, the relative \(L_2\) error in \(\operatorname{tr}\mathbf{A}\) is \(0.581\%\) over the 16,448 fluid cells and \(0.819\%\) over the 506 cells in the ROI band. Recovery of the conformation field shows that the fitted velocity trajectory constrains the nonlinear stress-storage and relaxation dynamics rather than only the final velocity solution.

The physical-scalar trajectories for all three trainings are shown in Supplementary Fig.~\ref{fig:sfig_giesekus_training}b. The solid curves are the values obtained during flow-field training, whereas the circles are independent parameter estimates obtained afterward by fitting a classical Giesekus model to each trained closure's rheometric response. The Giesekus mobility parameter \(\alpha\) is not an explicit training scalar: it is represented through the learned mobility and becomes available only through this constitutive readback. The recovered values are \(0.2954\), \(0.2988\), and \(0.3009\) for Trainings~1, 2, and 3, respectively, giving \(0.298\pm0.003\) across the three trainings.

For rheometric readback, each trained closure was evaluated under the twelve oscillatory shear-rate conditions described in Methods Sec.~\ref{sec:digital_rheometer}. Independent zero-mean Gaussian noise with standard deviation \(0.03\), expressed in the same nondimensional units as the total shear stress, was added to the TBNN-generated stress samples before fitting the candidate constitutive models. This is an absolute stress-noise scale rather than a fixed percentage of the response amplitude.

\begin{figure}[t!]
    \centering
    \includegraphics[width=0.75\linewidth]{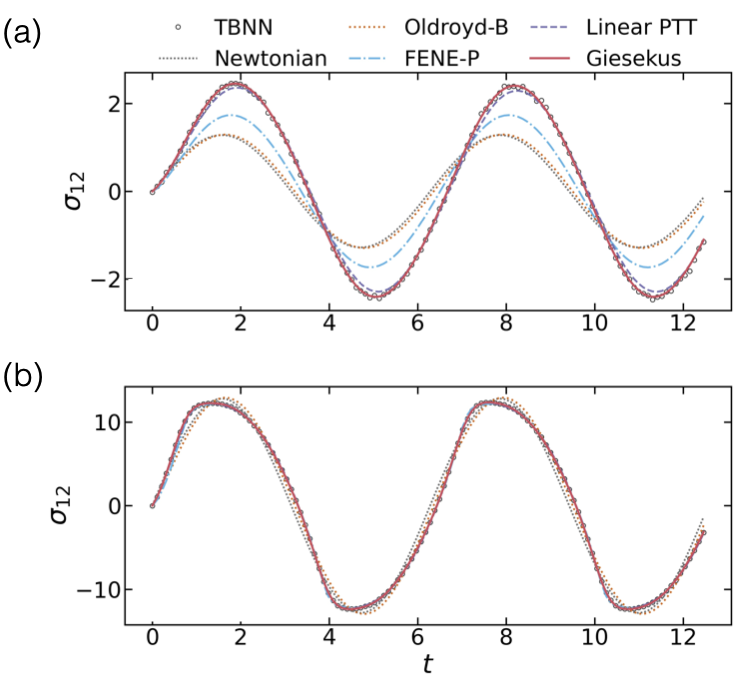}
    \caption{\textbf{Rheometric identification of the trained Giesekus memory TBNN.} Shear-stress response of the trained TBNN under oscillatory shear with \(\omega=1\) and strain-rate amplitudes (a) \(f=1\) and (b) \(f=10\). Two forcing cycles are shown. Open circles denote the TBNN response, and lines denote fits of the five candidate constitutive families. The selected Giesekus model reproduces the TBNN response at both amplitudes, whereas the rejected families show larger deviations, particularly at \(f=1\).}
    \label{fig:sfig_giesekus_rheometry}
\end{figure}

Rheometric model selection identifies Giesekus for every training. Relative to Giesekus, the closest competing family, linear PTT, has \(\Delta\mathrm{BIC}=6748\), \(7713\), and \(8402\) for Trainings~1, 2, and 3, respectively. For the Training~3 closure used in the main text, the remaining \(\Delta\mathrm{BIC}\) values are \(20{,}408\) for FENE-P, \(25{,}435\) for Oldroyd-B, and \(27{,}529\) for the Newtonian model. Supplementary Fig.~\ref{fig:sfig_giesekus_rheometry} displays two representative legs of the complete twelve-condition readback battery: \(\omega=1\) with strain-rate amplitudes \(f=1\) and \(10\). All twelve conditions entered each model fit; the displayed traces are only representative examples. The Giesekus fit follows the trained closure at both amplitudes, while the rejected families exhibit systematic waveform errors. The complete forcing, optimization, and BIC protocol is described in Methods Sec.~\ref{sec:digital_rheometer}.

\subsection{Transfer of the memory TBNN to the lid-driven cavity}\label{sec:si_cavity}

The memory TBNN trained only on the Giesekus contraction flow was frozen and transferred without further optimization to a square lid-driven cavity. For each Deborah number, the ground-truth Giesekus and learned-closure calculations used identical grids, timesteps, lid ramps, boundary conditions, and flow-solver settings, differing only in the constitutive closure. The Deborah number was defined using the ground-truth relaxation time \(\lambda=0.7\) and cavity side length \(L=1\), so that
\(\mathrm{De}=0.20\), \(0.35\), and \(0.50\)
corresponded to
\(U_{\mathrm{lid}}=0.286\), \(0.500\), and \(0.714\), respectively. The nominal Reynolds number was fixed at unity.

The \(\mathrm{De}=0.20\) calculations used a \(96\times96\) grid and a timestep of \(5\times10^{-5}\), whereas the \(\mathrm{De}=0.35\) and \(0.50\) calculations used a \(128\times128\) grid and a timestep of \(2\times10^{-6}\). In every case, the regularized lid velocity was increased from rest using the cosine ramp in Eq.~\eqref{eq:memory_inlet_ramp} with \(t_r=0.7\), and the flow was integrated to \(T=4\). All six trajectories satisfied the stress-inclusive steadiness criterion in Methods Sec.~\ref{subsec:visco_numerics}.

\begin{figure}[t!]
    \centering
    \includegraphics[width=\linewidth]{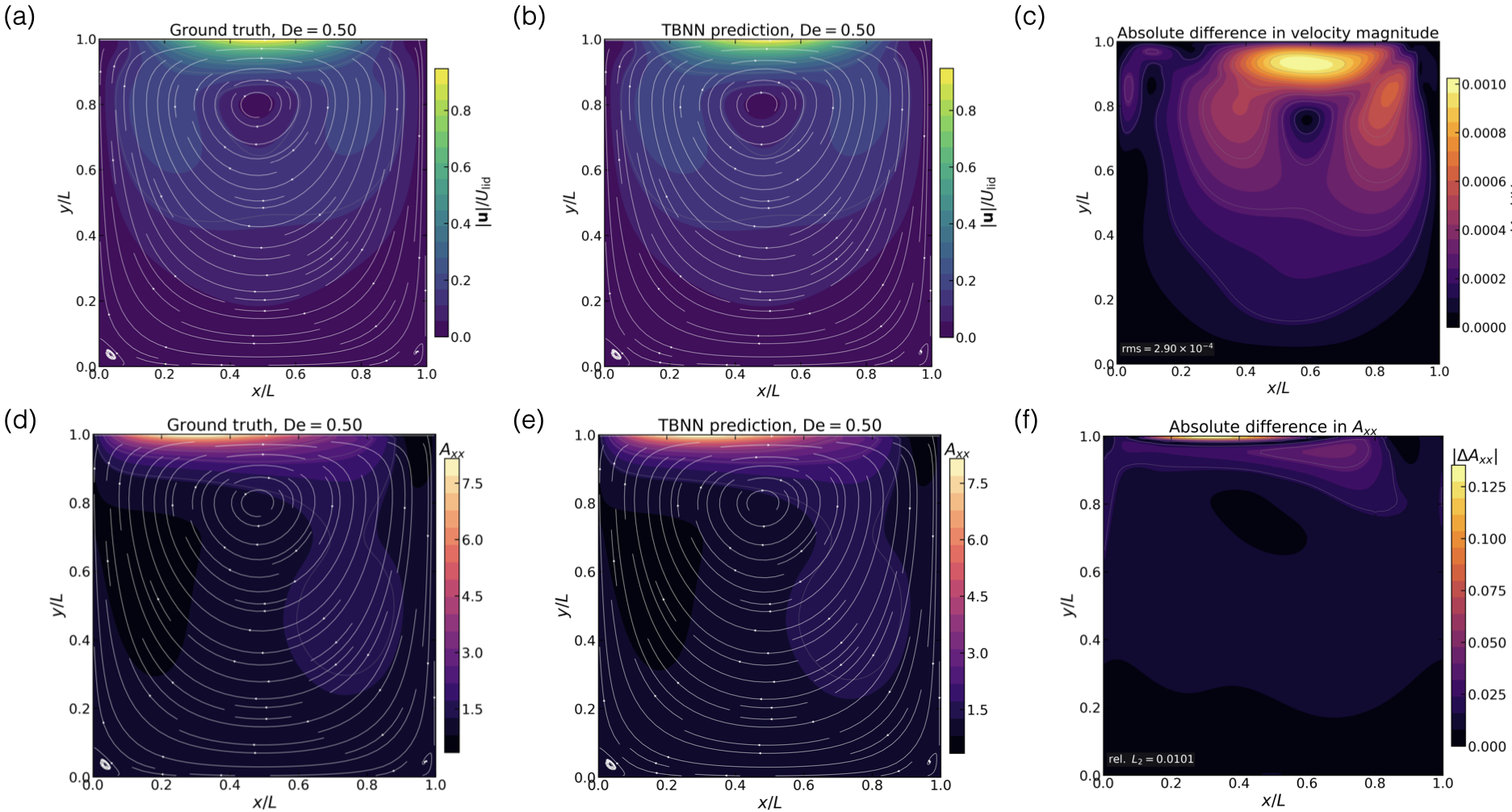}
    \caption{\textbf{Transfer of the contraction-trained memory TBNN to a lid-driven cavity.} Final steady-state fields at \(t=4\) and \(\mathrm{De}=0.50\), corresponding to \(U_{\mathrm{lid}}=0.714\). (a) Velocity magnitude \(\lVert\mathbf{u}\rVert/U_{\mathrm{lid}}\) and streamlines for the Giesekus ground truth and (b) TBNN prediction. (c) Normalized velocity error \(\lVert\Delta\mathbf{u}\rVert/U_{\mathrm{lid}}\), with rms error \(2.90\times10^{-4}\). (d) Conformation component \(A_{xx}\) and streamlines for the ground truth and (e) TBNN prediction. (f) Absolute difference in \(A_{xx}\), with relative \(L_2\) error \(0.0101\). The memory TBNN was trained only in the \(4{:}1\) contraction and was evaluated here without cavity training data.}
    \label{fig:sfig_cavity_fields}
\end{figure}

Figure~\ref{fig:memory_tbnn}d compares the steady vertical-centerline velocity profiles and transient maximum \(A_{xx}\) across the complete Deborah-number ladder. Supplementary Fig.~\ref{fig:sfig_cavity_fields} shows the corresponding two-dimensional velocity and \(A_{xx}\) fields at \(\mathrm{De}=0.50\). At this highest tested Deborah number, the domain-wide rms velocity difference was \(2.90\times10^{-4}U_{\mathrm{lid}}\), and the relative \(L_2\) error in the final \(A_{xx}\) field was \(1.01\%\).

\subsection{FENE-P family identification and parameter identifiability}\label{sec:si_fenep}

\begin{figure}[t!]
    \centering
    \includegraphics[width=\linewidth]{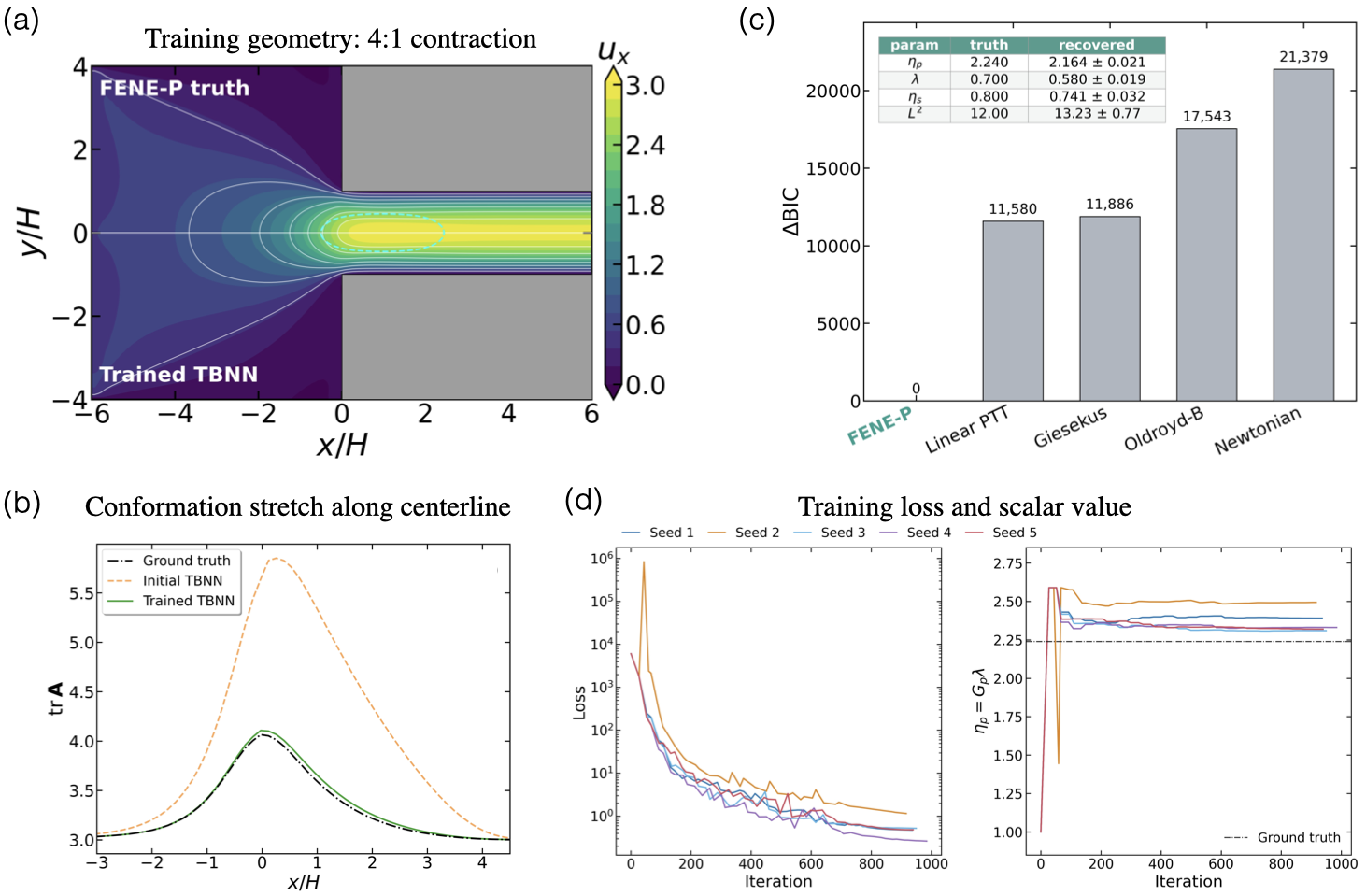}
    \caption{\textbf{Learning FENE-P rheology with memory TBNN.} (a) Steady streamwise velocity \(u_x\) in a \(4{:}1\) contraction at \(U=0.5\) (\(\mathrm{De}=0.35\)) for the FENE-P ground truth (upper half) and trained TBNN (lower half). (b) Centerline conformation stretch \(\mathrm{tr}\,\mathbf{A}\) for the ground truth, initial TBNN, and trained TBNN. Although conformation was not observed during training, the trained model reproduces the ground-truth profile with slightly larger deviations than in the Giesekus case (Fig.~\ref{fig:memory_tbnn}). (c) Difference in Bayesian Information Criterion, \(\Delta\mathrm{BIC}\), relative to the best-fitting constitutive family for a representative trained TBNN; FENE-P is selected over Linear PTT, Giesekus, Oldroyd-B, and Newtonian models. Inset: ground-truth parameters and values recovered by the rheometric readback, mean \(\pm\) sample standard deviation across five independently seeded TBNNs with identical training protocol. The correct family is identified, but the parameters are recovered less accurately than in the Giesekus case. (d) Training loss and the polymer viscosity \(\eta_p=G_p\lambda\) formed from the scalars fitted during training, for the five seeds; the dashed line marks the ground truth \(\eta_p=2.24\).}
    \label{fig:sfig_fenep_recovery}
\end{figure}

We next examine finite-extensibility recovery using a FENE-P ground truth with \(G_p=3.2\), \(\lambda=0.7\), \(\eta_s=0.8\), and \(L^2=12\), corresponding to \(\eta_p=G_p\lambda=2.24\). Five independently initialized memory TBNNs were trained using the same \(U=0.5\), velocity-only contraction protocol described in Methods Sec.~\ref{subsec:memory_contraction_training}. Each network began with the exact Oldroyd-B constitutive form but with \(G_p=\lambda=\eta_s=1\); both the correct material scales and the finite-extensibility response therefore had to be inferred from the velocity trajectories. Supplementary Fig.~\ref{fig:sfig_fenep_recovery}a,b shows that the trained closure reproduces the velocity field, while the unobserved centerline conformation stretch \(\operatorname{tr}\mathbf{A}\) is recovered with a maximum relative deviation well below \(2\%\), despite conformation being absent from the training loss.

Each trained closure was subjected to the same twelve-condition oscillatory-shear readback used for the Giesekus experiment. Independent Gaussian noise with standard deviation \(0.03\) in nondimensional total-stress units was added before fitting the five candidate families, with three optimization restarts per candidate. FENE-P was selected for all five trained closures. For the representative closure shown in Supplementary Fig.~\ref{fig:sfig_fenep_recovery}c, the \(\Delta\mathrm{BIC}\) values relative to FENE-P were \(11{,}580\) for linear PTT, \(11{,}886\) for Giesekus, \(17{,}543\) for Oldroyd-B, and \(21{,}379\) for the Newtonian model. During flow-field training, the explicit polymer stress scale is \(G_p\); for comparison with the rheometric parameterization, Supplementary Fig.~\ref{fig:sfig_fenep_recovery}d reports the corresponding polymer viscosity \(\eta_p=G_p\lambda\). Across the five readbacks, the recovered parameters were
\(\eta_p=2.164\pm0.021\),
\(\lambda=0.580\pm0.019\),
\(\eta_s=0.741\pm0.032\), and
\(L^2=13.23\pm0.77\),
where the uncertainties are sample standard deviations across the independently initialized networks. Thus, the constitutive family is consistently identified, although some physical parameters, most notably \(\lambda\), are recovered less precisely than for Giesekus.

\begin{figure}[t!]
    \centering
    \includegraphics[width=\linewidth]{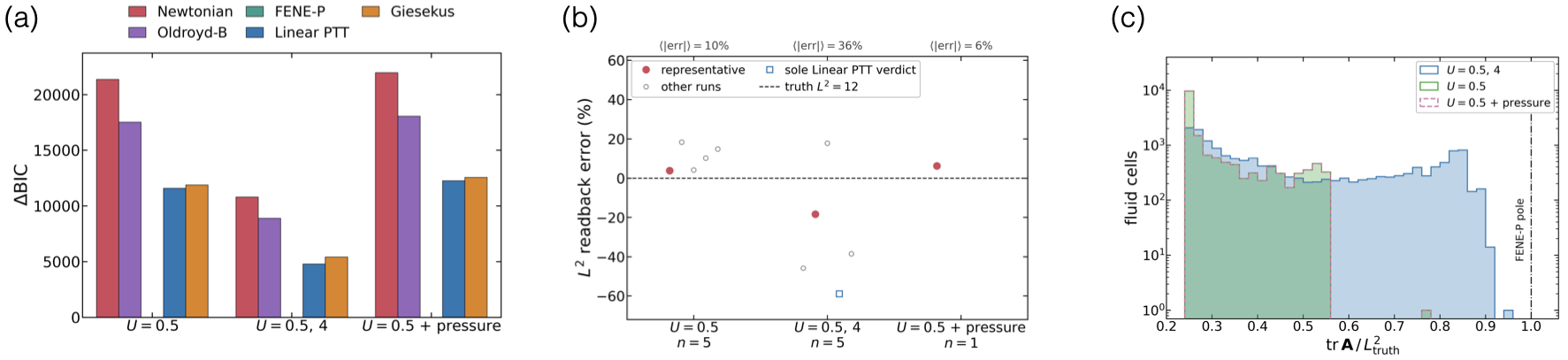}
    \caption{\textbf{FENE-P model selection and finite-extensibility identifiability across training conditions.} (a) \(\Delta\mathrm{BIC}\) for one representative run per training condition, selected by the lowest mean absolute relative error across the recovered \(\eta_p\), \(\lambda\), \(\eta_s\), and \(L^2\). Conditions are single-rate velocity data at \(U=0.5\), dual-rate velocity and pressure data at \(U=0.5\) and \(4\) with equal rate weighting, and single-rate velocity and pressure data at \(U=0.5\). The absent bar at zero identifies the selected family. For each representative run, FENE-P is selected over Linear PTT, Giesekus, Oldroyd-B, and Newtonian models. (b) Relative error in the recovered extensibility \(L^2\) for every run. Mean absolute errors are \(10\%\), \(36\%\), and \(6\%\), respectively. The open blue square marks the one run out of eleven for which the readback selected Linear PTT instead of FENE-P; this run also has the poorest constitutive reconstruction, with its best candidate retaining a mean-squared residual approximately \(1.68\times10^3\) times the injected-noise variance. (c) Distribution of the ground-truth \(\mathrm{tr}\,\mathbf{A}/L^2_{\mathrm{truth}}\) over the final-frame fluid cells for each representative condition, using \(L^2_{\mathrm{truth}}=12\). The FENE-P pole---the finite-extensibility limit---lies at one. The single-rate and single-rate-plus-pressure curves coincide because they sample the same truth field. The dual-rate condition reaches \(0.959\), but only \(15\) of \(16{,}448\) cells exceed \(0.9\); thus, its increased maximum stretch provides little probability mass near the pole and does not improve recovery of \(L^2\).}
    \label{fig:sfig_fenep_identifiability}
\end{figure}

The preceding results refer exclusively to the five single-rate, velocity-only trainings at \(U=0.5\). We then tested two extensions separately. First, to isolate the effect of adding another flow observable without changing the sampled flow, one additional TBNN was trained at \(U=0.5\) using both velocity and pressure. Second, to test whether sampling a more strongly stretched flow improved finite-extensibility recovery, five additional TBNNs were trained using velocity and pressure at both \(U=0.5\) and \(4\) (Supplementary Fig.~\ref{fig:sfig_fenep_identifiability}).

For each pressure-enriched flow condition, the velocity loss was augmented by
\begin{equation}
\begin{aligned}
\mathcal{L}_{\mathrm{vp}}
&=
\mathcal{L}_{\mathrm{vel}}
+w_p\mathcal{L}_{p},\\
\mathcal{L}_{p}
&=
\sum_{n\in\mathcal{S}}
\sum_{k=2}^{4}
\left[
\frac{
\left(p_{\theta,k}(t_n)-p_{\theta,1}(t_n)\right)
-
\left(p_k^{*}(t_n)-p_1^{*}(t_n)\right)
}{
\displaystyle
\max_{t>t_r}
\left|p_k^{*}(t)-p_1^{*}(t)\right|
}
\right]^2.
\end{aligned}
\label{eq:si_fenep_pressure_loss}
\end{equation}
Here, \(\mathcal{L}_{\mathrm{vel}}\) is the velocity loss in Eq.~\eqref{eq:memory_contraction_loss} evaluated for one inlet speed, while \(p_{\theta,k}\) and \(p_k^{*}\) are the predicted and ground-truth pressures at tap \(k\). The four taps, indexed by \(k=1,\ldots,4\), were located at
\(\mathbf{x}_k/H=(-2.917,3.583)\),
\((-0.750,3.583)\),
\((0.750,0.583)\), and
\((6.083,0.583)\).
Tap \(k=1\) is the reference, so the loss compares the three pressure differences \(p_k-p_1\) for \(k=2,3,4\). The set \(\mathcal{S}\) contains eight uniformly spaced post-ramp states, and \(w_p\) was set separately for each flow condition at initialization to balance its pressure and velocity contributions.

The single-rate pressure experiment used this loss only at \(U=0.5\). The dual-rate trainings also began at \(U=0.5\); after the first alternating optimization block, the \(U=4\) trajectory was activated and the losses for the two flow conditions were averaged, with their velocity contributions weighted equally.

Adding pressure to the single-rate training did not systematically improve parameter readback. For the five single-rate velocity-only trainings, the mean absolute relative error in \(L^2\) was \(10.28\%\), with individual errors ranging from \(3.83\%\) to \(18.34\%\). The single pressure-enriched training produced errors of \(3.78\%\), \(14.24\%\), \(8.39\%\), and \(6.21\%\) in \(\eta_p\), \(\lambda\), \(\eta_s\), and \(L^2\), respectively. Its mean absolute relative error across all four parameters was \(8.16\%\), compared with a velocity-only mean of \(9.55\%\) and best value of \(8.08\%\). Thus, pressure improved recovery relative to the velocity-only mean but not relative to the best velocity-only training, and the result remained within the observed seed-to-seed variation. Because only one pressure-enriched network was trained, it does not provide evidence of a systematic improvement from adding pressure. The velocity-only and pressure-enriched trainings sample the same constitutive states, as reflected by their coincident distributions in Supplementary Fig.~\ref{fig:sfig_fenep_identifiability}c; pressure supplies an additional observable of those states but does not expand the sampled state space. Importantly, this also reflects a structural property of the differentiable-physics framework: because the solver enforces mass and momentum conservation exactly, a closure that reproduces the velocity carries a dynamically consistent pressure at no additional cost, whereas methods that impose the governing equations only as soft penalties can fit velocity fields while yielding an inconsistent pressure.

Adding a second, high-rate trajectory likewise did not improve finite-extensibility readback: the five dual-rate velocity-and-pressure trainings had a mean absolute \(L^2\) error of \(36\%\), compared with \(10.28\%\) for the five single-rate velocity-only trainings (Supplementary Fig.~\ref{fig:sfig_fenep_identifiability}b). Across all eleven trained closures, ten readbacks selected FENE-P (Supplementary Fig.~\ref{fig:sfig_fenep_identifiability}a). The sole exception was one dual-rate closure for which linear PTT had the lowest BIC; however, even its best candidate fit retained a mean-squared residual of \(1.516\), compared with the injected-noise variance of \(9\times10^{-4}\). We therefore treated this result as an inconclusive readback rather than a meaningful Linear-PTT identification. The same prespecified rheometric optimization protocol was applied to every closure without case-specific retuning.

The lack of improvement and degradation under dual-rate training has two parts. The first is the limit of the upside: the Peterlin pole occurs as \(\operatorname{tr}\mathbf{A}/L^2\) approaches unity, and although the representative dual-rate flow reached \(0.959\), only \(15\) of \(16{,}448\) final-frame fluid cells exceeded \(0.9\) (Supplementary Fig.~\ref{fig:sfig_fenep_identifiability}c), so the added rate contributed little probability mass in the strongly nonlinear region where \(L^2\) is most identifiable. The degradation itself may be an optimization effect: the near-pole states are stiff, and emphasizing them appears to inject high-variance gradients through the unrolled solver that degrade the closure at the very states carrying the finite-extensibility signal.  This is consistent with the widening of the seed-to-seed spread against our expectation, and with the uniformly smaller \(\Delta\mathrm{BIC}\) margins of the dual-rate readbacks (Supplementary Fig.~\ref{fig:sfig_fenep_identifiability}a).

We thus tested whether the data themselves suffice by replacing the neural closure with the known FENE-P model and calibrating it directly against the same single-rate, velocity-only \(U=0.5\) data. We optimized \(\log G_p\), \(\log\lambda\), \(\log\eta_s\), and \(\log L^2\) using L-BFGS-B with at most 60 iterations and function and gradient tolerances of \(10^{-12}\) and \(10^{-10}\), respectively. The four initializations shown in Supplementary Fig.~\ref{fig:sfig_fenep_direct} were \((G_p,\lambda,\eta_s,L^2)=(2,1.5,0.5,50)\), \((3,0.4,1,8)\), \((4.5,0.5,0.9,20)\), and \((7.921,0.880,0.525,26.711)\). All four recovered the ground-truth parameters within \(20\)--\(26\) objective evaluations, with final losses near machine precision. An additional extreme initialization with \(L^2=200\) remained in a nearly Oldroyd-B basin far from the pole and never converged. These known-form fits show that, once the constitutive family is fixed and the relevant optimization basin is reached, the velocity data are enough to determine all four FENE-P parameters. The less precise readback from the model-agnostic TBNN thus reflects the difficulty in the broader task of learning both the constitutive form and its parameters simultaneously from the finite range of states visited by the training flow.

\subsection{Elastoviscoplastic yield-stress recovery}\label{sec:si_evp}

The plane channel had streamwise length \(L_x=1\), full height \(L_y=2\), and half-height \(H=1\), with periodic streamwise boundaries and no-slip walls. The domain was discretized using \(N_x\times N_y=32\times64\) cells with timestep \(\Delta t=2.5\times10^{-3}\); 10 solver steps were taken between saved states. The headline five-seed trainings used \(G_x=\{1.8,2.5,4\}\) and \(T=2.1=3\lambda\). To test sensitivity to the sampled forcing conditions and training duration, we additionally considered the Cartesian product of two drive sets, \(G_x=\{1.8,2.5,4\}\) and \(\{1.6,1.8,2.5\}\), and two time horizons, \(T=2.1=3\lambda\) and \(T=5.0=7.14\lambda\). These four protocol comparisons used a common network initialization and otherwise retained the loss and optimization schedule described in Methods Sec.~\ref{sec:memory_methods}.

\begin{figure}[t!]
    \centering
    \includegraphics[width=\linewidth]{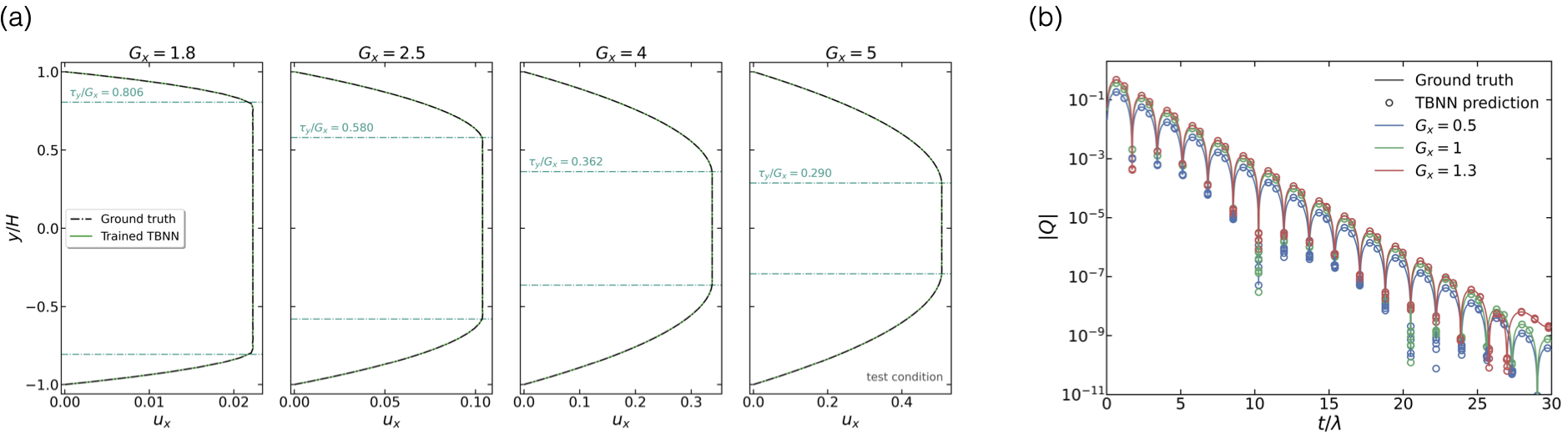}
    \caption{\textbf{Elastoviscoplastic velocity profiles and sub-yield arrest.} Results show the Saramito ground truth and five independently seeded memory TBNNs. (a) Streamwise-velocity profiles after \(15\lambda\) at the three training conditions \(G_x=1.8\), \(2.5\), and \(4\), and at the unseen test condition \(G_x=5\). Dash-dotted lines mark the yield surfaces \(y/H=\pm\tau_y/G_x\). The five trained predictions overlap at plot resolution. (b) Sub-yield flow-rate magnitude \(|Q|\) through \(30\lambda\) at \(G_x=0.5\), \(1\), and \(1.3\), all below the critical drive \(G_c=1.45\). Solid lines denote the ground truth and open markers the five TBNN predictions. The sub-yield response is a damped elastic ring-down rather than persistent creep: the truth and learned closures oscillate at the same frequency and phase while decaying through approximately eight decades.}
    \label{fig:sfig_evp_profiles_arrest}
\end{figure}

The reported five-seed evaluation ladder was \(G_x=\{0.5,1,1.3,1.45,1.6,1.8,2.5,4,4.5,5\}\). To test the trained closure, we evolved dynamics for \(15\lambda\) (extrapolating beyond the horizon we trained on). The five independently initialized memory TBNNs reproduced the Saramito velocity profiles and yield surfaces at the training drives as well as at the unseen test condition up to \(G_x=5\) (Supplementary Fig.~\ref{fig:sfig_evp_profiles_arrest}a). At the sub-yield drives \(G_x=0.5\), \(1\), and \(1.3\), the response is a damped elastic ring-down rather than persistent creep: the ground truth and learned closures oscillate at the same frequency and phase while the flow rate decays through approximately eight decades to order \(10^{-9}\) over \(30\lambda\) (Supplementary Fig.~\ref{fig:sfig_evp_profiles_arrest}b).

\begin{figure}[t!]
    \centering
    \includegraphics[width=\linewidth]{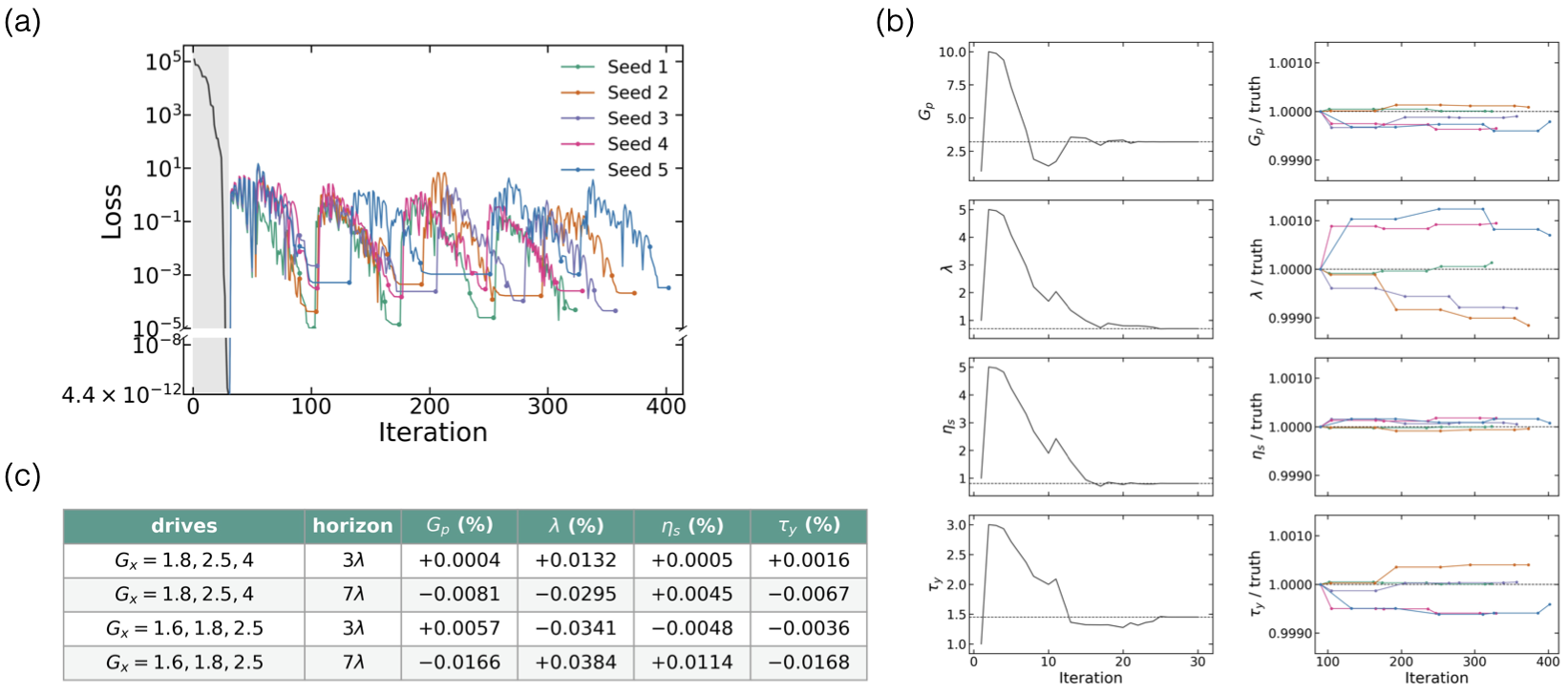}
    \caption{\textbf{Training convergence and robustness of elastoviscoplastic parameter recovery.} Results in (a,b) show five network seeds trained at \(G_x=1.8\), \(2.5\), and \(4\) over a \(3\lambda\) horizon. All four material scalars begin at \(1.0\), while the seeds differ in the initial hidden-layer weights. (a) Training loss. During the initial deterministic stage, the network is held at its Oldroyd-B representation while \(G_p\), \(\lambda\), \(\eta_s\), and \(\tau_y\) are optimized, reducing the loss from \(1.19\times10^5\) to \(4.4\times10^{-12}\). Subsequent blocks alternate optimization of the network weights and material scalars. (b) Material-parameter trajectories. The left column shows the shared initial scalar optimization in physical units, with dashed lines marking ground truth. The right column shows the accepted parameter values during the subsequent network-training blocks, normalized by ground truth for each seed. Despite releasing the network weights, every final scalar remains within \(0.116\%\) of its ground-truth value. (c) Signed recovery errors for four training protocols formed from two sets of driving conditions and training horizons of \(T=2.1\) (\(3\lambda\)) or \(T=5.0\) (\(7.14\lambda\)), with all scalars initialized at \(1.0\). No reported parameter differs from ground truth by more than \(0.0384\%\), demonstrating that recovery is insensitive to the chosen drive set and training horizon.}
    \label{fig:sfig_evp_training}
\end{figure}

During the initial training stage, the neural potential and mobility were held at their Oldroyd-B initialization while the explicit Saramito yield gate remained active and \(G_p\), \(\lambda\), \(\eta_s\), and \(\tau_y\) were optimized. Because the Saramito truth model is recovered within this initialized architecture when the four scalars take their correct values, this stage reduced the loss from \(1.19\times10^5\) to \(4.4\times10^{-12}\). Releasing the network weights therefore tested whether this physical solution remained stable within the broader TBNN hypothesis class. Across five network initializations, the recovered yield stress was \(\tau_y=1.44984\pm0.00057\), and every final material scalar remained within \(0.116\%\) of its ground-truth value. Across the four common-initialization drive-set and horizon protocols defined above, no recovered scalar differed from ground truth by more than \(0.0384\%\) (Supplementary Fig.~\ref{fig:sfig_evp_training}).

\subsection{Training considerations for differentiable TBNN closures}\label{sec:si_lessons}

Training through a fluid solver couples the explicit material scales to the functional shape represented by the network, creating directions in which the two can compensate for one another. For the instantaneous TBNN, we found a short staged optimization useful: holding \(\eta_0\) fixed removed one adjustable viscosity scale, while fitting \(\eta_\infty\) first established the viscosity contrast before the higher-dimensional network learned its shear-rate dependence. This was a pragmatic strategy for avoiding rapid convergence toward a nearby Newtonian response rather than a requirement of the architecture. In preliminary memory TBNN optimizations, simultaneously releasing the physical scalars and network weights produced compensating updates that impeded the learned potential and mobility from developing the nonlinear constitutive response. In these preliminary tests, Adam updates of the scalars allowed the relaxation time to drift, whereas joint L-BFGS-B optimization of the network and scalars remained near the Oldroyd-B initialization. The retained block-coordinate schedule therefore alternated bounded L-BFGS-B updates of the physical scalars with the network fixed and Adam updates of the network with the scalars fixed.

When several flow conditions contribute to one objective, their weights must balance the aggregate loss contributed by each trajectory rather than only its pointwise error. We therefore equalized the integrated velocity-loss contributions of the two FENE-P rates and weighted the Saramito drives inversely by their observed velocity energies. In the contraction geometry, a weighting function defined entirely from the geometry emphasized the mixed shear and extension surrounding the contraction without requiring the unobserved conformation or polymer stress. In preliminary comparisons, emphasizing this region led to faster and stronger recovery of the Giesekus nonlinearity than an unweighted velocity loss. The instantaneous trainings showed the same effect: their data term's support was restricted to the region surrounding the constriction, and in a controlled re-training of the seven Table~\ref{tab:tbnn-cy-extraction} fluids with the support extended to the full fluid domain (one seed per fluid, all other settings unchanged), recovery of \(n\) and \(k\) degraded, in some cases only very slightly but in the most strongly shear-thinning cases by up to $30\%$, consistent with the abundant low-shear core diluting the identifiability signal. Numerical stability provided a separate consideration when adding the high-rate FENE-P trajectory: \(U=0.5\) has one-eighth the Deborah number of \(U=4\), limiting the initial conformation stretch while the scalars and closure departed from their Oldroyd-B initialization. The \(U=4\) trajectory was therefore activated after the first optimization block. This curriculum was used as a stability safeguard and was not interpreted as improving finite-extensibility identifiability.

The stability of the learned closure outside the constitutive states sampled during training was also important. In early instantaneous implementations, an unconstrained viscosity network could fit the sampled shear-rate range while developing a sharp response immediately outside it; when even a small region of the evolving flow entered that regime, the forward simulation became unstable. This failure mode motivated the smooth, positive viscosity parameterization and slope and curvature safeguards used for the instantaneous TBNN. The memory networks were likewise initialized at the Oldroyd-B limit, providing a stable admissible state while leaving the Giesekus mobility anisotropy, FENE-P finite extensibility, and other nonlinear physics to be learned from the velocity trajectories. Frame invariance, non-negative dissipation, and positive-definite conformation evolution were then enforced through the architecture rather than inferred from the sampled data. For the yield-stress extension, the positive-semidefinite floor was applied to the base mobility before multiplication by the yield factor; applying the operations in the opposite order reintroduced a finite relaxation mobility below yield and produced persistent creep.

Finally, all gradient-based trainings used 64-bit floating-point arithmetic for the memory TBNN. Preliminary single-precision attempts became numerically unstable, and we therefore retained double precision throughout.

\subsection{Computational benchmarks and scaling considerations}
\label{sec:si_computational_benchmarks}

Supplementary Table~\ref{tab:si_computational_cost} separates the cost of one standardized objective evaluation from the total wall time of the published optimizations. Each TBNN was paired with its corresponding analytic closure while holding the grid, timestep sequence, flow conditions, rematerialization schedule, and scalar objective fixed. Timings were measured in double precision on a single NVIDIA A100-SXM4-80GB after compilation. Despite increasing the differentiated dimension from \(4\)--\(5\) analytic parameters to \(646\) instantaneous TBNN or \(3{,}651\) memory TBNN weights, the TBNNs increased forward time by factors of only \(1.23\)--\(1.60\) and value-and-gradient time by \(1.22\)--\(1.56\). Thus, the coupled fluid calculation, rather than the number of constitutive parameters alone, remained the dominant per-evaluation cost.

The larger overhead appeared in peak adjoint memory, which increased by a factor of \(5.57\) for the instantaneous TBNN and by \(2.61\)--\(2.75\) for the memory TBNNs. The larger instantaneous ratio does not reflect a larger network: its rematerialized blocks contained 400 physical timesteps, compared with 50 for the viscoelastic contraction and 10 for the elastoviscoplastic channel. Neural activations and their derivative graph are repeated over every cell and timestep within such a block, so closure complexity and rematerialization length jointly determine peak memory. Sampling additional observations from a trajectory already advanced by the solver adds only to loss accumulation and therefore does not meaningfully increase training time, whereas an additional flow condition requires another forward trajectory. Accordingly, evaluating two FENE-P flow rates sequentially doubled the forward and gradient times relative to one rate, while peak memory changed only from \(20.69\) to \(21.22\) GB because the two trajectories were not retained simultaneously.

Total optimization time hence reflects the number of objective evaluations and the portion of each trajectory included in the reverse pass. The recorded instantaneous training required \(3.72\) h for 70 accepted network updates and differentiated only the steady-state portion of the trajectory. The representative Giesekus training required \(34.40\) h for 600 network updates interleaved with 319 scalar value-and-gradient evaluations while differentiating the complete transient. The dual-rate FENE-P calculations required approximately twice the single-rate wall time because each optimization step advanced two trajectories sequentially. Direct calibration of the known FENE-P form required only 20--26 value-and-gradient evaluations in this synthetic test. This difference is not an intrinsic cost ratio between constitutive discovery and parameter fitting: the TBNN calculations used fixed optimization schedules rather than a shared discrepancy-based stopping criterion, and direct calibration presupposes the correct constitutive family and must be repeated for each candidate model. For experimental data, terminating optimization once the discrepancy reaches the measurement-noise floor could substantially reduce the number of TBNN evaluations, after which the learned closure can be interrogated against multiple constitutive families without repeating the inverse flow calculation.

At fixed timestep, solver iteration count, and rematerialization schedule, the leading cell-local work and state storage scale approximately with the number of grid cells and physical timesteps. Extending an \(N_x\times N_y\) planar calculation to \(N_x\times N_y\times N_z\) at comparable resolution therefore introduces an approximately \(N_z\)-fold increase in these local contributions, together with a third velocity component and the two additional off-plane conformation components \(A_{xz}\) and \(A_{yz}\). These estimates are structural rather than measured because a matched two- and three-dimensional benchmark was not performed. The dominant additions are nevertheless naturally distributable: the spatial state and pointwise constitutive evaluations can be partitioned across accelerators, while independent flow conditions can be evaluated concurrently. Peak reverse-mode memory can be controlled through shorter rematerialization intervals, which recompute intermediate states during the reverse pass instead of retaining them. Finally, the present calculations used conservative fixed optimization schedules and sequential flow conditions and were not optimized for throughput; parallel condition evaluation, distributed spatial execution, and preconditioned or multilevel flow solvers provide complementary routes to larger three-dimensional calculations. The instantaneous TBNN would require the complete three-dimensional invariant and tensor basis, whereas the memory TBNN already uses three-dimensional conformation invariants and isotropic tensor bases and therefore requires no larger parameterization merely to operate in three dimensions.

\subsection{Minimal forcing model fitting}\label{sec:si_minimal_forcing}

\begin{figure}[t!]
\centering
\includegraphics[width=0.5\linewidth]{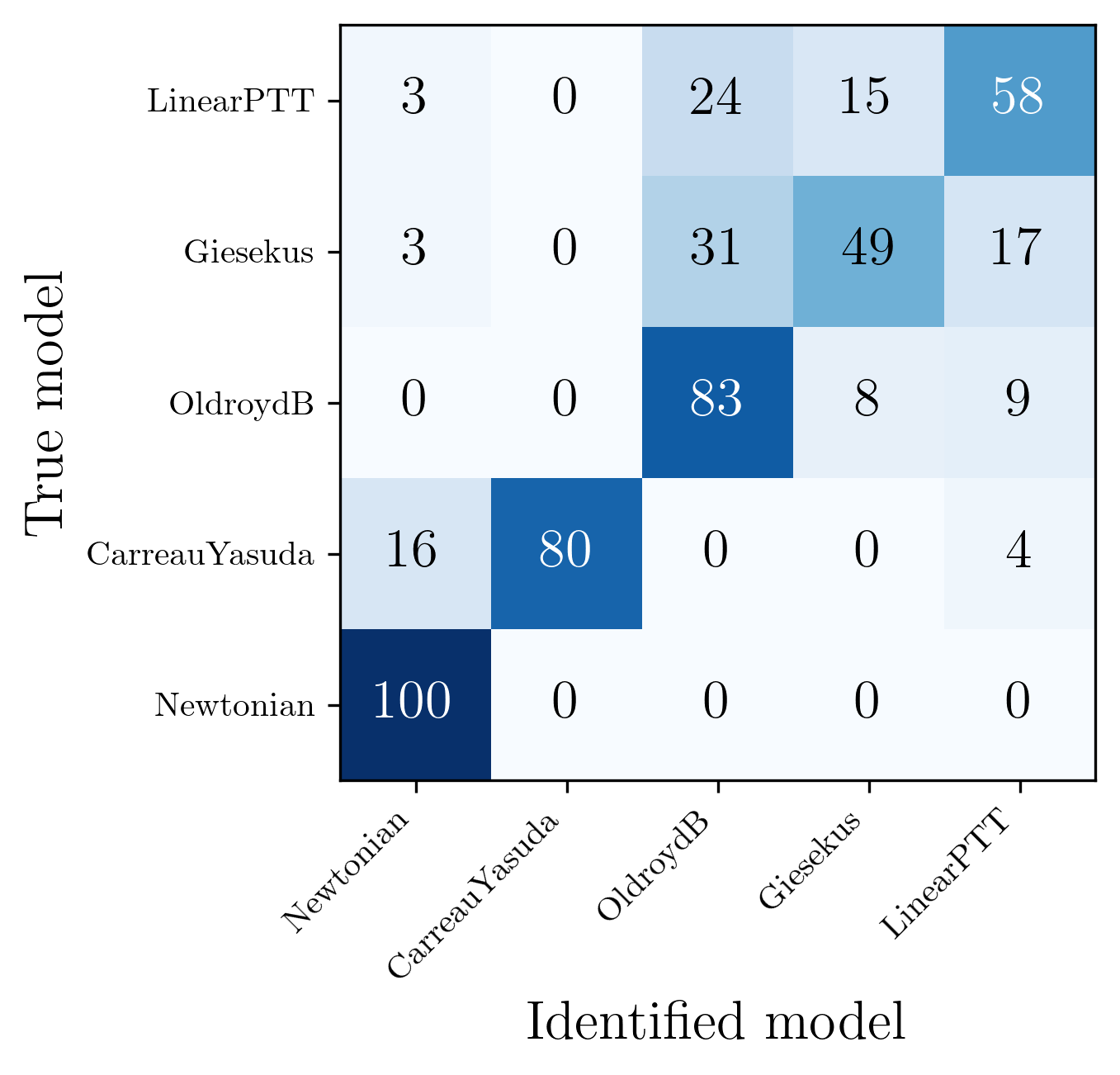}
\caption{\textbf{Model selection under minimal forcing.} Model selection results when tested against a single input forcing $\dot{\gamma} = \sin(t)$. Compared with the results in Fig.~\ref{fig:fittingresponse}, the true models are not able to be identified with nearly as much precision, due to the lack of low and high shear rates present in the data.
}
\label{fig:sfig7}
\end{figure}

All the results we present with regards to model selection from one-dimensional data are conditioned on the observed data. Therefore, they are implicitly conditioned on the experiments used to generate that data. In the main text, we fit the models simultaneously to 12 experimental traces generated by the combinations of forcing amplitude $f\in[0.01,0.1,1.0,10.0]$ and forcing frequency $\omega \in [0.33,1.0,2.0]$. These represent a large sweep in potential amplitude space and a smaller distribution of frequency space. To test the effects of the choice of experimental protocol on fitting results, we repeated the same random parameter identification experiment on a new set of samples with only one experiment: $f=1.0$ and $\omega = 1.0$. The results of this repeated experiment are shown in Fig. \ref{fig:sfig7} as a confusion matrix. We find that limiting the experimental data significantly reduces the model identification accuracy for all cases except the Newtonian baseline. This is likely due to the fact that the simplified experiments do not sweep sufficient shear rates to fully identify all components of any given ground-truth model, especially with parameters drawn from the distributions in Table~\ref{tab:parameter-ranges}. These results open the door to optimized experimental design, in which forcing amplitudes and frequencies are selected to maximally discriminate between competing constitutive models, which we leave to future work.

\subsection{Model identification of real gel experiments and comparison to RUDEs} \label{sec:gel_si}
To evaluate our model-identification framework on experimental data, we apply it to measurements from a metal-crosslinked polymer hydrogel previously used to benchmark RUDEs \cite{Lennon2023PNAS}. As stated in the original reference, the material has a single mode Maxwell-like response under SAOS (small amplitude oscillatory shear) experiments, with $\eta_s = 0$, $\eta_p = 20,934.85\ \textrm{Pa}\cdot\textrm{s}$ and $\lambda = 0.557\ \textrm{s}$ as fit from a SAOS experiment. The hydrogel data is given as four LAOS (large amplitude oscillatory shear) experiments in the form $\sigma_{12} = f\sin(\omega t)$ with $\omega = 1\ \textrm{s}^{-1}$ and $f \in [1,2,3,4]\ \textrm{kPa}$. Consistent with the original paper, we omit the $f = 3 \ \textrm{kPa}$ experiment as testing data to determine if the RUDE can interpolate between training examples. Further, the normal stress difference $N_1 = \sigma_{11} - \sigma_{22}$ is provided as a method of testing the RUDE's ability to extrapolate beyond its training. Unlike the other results in this paper, this data is stress controlled, where the stress is applied and the strain $\gamma$ measured. Further, in training the RUDE, Lennon et al. first fit the Maxwell parameters using their independent SAOS data to nondimensionalize the LAOS data before training the network.

\begin{figure}
    \centering
    \includegraphics[width=0.8\linewidth]{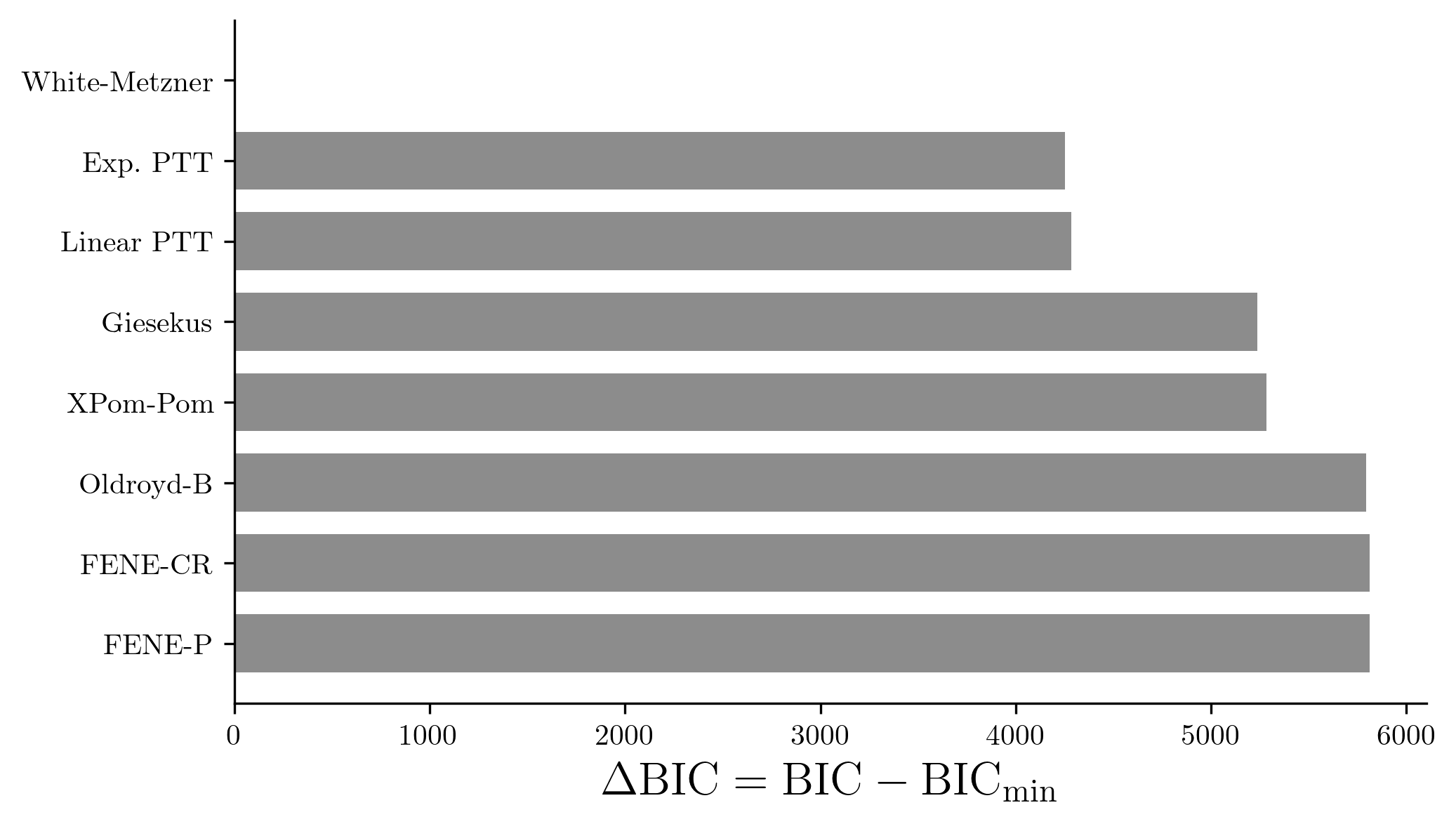}
    \caption{\textbf{Model selection for gel polymer.} BIC values of various viscoelastic models fit to gel polymer data, plotted relative to the BIC value of the best model, White-Metzner. Here, a lower score represents a better model agreement with the underlying data.}
    \label{fig:sfig8}
\end{figure}

The RUDE as implemented in \cite{Lennon2023PNAS} is instantiated as two networks, one to learn the linear tensor terms and the other to learn the non-linear terms. Rather than directly predict the strain given the stress inputs, the RUDE is trained by first approximating the strain rate $\dot{\gamma}$ by applying finite differences to the strain data. The RUDE then tries to recover the known stress input from this recovered strain rate. We follow this protocol to fit a series of viscoelastic models to the data using our differentiable code. The relative BIC values of the different models are shown in Fig.~\ref{fig:sfig8}. While none of the models from our synthetic experiment fit the data particularly well, when we expanded our experiment to include almost all of the viscoelastic models implemented in OpenFOAM, we found that a White-Metzner model \cite{white1963development} best described the data, with parameters $\lambda$ = 0.64 s, $\eta_{p}$ = $1.5 \times 10^4$ Pa·s, $K$ = 2.1 s, $L$ = 1.8 s, $n$ = 0.38, $m$ = 4.2, $a$ = 1.0, and $b$ = 6.8 (using notation consistent with the OpenFOAM RheoTool documentation).

\begin{figure}
    \centering
    \includegraphics[width=\linewidth]{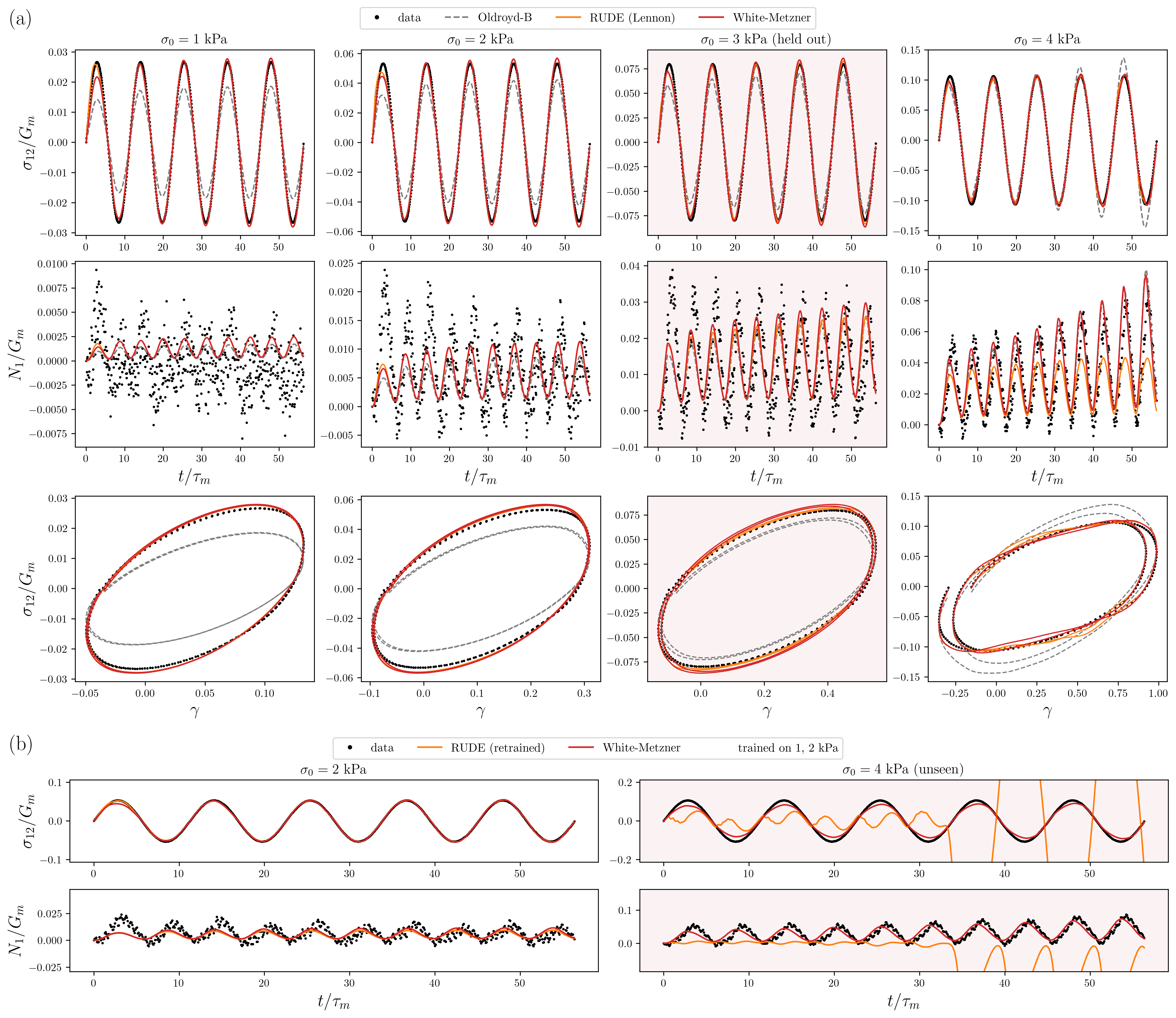}
    \caption{\textbf{Polymer gel model predictions.} Comparison of a RUDE and a White-Metzner model at describing the material response of a polymer gel. In (a) we show the response of our best-fit White-Metzner model, Lennon et al.'s RUDE, our best-fit Oldroyd-B model, and the experimental data. Note that for all of these models the $\sigma_0=3$ kPa experiment was withheld from the training to provide testing data. The White-Metzner model generally performs better than the RUDE at predicting the held-out $N_1$ data, especially at $\sigma_0=4$ kPa. In (b) we compare a retrained RUDE (which we train according to Lennon et al.'s protocol) to a White-Metzner model where both models are trained only on $\sigma_0 = 1$, 2 kPa and are asked to extrapolate to predict the material response at $\sigma_0 = 4$ kPa. Note that the RUDE is unable to extrapolate beyond the training data and provides an unphysical prediction, while the White-Metzner model does a reasonable job at predicting this unseen data.}
    \label{fig:sfig9}
\end{figure}

A comparison of the best Oldroyd-B model, the best White-Metzner model, and Lennon et al.'s trained RUDE at fitting the experimental data (both $\sigma_{12}$ and $N_1$) is provided in Fig.~\ref{fig:sfig9}(a). In this case, the White-Metzner model fit the Maxwell parameters directly from the LAOS data and did not have access to the SAOS fit. Both the White-Metzner model and the RUDE do a good job at tracking the shear stress, including in the withheld testing sample, but the White-Metzner model does a better job at predicting normal stress difference at higher input stresses. The stress--strain Lissajous curves at 4 kPa also exhibit fewer oscillations than those predicted by the RUDE.

We also tested the RUDE architecture's ability to extrapolate to unseen forcing conditions. To do so, we retrained a new RUDE based on Lennon et al.'s protocol that only has access to the 1 and 2 kPa data. We also fit a White-Metzner model to the same limited data set and then plotted how both the RUDE and model did at predicting the material response at 4 kPa in Fig.~\ref{fig:sfig9}(b). The network is unable to extrapolate beyond its training data, producing unphysical blow up solutions when the stress is doubled from the region in which it was trained, while the standard viscoelastic model still predicts the material response.

\begin{figure}
    \centering
    \includegraphics[width=\linewidth]{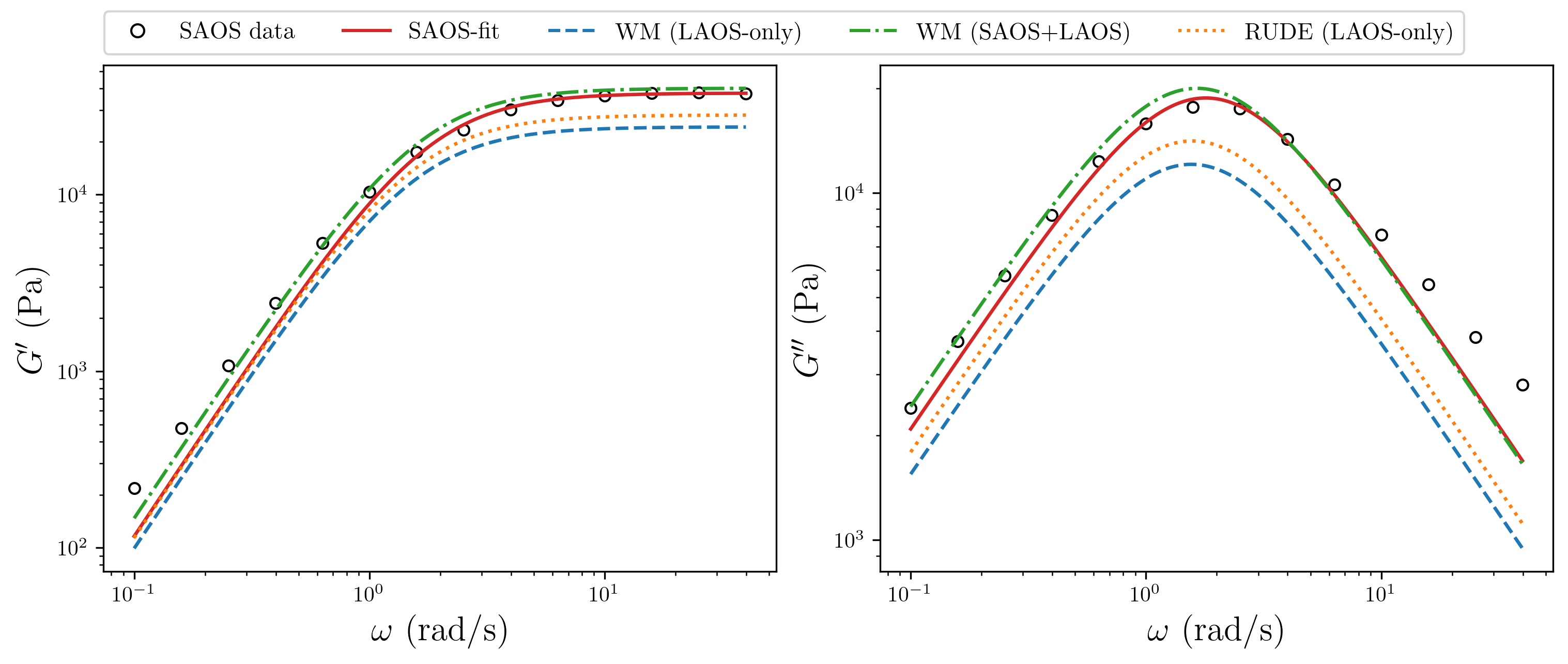}
    \caption{\textbf{Effect of training on SAOS data.} We plot a comparison of the experimental SAOS data for the gel polymer with the equivalent SAOS parameterization obtained in several ways, including a traditional SAOS-fit as used by the RUDE, the parameter results predicted by White-Metzner fitting only on the LAOS data, the White-Metzner results obtained by fitting jointly on the SAOS and LAOS data, and the parameters obtained by a retrained RUDE where the Maxwell parameters are left as unknowns during the fit. In general, the joint fit appears to track the data the best, consistent with the fit being performed with the maximum possible information available.}
    \label{fig:sfig10}
\end{figure}

Finally, we wanted to compare the effects of pre-fitting the Maxwell parameters from SAOS data and learning them directly from the LAOS data. While both protocols produced nearly identical predictions for the LAOS experiments, they differed when used to predict the SAOS data, as shown in Fig.~\ref{fig:sfig10}. The LAOS-only fitting protocol when used both with White-Metzner and with a RUDE where the Maxwell parameters are left free to fit from the LAOS data underpredict the SAOS experimental results. However, this can easily be rectified by including the SAOS experimental data in the White-Metzner loss function and jointly fitting both the SAOS data and LAOS data. Including the full set of available experimental information pins the solve onto the SAOS data while still fitting the LAOS experiments well all run in a single optimization loop, forgoing the need for multi-step fitting and enabling better fits at low frequencies than the pure SAOS analysis.

\clearpage
\section*{Supplementary Figures}
\FloatBarrier   

\begin{figure}[!ht]
\centering
\includegraphics[width=\linewidth]{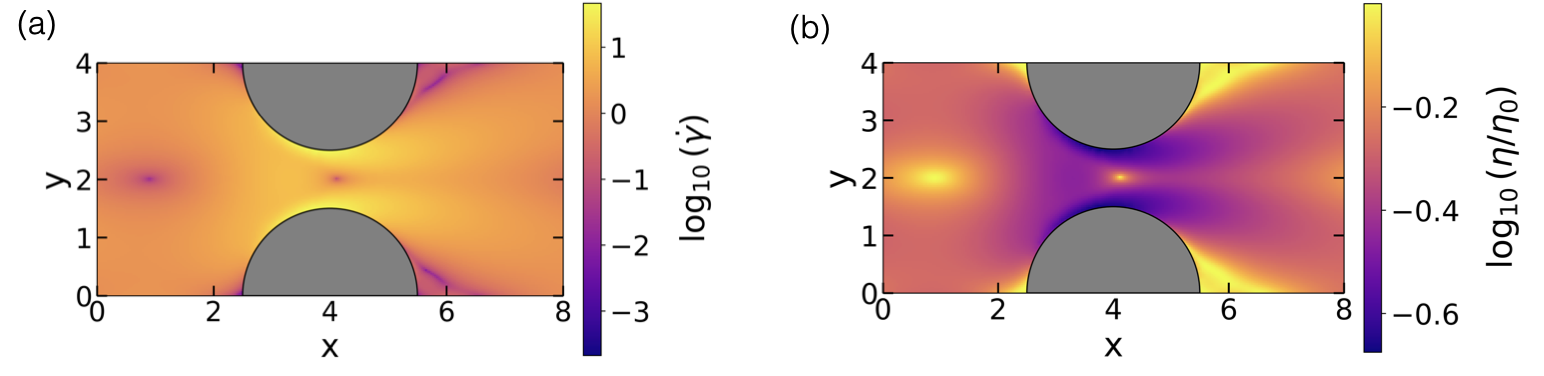}
\caption{\textbf{Example kinematic and rheological fields used for training.}
(a) Local strain-rate magnitude $\log_{10}(\dot{\gamma})$ showing the broad range of kinematic conditions sampled within the constriction geometry for a shear-thinning fluid with parameters $\eta_0 = 1.0$, $\eta_\infty = 0.02$, $k = 5.0$, $n = 0.7$, and $a = 2.0$.
(b) Corresponding viscosity field $\log_{10}(\eta/\eta_0)$ from the Carreau--Yasuda model, demonstrating a wide range of local viscosities and strong shear thinning in the throat region.}
\label{fig:sfig1}
\end{figure}

\begin{figure}[!ht]
\centering
\includegraphics[width=\linewidth]{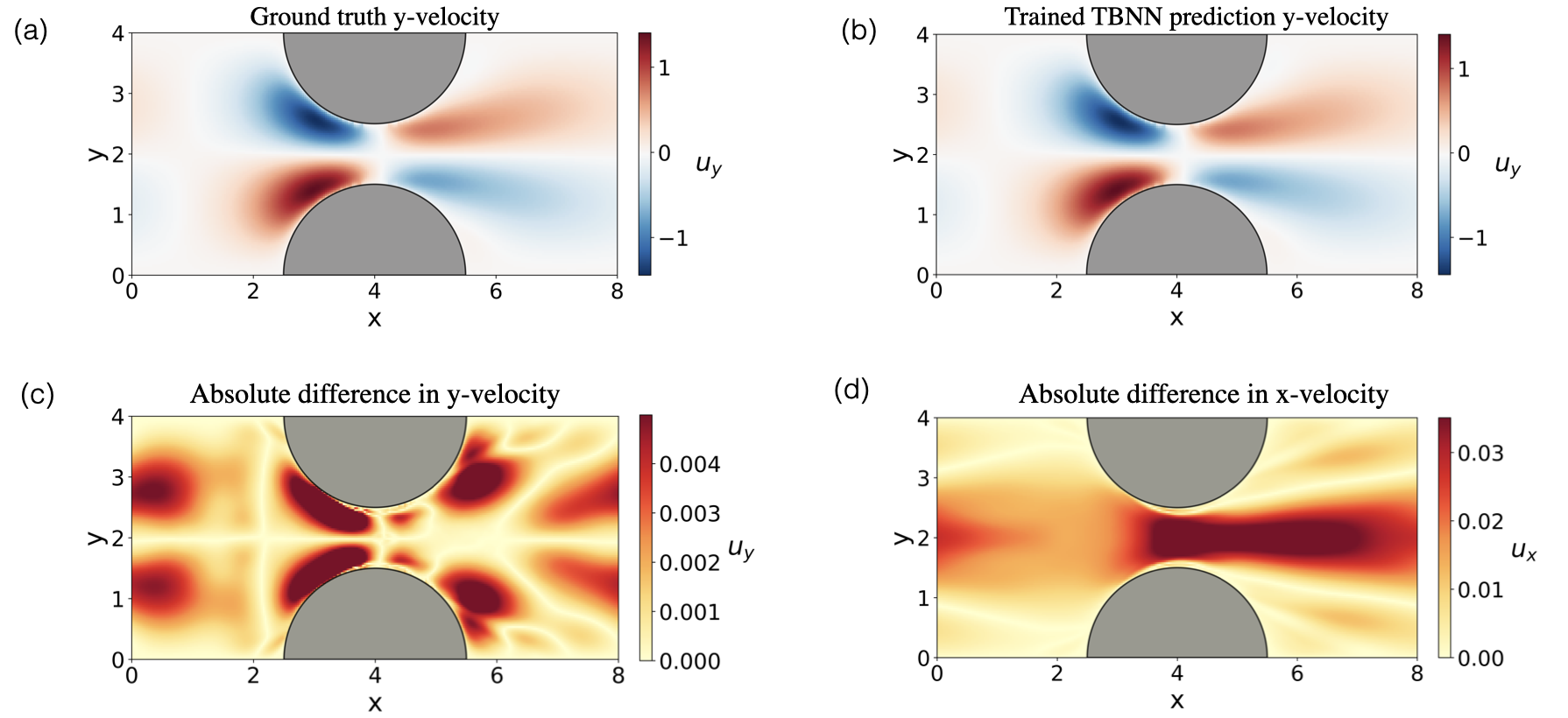}
\caption{\textbf{Velocity-field reconstruction and errors.}
Comparison between ground-truth and TBNN-predicted velocity components for the constriction geometry.
(a,b) $y$-velocity fields showing near-perfect recovery of the cross-stream flow structure ($u_y$).
(c,d) Absolute differences for $u_y$ and $u_x$ (the corresponding ground-truth and TBNN $u_x$ fields are shown in Fig.~\ref{fig:tbnn_learning} of the main text), indicating very low absolute errors compared with the flow magnitude.}
\label{fig:sfig2}
\end{figure}

\begin{figure}[t!]
\centering
\includegraphics[width=\linewidth]{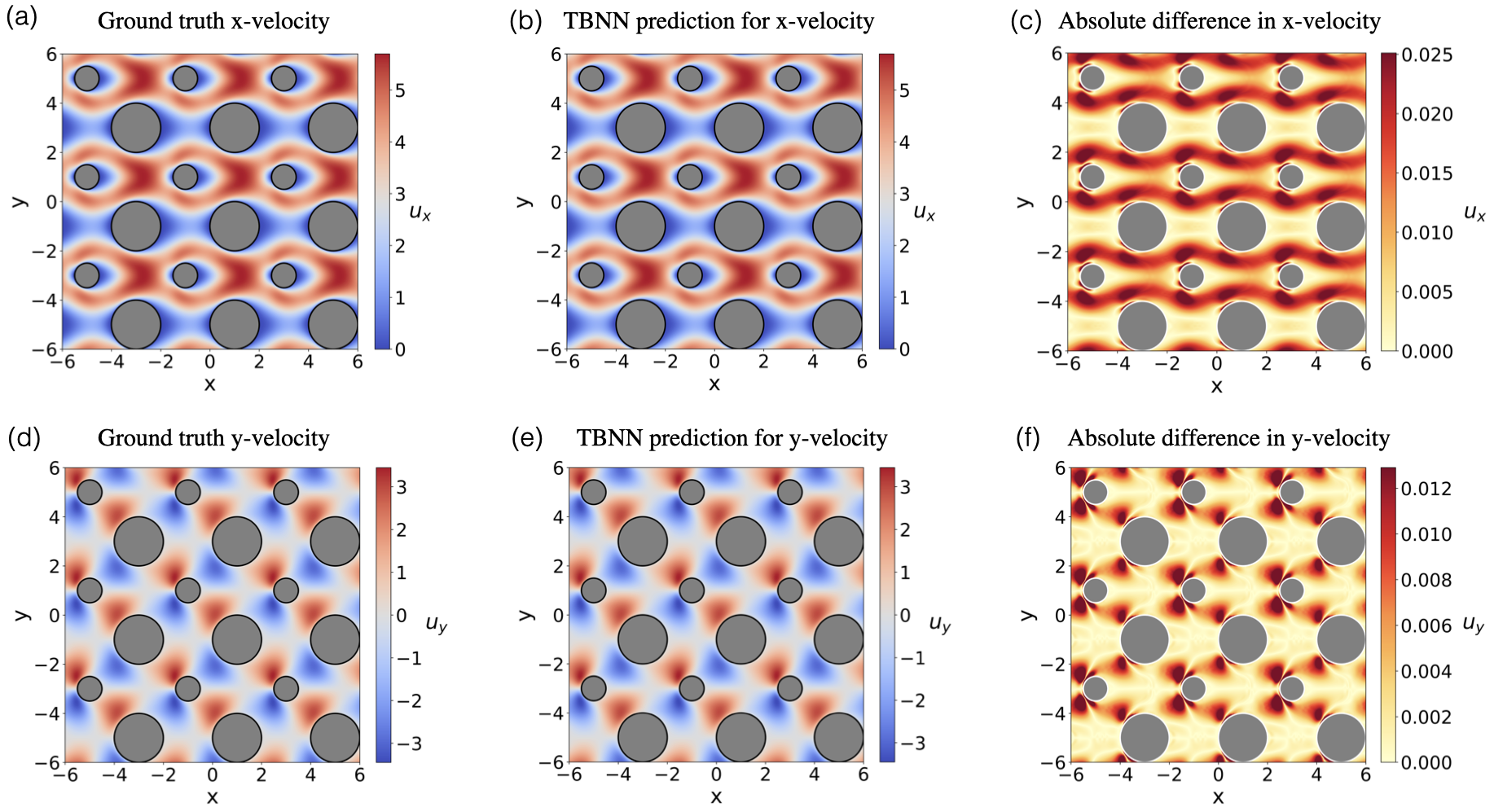}
\caption{\textbf{Generalization to unseen geometry.}
Comparison between TBNN predictions and ground-truth fields for flow through a bidisperse porous array.
(a,b) Ground-truth and predicted $u_x$ fields.
(c) Absolute error in $u_x$.
(d--f) Analogous comparison for $u_y$.
The learned closure transfers successfully to an out-of-training pressure drive in a geometry with different boundary conditions, preserving spatial structure and amplitude of both velocity components.}
\label{fig:sfig3}
\end{figure}

\begin{figure}[t!]
    \centering
    \includegraphics[width=\linewidth]{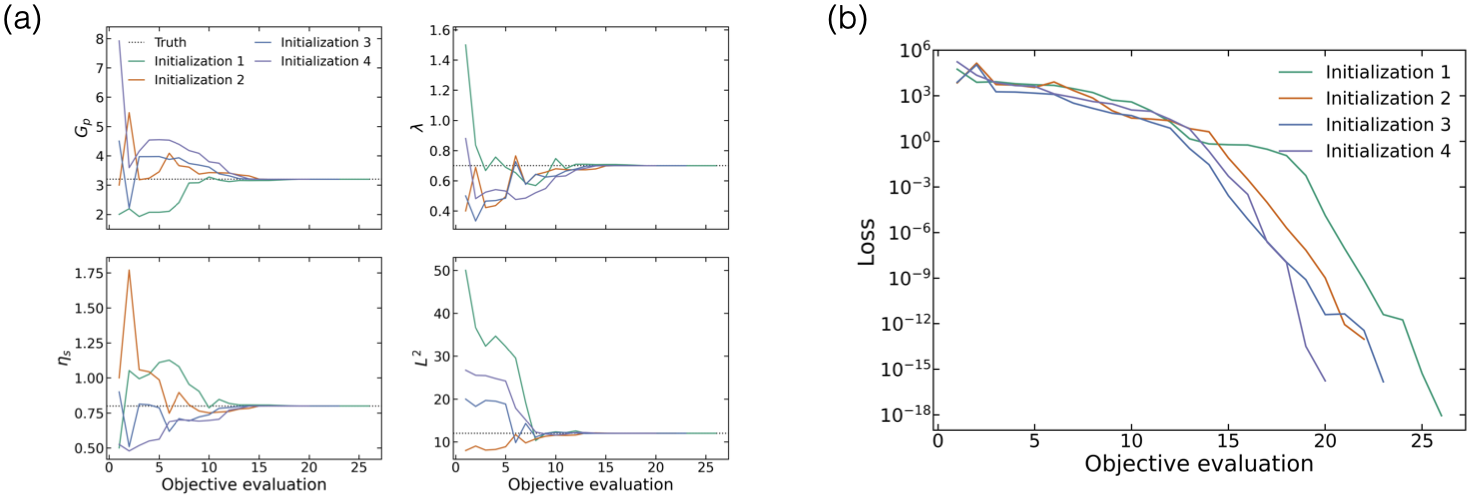}
    \caption{\textbf{Direct known-form calibration of the FENE-P model.} The four FENE-P scalars are fitted to the same single-rate, velocity-only \(U=0.5\) data used for the representative memory TBNN by gradient-based optimization through the same differentiable flow solver, with the FENE-P model in place of the neural-network closure. Optimization is performed in the logarithms of \(G_p\), \(\lambda\), \(\eta_s\), and \(L^2\), while the panels report their physical values. (a) Parameter trajectories from four distinct initializations; dashed lines mark the ground-truth values \(G_p=3.2\), \(\lambda=0.7\), \(\eta_s=0.8\), and \(L^2=12\). All four initializations recover every parameter within \(20\)--\(26\) objective evaluations. (b) Corresponding losses, which decrease to near machine precision. When the constitutive family is known, the velocity data therefore determine all four parameters; the less accurate readback from the model-agnostic TBNN instead reflects the broader task of learning both the constitutive form and its parameters from the finite range of states visited by the training flow, rather than insufficient information in the data.}
    \label{fig:sfig_fenep_direct}
\end{figure}

\begin{figure}
    \centering
    \includegraphics[width=\linewidth]{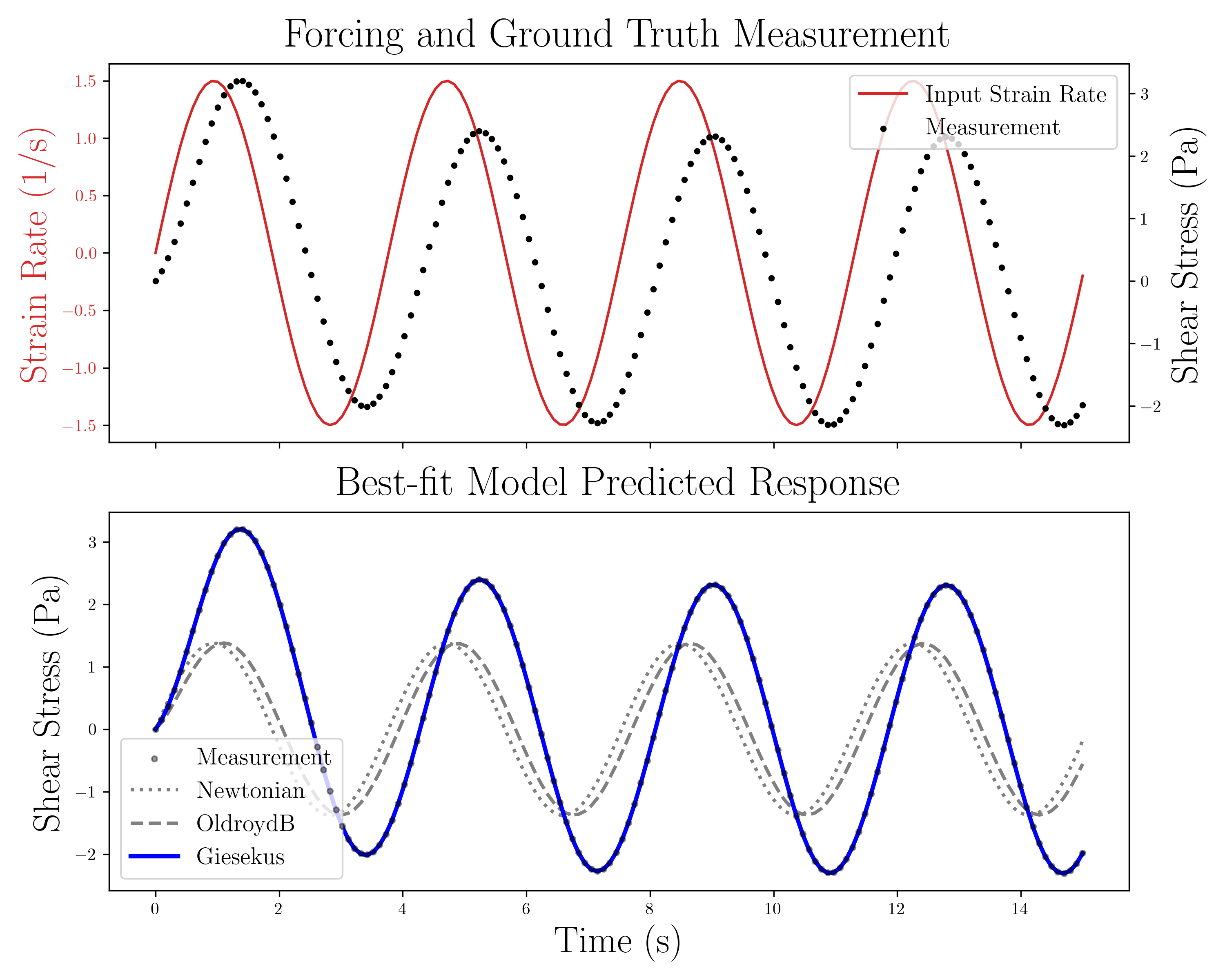}
    \caption{\textbf{Demonstration of fitting different constitutive models to the same ground-truth data.} Several candidate constitutive models are fit jointly to all 12 forcing conditions of a synthetic Giesekus rheometer dataset (Methods Sec.~\ref{sec:digital_rheometer}) and evaluated on their ability to predict the stress response under an unseen forcing (top); the corresponding best-fit predictions are shown in the bottom panel. The correctly identified Giesekus model reproduces the held-out response, while incorrect models diverge.}    \label{fig:demonstration}
\end{figure}

\clearpage
\section*{Supplementary Tables}
\FloatBarrier   

\sisetup{
  detect-weight=true,
  detect-family=true,
  round-mode=places,
  round-precision=3
}

\begin{table}[h!]
\centering
\small
\setlength{\tabcolsep}{5pt}
\renewcommand{\arraystretch}{1.15}
\begin{tabular}{@{}l l
                S[table-format=1.3]  
                S[table-format=1.3]  
                S[table-format=2.3]  
                S[table-format=1.3]  
                S[table-format=1.3]  
                c                     
                c                     
                @{}}
\toprule
\multirow{2}{*}{Run} & \multirow{2}{*}{Source}
& \multicolumn{5}{c}{Carreau--Yasuda parameters}
& \multirow{2}{*}{$\eta_N$}
& \multirow{2}{*}{$\Delta\mathrm{BIC}$} \\
\cmidrule(lr){3-7}
& & {$\eta_0$} & {$\eta_\infty$} & {$k$} & {$n$} & {$a$} & & \\
\midrule
\multirow{2}{*}{1}
 & GT      & 1.000 & 0.020 & 5.000  & 0.700 & 2.000 & \multirow{2}{*}{0.340} & \multirow{2}{*}{915.35} \\
 & Learned & 0.985 & 0.021 & 4.941  & 0.701 & 1.615 &                         &                          \\
\addlinespace
\multirow{2}{*}{2}
 & GT      & 1.000 & 0.015 & 3.000  & 0.600 & 2.000 & \multirow{2}{*}{0.295} & \multirow{2}{*}{814.99} \\
 & Learned & 0.972 & 0.017 & 3.031  & 0.615 & 2.360 &                         &                                    \\
\addlinespace
\multirow{2}{*}{3}
 & GT      & 1.000 & 0.012 & 1.000  & 0.600 & 2.000 & \multirow{2}{*}{0.456} & \multirow{2}{*}{434.86} \\
 & Learned & 0.992 & 0.013 & 0.983  & 0.639 & 1.011 &                         &                          \\
\addlinespace
\multirow{2}{*}{4}
 & GT      & 1.000 & 0.010 & 7.000  & 0.600 & 2.000 & \multirow{2}{*}{0.231} & \multirow{2}{*}{221.63} \\
 & Learned & 1.037 & 0.010 & 6.626  & 0.602 & 2.067 &                         &                          \\
\addlinespace
\multirow{2}{*}{5}
 & GT      & 1.000 & 0.040 & 5.000  & 0.800 & 2.000 & \multirow{2}{*}{0.495} & \multirow{2}{*}{526.63} \\
 & Learned & 1.018 & 0.040 & 4.820  & 0.792 & 2.018 &                         &                          \\
\addlinespace
\multirow{2}{*}{6}
 & GT      & 1.000 & 0.045 & 10.000 & 0.800 & 2.000 & \multirow{2}{*}{0.441} & \multirow{2}{*}{670.03} \\
 & Learned & 1.020 & 0.045 & 9.623  & 0.793 & 2.011 &                         &                          \\
\addlinespace
\multirow{2}{*}{7}
 & GT      & 1.000 & 0.050 & 15.000 & 0.800 & 2.000 & \multirow{2}{*}{0.416} & \multirow{2}{*}{555.15} \\
 & Learned & 1.022 & 0.050 & 14.404 & 0.795 & 2.021 &                         &                          \\
\bottomrule
\end{tabular}
\caption{\textbf{Model extraction from an instantaneous TBNN.} For each run, ground-truth (GT) Carreau--Yasuda (CY) parameters and CY parameters fitted to the instantaneous TBNN output (Learned). $\eta_{\mathrm{N}}$ is the best-fit Newtonian viscosity to the same instantaneous TBNN output. $\Delta\mathrm{BIC} \equiv \mathrm{BIC}_{\text{Newtonian}} - \mathrm{BIC}_{\text{CY}}$ (positive favors CY).}
\label{tab:tbnn-cy-extraction}
\end{table}


\addtolength{\tabcolsep}{-0.4em}
\begin{table*}[t!]
\centering
\setlength{\tabcolsep}{6pt}  
\begin{tabular}{cccccc}
\toprule
 & Newtonian & Carreau-Yasuda & OldroydB & Giesekus & LinearPTT \\
\midrule
$\eta_s$ & 1.000 & --- & 1.000 & 1.000 & 1.000 \\
$\eta_p$ & --- & --- & 1.000 & 1.000 & 0.919 \\
$\lambda$ & --- & --- & 1.000 & 1.000 & 0.910 \\
$\alpha$ & --- & --- & --- & 1.000 & --- \\
$\eta_0$ & --- & 1.042 & --- & --- & --- \\
$\eta_\infty$ & --- & 1.221 & --- & --- & --- \\
$a$ & --- & 0.882 & --- & --- & --- \\
$k$ & --- & 1.100 & --- & --- & --- \\
$n$ & --- & 1.006 & --- & --- & --- \\
$\zeta$ & --- & --- & --- & --- & 1.001 \\
$\epsilon$ & --- & --- & --- & --- & 0.997 \\
\bottomrule
\end{tabular}
\caption{\textbf{Geometric median factor (estimate/true) for each parameter by model.} Values $<1$ indicate median underestimation; $>1$ overestimation. For almost all models our method came within one percent of the true value. In the case of Carreau-Yasuda, our method produced reasonable parameter estimates, with deviations consistent with limited sampling of the high-shear regime by the applied forcing protocol.}
\label{tab:median-factors}
\end{table*}

\clearpage

\begin{table}[t!]
\centering
\setlength{\tabcolsep}{6pt}  
\begin{tabular}{llcc}
\toprule
\textbf{Model} & \textbf{Parameter} & \textbf{Range} & \textbf{Distribution} \\
\midrule
Newtonian & Viscosity ($\eta$) & [0.1, 10.0] & Log \\
\midrule
\multirow{5}{*}{Carreau-Yasuda} & Zero-shear viscosity ($\eta_0$) & [1.0, 100.0] & Log \\
 & Infinite-shear viscosity ($\eta_\infty$) & [0.01, 0.1] & Log \\
 & Consistency index ($k$) & [0.1, 10.0] & Uniform \\
 & Power-law index ($n$) & [0.2, 0.7] & Uniform \\
 & Transition parameter ($a$) & [0.5, 3.0] & Uniform \\
\midrule
\multirow{3}{*}{Oldroyd-B} & Polymer viscosity ($\eta_p$) & [1.0, 10.0] & Uniform \\
 & Relaxation time ($\lambda$) & [1.0, 10.0] & Uniform \\
 & Solvent viscosity ($\eta_s$) & [0.1, 10.0] & Uniform \\
\midrule
\multirow{4}{*}{Giesekus} & Polymer viscosity ($\eta_p$) & [0.1, 10.0] & Log \\
 & Relaxation time ($\lambda$) & [1.0, 10.0] & Log \\
 & Solvent viscosity ($\eta_s$) & [0.1, 10.0] & Log \\
 & Mobility factor ($\alpha$) & [0.01, 0.5] & Uniform \\
\midrule
\multirow{5}{*}{Linear PTT} & Polymer viscosity ($\eta_p$) & [0.1, 10.0] & Log \\
 & Relaxation time ($\lambda$) & [1.0, 10.0] & Log \\
 & Solvent viscosity ($\eta_s$) & [0.1, 10.0] & Log \\
 & Elongational parameter ($\zeta$) & [0.01, 0.2] & Uniform \\
 & Mobility parameter ($\epsilon$) & [0.01, 0.5] & Uniform \\
\bottomrule
\end{tabular}
\caption{\textbf{Parameter ranges and distributions for the random generation of ground truth models.} For each parameter, we list the range of values and the distribution from which it is sampled.}
\label{tab:parameter-ranges}
\end{table}

\clearpage

\begin{table*}[p]
\centering
\begingroup
\footnotesize
\setlength{\tabcolsep}{3.9pt}
\renewcommand{\arraystretch}{1.18}

\begin{tabular}{lrrrrrr}
\toprule
\multicolumn{7}{l}{\textbf{(a) Standardized objective evaluation}} \\
\addlinespace[2pt]
Model
&
Parameters
&
Remat (steps \(\times\) cells)
&
Conditions
&
Forward (s)
&
\(\nabla\) objective (s)
&
Peak memory (GB)
\\
\midrule
Carreau--Yasuda
& 5
& \(400\times32{,}768\)
& 1
& 194.8
& 218.5
& 4.38
\\
Instantaneous TBNN
& 646
& \(400\times32{,}768\)
& 1
& 240.6
& 271.9
& 24.42
\\

\addlinespace
FENE-P
& 4
& \(50\times32{,}768\)
& 1
& 14.63
& 88.45
& 7.60
\\
Memory TBNN (FENE-P data)
& 3651
& \(50\times32{,}768\)
& 1
& 21.50
& 136.7
& 20.69
\\

\addlinespace
Giesekus
& 4
& \(50\times32{,}768\)
& 1
& 13.49
& 86.60
& 7.51
\\
Memory TBNN (Giesekus data)
& 3651
& \(50\times32{,}768\)
& 1
& 21.64
& 134.9
& 20.69
\\

\addlinespace
FENE-P, dual rate
& 4
& \(50\times32{,}768\)
& 2
& 28.52
& 179.3
& 8.13
\\
Memory TBNN, dual rate
& 3651
& \(50\times32{,}768\)
& 2
& 42.89
& 273.1
& 21.22
\\

\addlinespace
Saramito, three drives
& 4
& \(10\times2{,}048\)
& 3
& 4.362
& 37.06
& 0.0967
\\
Memory TBNN, three drives
& 3651
& \(10\times2{,}048\)
& 3
& 5.380
& 45.21
& 0.2659
\\
\bottomrule
\end{tabular}

\vspace{1.0em}

\begin{tabular}{lrrr}
\toprule
\multicolumn{4}{l}{\textbf{(b) Production optimization}} \\
\addlinespace[2pt]
Training
&
Network updates
&
Scalar V\&G evaluations
&
Wall time (h)
\\
\midrule
Instantaneous TBNN
& 70
& 0
& \(3.72^{*}\)
\\
Memory TBNN, FENE-P
& 600
& 336
& 35.43
\\
Memory TBNN, Giesekus
& 600
& 319
& 34.40
\\
Memory TBNN, dual-rate FENE-P
& 600
& 295
& 65.88
\\
Memory TBNN, Saramito
& 240
& 83
& 4.28
\\
Direct FENE-P, single rate
& 0
& 20--26
& 0.476--0.640
\\
\bottomrule
\end{tabular}

\caption{\textbf{Computational cost of constitutive learning and matched analytic-model evaluations.}
(a) One complete production objective evaluated in double precision on a single NVIDIA A100-SXM4-80GB. Forward and gradient times are medians of five synchronized post-compilation evaluations of the same scalar objective. Parameters denotes the number differentiated in the corresponding gradient evaluation. Remat reports the physical timesteps and grid cells within each checkpointed block. Multiple conditions were evaluated sequentially. Peak memory is the clean-process JAX allocator peak, in decimal GB, through compilation and one value-and-gradient evaluation with XLA preallocation disabled. The instantaneous objective advanced 2,000 blocks and differentiated the final ten, when the flow profile was steady; the memory objectives differentiated all 400 blocks per contraction condition or all 84 blocks per Saramito drive. (b) Published production optimizations. Network updates are accepted Adam steps; scalar V\&G evaluations are L-BFGS-B function-and-gradient calls with the network fixed. Wall times exclude compilation for the memory and direct-model optimizations. The instantaneous value marked by an asterisk is the complete recorded runtime, which also includes compilation, Carreau--Yasuda reference generation, additional forward evaluations, and plotting. The direct FENE-P range covers four successful initializations.}
\label{tab:si_computational_cost}

\endgroup
\end{table*}

\clearpage

\end{document}